\documentclass[preprint,prd,aps,showpacs,showkeys,nofootinbib]{revtex4}
\usepackage{graphicx}
\usepackage{makecell}
\usepackage{color}
\usepackage{bm}
\usepackage{url}
\usepackage{float}
\usepackage{subfigure}
\begin{document}


\title{The Higgs boson in the CP-violating NB-LSSM}

\author{Xing-Xing Dong$^{1,2,3,4}$\footnote{dongxx@hbu.edu.cn},
Wen-Hui Zhang$^{1,2,3}$,
Cai Guo$^{1,2,3}$,
Shu-Min Zhao$^{1,2,3}$\footnote{zhaosm@hbu.edu.cn},
Tai-Fu Feng$^{1,2,3,5}$\footnote{fengtf@hbu.edu.cn}}
\affiliation{$^1$ College of Physics Science $\&$ Technology, Hebei University, Baoding, 071002, China\\
$^2$ Hebei Key Laboratory of High-precision Computation and Application of Quantum Field Theory, 071002, China\\
$^3$ Hebei Research Center of the Basic Discipline for Computational Physics, Baoding, 071002, China\\
$^4$ Departamento de F\'{i}sica and CFTP, Instituto Superior T\'{e}cnico, Universidade de Lisboa,
Av. Rovisco Pais 1, 1049-001 Lisboa, Portugal\\
$^5$ Department of Physics, Chongqing University, Chongqing, 401331, China}
\begin{abstract}
This study investigates the lightest and second-lightest Higgs bosons at around 95 GeV (which is only a hypothetical scenario) and 125 GeV respectively within the CP-violating next to minimum B-L supersymmetric model(NB-LSSM). In the NB-LSSM, the CP violation in the Higgs potential leads to the mixing mass terms between the CP-even and CP-odd Higgs fields. Thus, one has to consider a $(10 \times 10)$-dimensional mass matrix for the neutral Higgs boson. These potential mixing effects may lead to drastic variations on the neutral Higgs boson masses. Besides, the neutral Higgs bosons predicted in the NB-LSSM may strongly mix with one another, thereby significantly modifying the couplings of the Higgs bosons to the fermions or gauge bosons. It is found that the specific parameters $g_{YB}$, $\tan\beta$, $T_\kappa$, $T_\lambda, \cdot\cdot\cdot$ and CP-violating phases $\theta_{1,2,3,4,6,7,8}$ in the NB-LSSM affect the theoretical predictions on the Higgs boson mass and corresponding signal strengthes significantly. And the theoretical predictions on the signal strengthes of SM-like Higgs decay channels and excess signals at around 95 GeV are fitted well to the observed experimental data.
\end{abstract}
\keywords{Supersymmetric model, Higgs boson mass and decay, CP violation}
\pacs{12.60.Jv, 14.80.Da, 11.30.Er}

\maketitle

\section{Introduction\label{sec1}}
In 2012, the Large Hadron Collider (LHC) discovered the 125 GeV Higgs boson\cite{h0CMS,h0ATLAS}, whose properties are consistent well with the Standard Model (SM) predictions. This discovery is one of the most remarkable achievements in theoretical physics because it indicates that all of the
fundamental particles predicted by the SM are observed.

At present, the measured Higgs boson mass is $m_h^{125}=125.20\pm0.11$ GeV\cite{PDG2024}. And one of the prime goals of the current LHC programme is to investigate whether the detected Higgs boson is the only fundamental scalar particle or the part of new physics (NP). Therefore, it is necessary to perform precise measurements on various Higgs decay modes. We have learnt about the properties of the Higgs boson much more with the experimental data corresponding to the production of Higgs bosons. For example, the CMS and ATLAS collaborations have observed the Higgs boson in numerous bosonic and fermionic decay channels\cite{PDG2024,h02gamma,h02gamma2W2Z2b2tau1,h02gamma2W2Z2b2tau2}, established its spin-parity quantum numbers, determined its mass and measured its production cross-sections in various modes. The inclusive Higgs boson signal strength in the diphoton channel is measured to be $\mu_{\gamma\gamma}^{exp}(125)=1.10\pm0.06$\cite{h02gamma,h02gamma2W2Z2b2tau1,h02gamma2W2Z2b2tau2,h02gamma2W2b2tau}. The inclusive Higgs boson signal strengths for vector-boson fusion, associated production with the $W$ or $Z$ boson, are measured to be $\mu_{WW^*}^{exp}(125)=1.00\pm0.08$\cite{h02gamma2W2Z2b2tau1,h02gamma2W2Z2b2tau2,h02gamma2W2b2tau}, $\mu_{ZZ}^{exp}(125)=1.02\pm0.08$\cite{h02gamma2W2Z2b2tau1,h02gamma2W2Z2b2tau2,h02Z}, and bottom, tauon associated production processes are reported $\mu_{b\bar{b}}^{exp}(125)=0.99\pm0.12$\cite{h02gamma2W2Z2b2tau1,h02gamma2W2Z2b2tau2,h02gamma2W2b2tau,h02b1,h02b2}, $\mu_{\tau\bar{\tau}}^{exp}(125)=0.91\pm0.09$\cite{h02gamma2W2Z2b2tau1,h02gamma2W2Z2b2tau2,h02gamma2W2b2tau,h02tau}.

Besides, although we have not discovered any other new scalar particle at the LHC, the searches for the Higgs boson with mass below 125 GeV have been performed at the LEP, Tevatron and LHC at CERN\cite{LEP1,LEP2,LEP3,Tevatron,95expCMS1,95expCMS2,95expCMS3,95expCMS4,95expATLAS}, which will help us to explore the fundamental physics mechanisms of electroweak symmetry breaking. CMS has performed searches for scalar diphoton resonances at 8 and 13 TeV, and revealed a $2.8\sigma$ local excess at 95.3 GeV\cite{95expCMS3}, which is compatible with the latest ATLAS result: $\mu_{\gamma\gamma}^{exp}(95)=\mu_{\gamma\gamma}^{CMS+ATLAS}(95)=0.24_{-0.08}^{+0.09}$\cite{95expATLAS}. LEP reported a $2.3\sigma$ local excess in the $b\bar{b}$ final state with the mass around 95.4 GeV, the relevant signal strength is $\mu_{b\bar{b}}^{exp}(95)=0.117\pm0.057$\cite{LEP2}. The above experiments give us direction to study Higgs boson with mass below 125 GeV, which will help us explore the fundamental physics mechanisms of electroweak symmetry breaking.

Since the excess is performed, it has received considerable attention, specially in numerous NP frameworks. Refs.\cite{CPMSSM1,CPMSSM2} have performed systematic studies of SM-like Higgs boson in the minimum supersymmetric Standard Model (MSSM)\cite{MSSM1,MSSM2,MSSM3,MSSM4} with CP violation. The authors of Refs.\cite{NMSSM1,NMSSM2,NMSSM3,NMSSM4,NMSSM5,NMSSM6,NMSSM7,NMSSM8,NMSSM9,NMSSM10,NMSSM11,NMSSM12,NMSSM13,NMSSM14,NMSSM15,95NMSSM1,95NMSSM2} have researched the SM-like Higgs boson along with the presence of a light Higgs which can account for the diphoton and $b\bar{b}$ excess in the next-to-minimal supersymmetric standard model (NMSSM). The authors of Refs.\cite{munuSSM1,munuSSM2,munuSSM3} have simultaneously accommodated two excesses measured at the LEP and LHC within CP-conserving and CP-violating $\mu\nu$SSM respectively. In some Two-Higgs doublet model (2HDM) with an additional Higgs singlet\cite{95N2HDM1,95N2HDM2,95N2HDM3,95N2HDM4,95N2HDM5,95N2HDM6,95N2HDM7,95N2HDM8,95N2HDM9}, the authors have discussed the parameter space where the lightest and second-lightest Higgs boson can yield a perfect fit to the excesses at around 95 GeV and SM-like Higgs boson measurements simultaneously. Some $U(1)$ extension models such as $U(1)_XSSM$, flavor-dependent $U(1)_X$ model have also explained the great potential for the excesses at around 95 GeV through extra Higgs singlets\cite{U1X,U1X1}.

There are some $U(1)_{B-L}$ extension models such as the minimal SUSY B-L theory\cite{mB-L1,mB-L2,mB-L3,mB-L4,mB-L5}, which only introduces three generations of right-handed neutrinos required for anomaly cancellation and without any other fields. Once the right-handed sneutrino that has both B-L and $SU(2)_R$ quantum numbers acquires a vacuum expectation value(VEV), both $SU(2)_R\otimes U(1)_{B-L}$ and R-parity are spontaneously broken with the relevant scales determined by the soft SUSY mass scale. Neutrino masses are generated by the type I seesaw mechanism. The $Z'$ decays, the usual R-parity violating decays of neutralinos and charginos, as well as the SUSY induced mass degeneracy between the right-handed sneutrino and the $Z'$ gauge boson can lead the model to be simple and predictive. This minimal SUSY B-L theory possesses a local $U(1)_{B-L}$ where the Majoron becomes the longitudinal component of $Z'$ boson, therefore the Majoron problem associated with spontaneous lepton number breaking is no longer present. Furthermore, all of this could be accomplished with the same Higgs sector of the MSSM to make this model the simplest left-right symmetric theory without R-parity. The left-right discrete symmetry in this theory is broken only by the soft terms in order to have a consistent mechanism for R-parity violation and avoid the domain wall problem.

The next to minimum B-L supersymmetric model (NB-LSSM)\cite{mB-L1,mB-L4,NB-LSSM1,NB-LSSM2,B-LSSM3,B-LSSM4} has some different characters as above minimal SUSY B-L theory.  Based on the MSSM\cite{MSSM1,MSSM2,MSSM3,MSSM4}, NB-LSSM extends the gauge symmetry group to $SU(3)_C\otimes{SU(2)_L}\otimes{U(1)_Y}\otimes{U(1)_{B-L}}$, where $B$ represents the baryon number and $L$ stands for the lepton number. Compared with the MSSM, the NB-LSSM adds three singlet Higgs superfields $\hat{\eta}$, $\hat{\bar{\eta}}$, $\hat{S}$ and three generations of right-handed neutrinos superfields $\hat{\nu}_i^c$. In the NB-LSSM, the invariance under $U(1)_{B-L}$ gauge groups impose the R-parity conservation assumed in the MSSM to avoid proton decay\cite{B-L R Parity}. And R-parity conservation can be maintained if $U(1)_{B-L}$ symmetry is spontaneously broken. The singlet scalar $S$ can obtain a VEV $\langle S\rangle =v_S/\sqrt{2} \sim$ TeV after breaking, which is motivated to explain the $\mu$ problem in MSSM naturally since the $\mu$ term is effectively obtained as $\mu\equiv \lambda v_S/\sqrt{2}$. Besides, through the additional singlet Higgs states $\eta, \bar{\eta}, S$ and right-handed (s)neutrinos, additional parameter space in the NB-LSSM is released from the LEP, Tevatron and LHC constraints to alleviate the hierarchy problem of the MSSM\cite{B-L hierarchy1,B-L hierarchy2}. The neutrinos in the NB-LSSM acquire tiny mass at tree-level through the seesaw mechanism, facilitated by the introduction of right-handed neutrino fields. The extended Higgs sector in the NB-LSSM leads to a $5\times5$ neutral CP-even Higgs mass matrix. After considering CP violation, the CP-even and CP-odd Higgs sectors mix together and  provide a rich theoretical framework for exploring the fundamental Higgs boson. In addition, the fermionic components $\tilde{S}$, $\tilde{\eta}$, $\tilde{\bar{\eta}}$ (singlinos), can mix with the MSSM neutralinos after the symmetry breaking, and the superpartner of $Z'$ boson can also mix with the other neutralinos, therefore the neutralino sector is enriched. Thus in the basis $(\tilde{B},\tilde{W}^0,\tilde{H}^0_d,
\tilde{H}^0_u,\tilde{B}',\tilde{\eta},\tilde{\bar{\eta}},\tilde{S})$, the neutralino mass matrix becomes a $8\times 8$ matrix. The variety of neutralinos can result in singlino-like and blino-like lightest supersymmetric particles (LSPs) in addition to bino-dominated LSP, which lead NB-LSSM to provide much more dark matter (DM) candidates comparing with the MSSM\cite{B-LDM1,B-LDM2,B-LDM3,B-LDM4}. All of these are the motivations why we consider this extension model in this work.

In 1964, the decay experiments of neutral kaons into dipion did not exhibit perfect CP symmetry, which demonstrates that the CP violation was first observed experimentally and the nature does not always follow the CP symmetry\cite{CP1,CP2,CP3}. One of the most intriguing aspects of CP violation is that it connects to the matter-antimatter asymmetry in the universe. There are equal amounts of matter and antimatter after the Big Bang, but the universe ends up with a slight excess of matter. If the CP violation occurs in a way that affects the production and decay of particles, it could help explain why more matter than antimatter survived in the early universe. As we know, the observed CP violation is relatively small in the SM, which indicates that SM cannot fully explain the matter-antimatter asymmetry in the universe. Therefore, additional sources of CP violation beyond the SM should be considered. In the NB-LSSM, the CP violation in the Higgs potential leads to the mixing mass terms between the CP-even and CP-odd Higgs fields. Thus, one has to consider a $(10 \times 10)$-dimensional mass matrix for the neutral Higgs boson. These potential mixing effects may lead to drastic variations in the definition of the neutral Higgs boson masses. Besides, the neutral Higgs particles predicted by the theory may strongly mix with one another, thereby significantly modifying the couplings of the Higgs bosons to the fermions or gauge bosons\cite{CPMSSM1,CPMSSM2}. Therefore, we study the Higgs masses, SM-like Higgs decays and the diphoton, $b\bar{b}$ excesses in the NB-LSSM with CP violation.

This work is organized as follows. In Sec.II, we introduce the scalar sector of the NB-LSSM and the CP-violating effective potential to the neutral Higgs boson mass squared matrix, and we derive the signal strengths of SM-like Higgs boson with mass around 125 GeV and excess signals with mass around 95 GeV. The numerical analyses are given out in Sec.III, and the conclusion is discussed in Sec.IV. The tedious formulae are collected in the Appendix.
\section{The CP-violating NB-LSSM and Higgs sector}
\subsection{The CP-violating NB-LSSM}
The NB-LSSM extends the superfields of the MSSM by introducing a $U(1)_{B-L}$ gauge superfield. Therefore, the local gauge group of the NB-LSSM is defined as $SU(3)_C\otimes{SU(2)_L}\otimes{U(1)_Y}\otimes{U(1)_{B-L}}$. Compared with the MSSM, the NB-LSSM adds three singlet Higgs superfields $\hat{\eta}$, $\hat{\bar{\eta}}$, $\hat{S}$ and three generations of right-handed neutrinos superfields $\hat{\nu}_i^c$. The NB-LSSM superpotential is deduced as
\begin{eqnarray}
&&{W_{NB-L}}=Y_u^{ij}\hat U_i\hat Q_j\hat H_u-Y_d^{ij}\hat D_i\hat Q_j\hat H_d-Y_e^{ij}\hat E_i\hat L_j\hat H_d+\lambda \hat S\hat H_u\hat H_d\nonumber\\&&\hspace{1.5cm}-\lambda_2 \hat S {\hat\eta}\hat{\bar{\eta}}+\frac{1}{3}\kappa \hat S\hat S\hat S+Y_x^{ij}\hat{\nu}_i\hat{\eta}\hat{\nu}_j+Y_\nu^{ij}\hat{L}_i\hat{H}_u\hat{\nu}_j,
\label{Wb}
\end{eqnarray}
where $i,j$ are the generation indices, $Y_{u}^{ij},Y_{d}^{ij},Y_{e}^{ij},Y_{x}^{ij},Y_{\nu}^{ij},\lambda,\lambda_2$ and $\kappa$ are the corresponding Yukawa coupling coefficients in the NB-LSSM. The chiral superfields and quantum numbers of the NB-LSSM are in the TABLE \ref{NB-Lquantum numbers}.
\begin{table}[t]
\caption{ \label{NB-Lquantum numbers}  The chiral superfields and quantum numbers in the NB-LSSM.}
\footnotesize
\begin{tabular}{|c|c|c|c|c|c|c|}
\hline
Superfield & Spin 0 & Spin \(\frac{1}{2}\) & Generations &\(U(1)_Y\otimes\, \text{SU}(2)_L\otimes\, \text{SU}(3)_C\otimes\, U(1)_{B-L}\) \\
\hline
\(\hat{H}_d\) & \(H_d\) & \(\tilde{H}_d\) & 1 & \((-\frac{1}{2},{\bf 2},{\bf 1},0) \) \\
\(\hat{H}_u\) & \(H_u\) & \(\tilde{H}_u\) & 1 & \((\frac{1}{2},{\bf 2},{\bf 1},0) \) \\
\(\hat{Q}_i\) & \(\tilde{Q}_i\) & \(Q_i\) & 3 & \((\frac{1}{6},{\bf 2},{\bf 3},\frac{1}{6}) \) \\
\(\hat{L}_i\) & \(\tilde{L}_i\) & \(L_i\) & 3 & \((-\frac{1}{2},{\bf 2},{\bf 1},-\frac{1}{2}) \) \\
\(\hat{D}_i\) & \(\tilde{D}_i\) & \(D_i\) & 3 & \((\frac{1}{3},{\bf 1},{\bf \overline{3}},-\frac{1}{6}) \) \\
\(\hat{U}_i\) & \(\tilde{U}_i\) & \(U_i\) & 3 & \((-\frac{2}{3},{\bf 1},{\bf \overline{3}},-\frac{1}{6}) \) \\
\(\hat{E}_i\) & \(\tilde{E}_i\) & \(E_i\) & 3 & \((1,{\bf 1},{\bf 1},\frac{1}{2}) \) \\
\(\hat{\nu}_i\) & \(\tilde{\nu}_i\) & \(\nu_i\) & 3 & \((0,{\bf 1},{\bf 1},\frac{1}{2}) \) \\
\(\hat{\eta}\) & \(\eta\) & \(\tilde{\eta}\) & 1 & \((0,{\bf 1},{\bf 1},-1) \) \\
\(\hat{\bar{\eta}}\) & \(\bar{\eta}\) & \(\tilde{\bar{\eta}}\) & 1 & \((0,{\bf 1},{\bf 1},1) \) \\
\(\hat{S}\) & \(S\) & \(\tilde{S}\) & 1 & \((0,{\bf 1},{\bf 1},0) \) \\
\hline
\end{tabular}
\end{table}
In the NB-LSSM, the Higgs doublets $H_d,H_u$ and Higgs singlets $\eta,\bar{\eta},S$ obtain the nonzero VEVs $v_d,v_u$ and $v_\eta,v_{\bar{\eta}},v_s$, then the $SU(2)_L\otimes U(1)_Y\otimes U(1)_{B-L}$ gauge group breaks to $U(1)_{em}$. The CP can be violated by the VEVs of the scalar fields. $\theta_1$ represents the relative phase of $v_d,v_u$, $\theta_2$ denotes the relative phase of $v_\eta,v_{\bar{\eta}}$, and $\theta_3$ is the relative phase of $v_s$.
\begin{eqnarray}
&&H_d=\left(
    \begin{array}{cc}
        H_d^- & \\
        \frac{1}{\sqrt{2}}\Big( \phi_d + v_d+ i\sigma_d\Big) & \\
    \end{array}
\right) , \quad  H_u=e^{i \theta_1}\left(
    \begin{array}{cc}
        \frac{1}{\sqrt{2}}\Big (\phi_u + v_u + i \sigma_u\Big) & \\
        H_u^+ & \\
    \end{array}
\right),\nonumber\\&&\eta = \frac{1}{\sqrt{2}} \Big( \phi_{\eta} + v_{\eta} + i \sigma_{\eta}\Big )  , \;
\bar{\eta} = e^{i \theta_2}\frac{1}{\sqrt{2}} \Big( \phi_{\bar{\eta}}+ v_{\bar{{\eta}}} + i \sigma_{\bar{{\eta}}}\Big),\;
S = e^{i \theta_3}\frac{1}{\sqrt{2}} \Big( \phi_s+ v_s + i \sigma_s\Big).
\end{eqnarray}
Here, we define $u^2=v_{\eta}^2+v_{\bar{\eta}}^2,\;v^2=v_d^2+v_u^2$ and $\tan\beta'=\frac{v_{\bar{\eta}}}{v_\eta}$ in analogy to the definition $\tan\beta=\frac{v_u}{v_d}$ in the MSSM.

\subsection{The CP-violating Higgs mass}
The NB-LSSM Higgs potential at the tree-level is written as follows:
\begin{eqnarray}
&&V^0 = \frac{1}{8}G^2  ( H_d^\dagger H_d)^2 \hspace{-0.1cm}+\hspace{-0.1cm}\frac{1}{8} G^2 ( H_u^\dagger H_u)^2 \hspace{-0.1cm}+\hspace{-0.1cm}\frac{1}{2} g_B^2 ( \eta^\dagger\eta )^2 \hspace{-0.1cm}+\hspace{-0.1cm}\frac{1}{2} g_B^2 (\bar{\eta}^\dagger\bar{\eta})^2\hspace{-0.1cm}-\hspace{-0.1cm}\frac{1}{4} G^2 ( H_d^\dagger H_d) ( H_u^\dagger H_u)
\nonumber\\
 &&\hspace{0.65cm}- g_B^2 ( \eta^\dagger\eta ) (\bar{\eta}^\dagger \bar{\eta} )\hspace{-0.1cm}+\hspace{-0.1cm}\frac{1}{2} g_B  g_{YB} ( \bar{\eta}^\dagger \bar{\eta} ) ( H_u^\dagger H_u)  \hspace{-0.1cm}+\hspace{-0.1cm}\frac{1}{2} g_B  g_{YB} ( \eta ^\dagger\eta ) ( H_d^\dagger H_d )\hspace{-0.1cm}-\hspace{-0.1cm}\frac{1}{2} g_B  g_{YB} (\bar{\eta}^\dagger \bar{\eta} ) ( H_d^\dagger H_d)\nonumber\\
&&\hspace{0.65cm}-\frac{1}{2} g_B  g_{YB} (\eta ^\dagger\eta)( H_u^\dagger H_u )\hspace{-0.1cm}+\hspace{-0.1cm}|\lambda|^2[(S^\dagger S)(H_u^\dagger H_u)+(S^\dagger S)(H_d^\dagger H_d)+(H_d^\dagger H_d)(H_u^\dagger H_u)]\nonumber\\
&&\hspace{0.65cm}+|\lambda_2|^2[(S^\dagger S)(\eta^\dagger \eta)\hspace{-0.1cm}+\hspace{-0.1cm}(S^\dagger S)(\bar{{\eta}}^\dagger \bar{{\eta}})\hspace{-0.1cm}+\hspace{-0.1cm}(\eta^\dagger \eta)(\bar{{\eta}}^\dagger \bar{{\eta}})]\hspace{-0.1cm}+\hspace{-0.1cm}|\kappa|^2(S^\dagger S)^2\hspace{-0.1cm}+\hspace{-0.1cm}\lambda_2^*\lambda \eta^\dagger\bar{{\eta}}^\dagger H_d H_u\hspace{-0.1cm}-\hspace{-0.1cm}\lambda_2^\dagger\kappa\eta^\dagger \bar{{\eta}}^\dagger S S\nonumber\\
&&\hspace{0.65cm}+\lambda^*\lambda_2 H_d^\dagger H_u^\dagger\eta \bar{{\eta}}-\lambda^*\kappa H_d^\dagger H_u^\dagger S S-\kappa^*\lambda_2 S^\dagger S^\dagger\eta \bar{{\eta}}-\kappa^* \lambda S^\dagger S^\dagger H_d H_u +m_\eta^2|\eta|^2+m_{\bar{\eta}}^2 |\bar{{\eta}}|^2\nonumber\\
&&\hspace{0.65cm}+m_{H_d}^2|{H_d}|^2+m_{H_u}^2|{H_u}|^2+m_S^2|S|^2+(\frac{1}{3}S^3 T_\kappa-H_d H_u ST_\lambda-\eta \bar{{\eta}} ST_2+h.c. ),
\label{Lv0}
\end{eqnarray}
where $G^2=g_1^2+g_2^2+g_{YB}^2$. Using the tadpole conditions $T_{\phi_{d}}^{(0)}=T_{\phi_{u}}^{(0)} =T_{\phi_{\eta}}^{(0)} =T_{\phi_{\bar{\eta}}}^{(0)}=T_{\phi_{s}}^{(0)}  =0 $ and $T_{\sigma_{d}}^{(0)}=T_{\sigma_{u}}^{(0)}=T_{\sigma_{\eta}}^{(0)}=T_{\sigma_{\bar{\eta}}}^{(0)}=T_{\sigma_{s}}^{(0)}=0$, which can be found in Appendix A, we present the neutral tree-level Higgs boson mass squared matrix in the basis $\{\phi_d,\phi_u,\phi_\eta,\phi_{\bar{\eta}},\phi_{s},\sigma_d,\sigma_u,\sigma_\eta,\sigma_{\bar{\eta}},\sigma_{s}\}$.
\begin{eqnarray}
{{{{\cal M}^2_h}^{(0)}}}=\left(
    \begin{array}{cccccc}
       {{\cal M}^2_S}^{(0)} & {{{\cal M}^2_{SY}}}^{(0)}&{{\cal M}^2_{BB'}}^{(0)}  & 0_{2\times2} & -{({{\cal M}^2_{YP}}^{(0)})^T}&  {{\cal M}^2_{EE}}^{(0)}  \\
       {({{\cal M}^2_{SY}}^{(0)})^T} & {{\cal M}^2_Y}^{(0)}& {{\cal M}^2_{CC'}}^{(0)}  &{{\cal M}^2_{YP}}^{(0)} & 0_{2\times2} &{{\cal M}^2_{FF}}^{(0)} \\
       {({{\cal M}^2_{BB'}}^{(0)})^T} & {({{\cal M}^2_{CC'}}^{(0)})^T}&{{\cal M}^2_{SS}}^{(0)} &{{\cal M}^2_{EE'}}^{(0)}   &{{\cal M}^2_{FF'}}^{(0)}  &{{\cal M}^2_{SS'}}^{(0)} \\
       0_{2\times2} &  {({{\cal M}^2_{YP}}^{(0)} )^T} & {({{\cal M}^2_{EE'}}^{(0)})^T} & {{\cal M}^2_{P}}^{(0)} & {{\cal M}^2_{PR}}^{(0)} & {{\cal M}^2_{BB}}^{(0)}  \\
        -{{\cal M}^2_{YP}}^{(0)} & 0_{2\times2} & {({{\cal M}^2_{FF'}}^{(0)})^T}&  {({{\cal M}^2_{PR}}^{(0)})^T}& {{\cal M}^2_{R}}^{(0)} &{{\cal M}^2_{CC}}^{(0)}  \\
      {({{\cal M}^2_{EE}}^{(0)})^T} & {({{\cal M}^2_{FF}}^{(0)})^T} &  {({{\cal M}^2_{SS'}}^{(0)})^T}& {({{\cal M}^2_{BB}}^{(0)})^T} & {({{\cal M}^2_{CC}}^{(0)})^T} &{{\cal M}^2_{S'S'}}^{(0)}
    \end{array}
\right),
\label{tree-level Higgs}
\end{eqnarray}
where ${{\cal M}^2_S}^{(0)}$, $({{\cal M}^2_{SY}})^{(0)}$, ${{\cal M}^2_Y}^{(0)}$, ${{\cal M}^2_{BB'}}^{(0)}$, ${{\cal M}^2_{CC'}}^{(0)}$, ${{\cal M}^2_{SS}}^{(0)}$, ${{\cal M}^2_P}^{(0)}$, ${{\cal M}^2_R}^{(0)}$, ${{\cal M}^2_{PR}}^{(0)}$, ${{\cal M}^2_{BB}}^{(0)}$, ${{\cal M}^2_{CC}}^{(0)}$, ${{\cal M}^2_{S'S'}}^{(0)}$ and ${{\cal M}^2_{YP}}^{(0)}$, ${{\cal M}^2_{EE}}^{(0)}$, ${{\cal M}^2_{FF}}^{(0)}$, ${{\cal M}^2_{EE'}}^{(0)}$, ${{\cal M}^2_{FF'}}^{(0)}$, ${{\cal M}^2_{SS'}}^{(0)}$ denote the CP-even scalar parts, CP-odd scalar parts, the mixing parts of CP-even and CP-odd scalar of the neutral Higgs boson mass squared matrices at the tree-level, respectively.
\begin{eqnarray}
&&{{\cal M}^2_P}^{(0)}=[\frac{ v _s}{\sqrt{2}}\Re(T_\lambda e^{i(\theta_1+\theta_3)})-\frac{ v_{\eta}v_{\bar\eta}}{2}\Re(\lambda_2^*\lambda e^{i(\theta_1-\theta_2)})+\frac{v_s^2 }{2}\Re(\lambda^* \kappa e^{i(2\theta_3-\theta_1)})] \left(
    \begin{array}{cc}
        \frac{v_u}{v_d}& 1 \\
       1 &  \frac{v_d}{v_u} \\
    \end{array}
\right),\nonumber\\
&&{{\cal M}^2_R}^{(0)}=[\frac{ v _s}{\sqrt{2}}\Re(T_2 e^{i(\theta_2+\theta_3)})-\frac{ v_{d}v_{u}}{2}\Re(\lambda_2^*\lambda e^{i(\theta_1-\theta_2)})+\frac{v_s^2 }{2}\Re(\lambda_2^* \kappa e^{i(2\theta_3-\theta_2)})] \left(
    \begin{array}{cc}
        \frac{v_{\bar\eta}}{v_\eta}& 1 \\
       1 &  \frac{v_\eta}{v_{\bar\eta}} \\
    \end{array}
\right),\nonumber\\
&&{{\cal M}^2_{PR}}^{(0)}\hspace{-0.1cm}=\hspace{-0.1cm}\frac{1}{2}\Re(\lambda_2^*\lambda e^{i(\theta_1-\theta_2)}) \hspace{-0.1cm}\left(\hspace{-0.1cm}
    \begin{array}{cc}
        v_u v_{\bar\eta}&v_u v_{\eta} \\
       v_d v_{\bar\eta} & v_d v_{\eta} \\
    \end{array}
\hspace{-0.1cm}\right),\;
{{\cal M}^2_{SY}}^{(0)}\hspace{-0.1cm}=\hspace{-0.1cm}\frac{g_B g_{YB}}{2}  \hspace{-0.1cm}\left(\hspace{-0.1cm}
    \begin{array}{cc}
        v_d v_\eta & -v_d v_{\bar\eta} \\
       - v_u v_\eta & v_u v_{\bar\eta} \\
    \end{array}
\hspace{-0.1cm}\right)\hspace{-0.1cm}+\hspace{-0.1cm}{{\cal M}^2_{PR}}^{(0)},\nonumber\\
&&{{\cal M}^2_S}^{(0)}\hspace{-0.1cm}=\hspace{-0.1cm}{{\cal M}^2_P}^{(0)}\hspace{-0.1cm}+\hspace{-0.1cm} \left(\hspace{-0.1cm}
    \begin{array}{cc}
       \frac{1}{4} G^2 v_d^2& (-\frac{1}{4} G^2\hspace{-0.1cm}+\hspace{-0.1cm}|\lambda|^2)v_u v_d \\
        (-\frac{1}{4} G^2\hspace{-0.1cm}+\hspace{-0.1cm}|\lambda|^2)v_u v_d & \frac{1}{4} G^2 v_u^2 \\
    \end{array}
\hspace{-0.1cm}\right),\;\nonumber\\
&&{{\cal M}^2_Y}^{(0)}\hspace{-0.1cm}=\hspace{-0.1cm}{{\cal M}^2_R}^{(0)} \hspace{-0.1cm}+\hspace{-0.1cm}  \left(\hspace{-0.1cm}
    \begin{array}{cc}
       g_B^2 v_\eta^2& (-g_B^2\hspace{-0.1cm}+\hspace{-0.1cm}|\lambda_2|^2)v_\eta v_{\bar\eta} \\
        (-g_B^2\hspace{-0.1cm}+\hspace{-0.1cm}|\lambda_2|^2)v_\eta v_{\bar\eta} & g_B^2v_{\bar\eta}^2 \\
    \end{array}
\hspace{-0.1cm}\right),\nonumber\\
&&{{\cal M}^2_{YP}}^{(0)}=-\frac{1}{2}\Im(\lambda_2^*\lambda e^{i(\theta_1-\theta_2)}) \left(
    \begin{array}{cc}
        v_u v_{\bar\eta}&v_d v_{\bar\eta} \\
      v_u v_{\eta} & v_d v_{\eta} \\
    \end{array}
\right),\nonumber\\
&&{{\cal M}^2_{BB}}^{(0)}=[\frac{ v}{\sqrt{2}}\Re(T_\lambda e^{i(\theta_1+\theta_3)})-\frac{v_s v }{2}\Re(\lambda^* \kappa e^{i(2\theta_3-\theta_1)})]\left(\begin{array}{c}
        \sin\beta \\
       \cos\beta \\ \end{array}\right),\nonumber\\
&&{{\cal M}^2_{CC}}^{(0)}=[\frac{ u}{\sqrt{2}}\Re(T_2 e^{i(\theta_2+\theta_3)})-\frac{v_s u }{2}\Re(\lambda_2^* \kappa e^{i(2\theta_3-\theta_2)})]\left( \begin{array}{c}
        \sin\beta' \\
       \cos\beta' \\\end{array}\right),\nonumber\\
&&{{\cal M}^2_{BB'}}^{(0)}=|\lambda|^2 v_s v \left(\begin{array}{c}
        \cos\beta \\
       \sin\beta \\ \end{array}\right)-[\frac{ v}{\sqrt{2}}\Re(T_\lambda e^{i(\theta_1+\theta_3)})+\frac{v_s v}{2}\Re(\lambda^* \kappa e^{i(2\theta_3-\theta_1)})]\left(\begin{array}{c}
        \sin\beta \\
       \cos\beta \\ \end{array}\right),\nonumber\\
&&{{\cal M}^2_{CC'}}^{(0)}=|\lambda_2|^2 v_s u \left(\begin{array}{c}
        \cos\beta' \\
       \sin\beta' \\ \end{array}\right)-[\frac{u}{\sqrt{2}}\Re(T_2 e^{i(\theta_2+\theta_3)})+\frac{v_s u}{2}\Re(\lambda_2^* \kappa e^{i(2\theta_3-\theta_2)})]\left( \begin{array}{c}
        \sin\beta' \\
       \cos\beta' \\\end{array}\right),\nonumber\\
&&{{\cal M}^2_{EE}}^{(0)}=[\frac{ v}{\sqrt{2}}\Im(T_\lambda e^{i(\theta_1+\theta_3)})+\frac{v_s v }{2}\Im(\lambda^* \kappa e^{i(2\theta_3-\theta_1)})]\left(\begin{array}{c}
        \sin\beta \\
       \cos\beta \\ \end{array}\right),\nonumber\\
&&{{\cal M}^2_{FF}}^{(0)}=[\frac{ u}{\sqrt{2}}\Im(T_2 e^{i(\theta_2+\theta_3)})+\frac{v_s u }{2}\Im(\lambda_2^* \kappa e^{i(2\theta_3-\theta_2)})]\left( \begin{array}{c}
        \sin\beta' \\
       \cos\beta' \\\end{array}\right),\nonumber\\
&&{{\cal M}^2_{EE'}}^{(0)}=[\frac{ v}{\sqrt{2}}\Im(T_\lambda e^{i(\theta_1+\theta_3)})-\frac{v_s v }{2}\Im(\lambda^* \kappa e^{i(2\theta_3-\theta_1)})]\left(\begin{array}{c}
        \sin\beta \\
       \cos\beta \\ \end{array}\right),\nonumber\\
&&{{\cal M}^2_{FF'}}^{(0)}=[\frac{ u}{\sqrt{2}}\Im(T_2 e^{i(\theta_2+\theta_3)})-\frac{v_s u }{2}\Im(\lambda_2^* \kappa e^{i(2\theta_3-\theta_2)})]\left( \begin{array}{c}
        \sin\beta' \\
       \cos\beta' \\\end{array}\right),\nonumber\\
&&{{\cal M}^2_{SS}}^{(0)}=2|\kappa|^2 v_s^2+\frac{ v_s}{\sqrt{2}}\Re\Big(T_\kappa e^{i3\theta_3}\Big)+\frac{ v_d v_u}{\sqrt{2}v_s} \Re\Big(T_\lambda e^{i(\theta_1+\theta_3)}\Big)+\frac{v_\eta v_{\bar{{\eta}}}}{\sqrt{2}v_s}\Re\Big(T_2 e^{i(\theta_2+\theta_3)}\Big),\nonumber\\
&&{{\cal M}^2_{S'S'}}^{(0)}=-2\sqrt{2}v_s\Re\Big(T_\kappa e^{i3\theta_3}\Big)+2\Re\Big(\lambda_2^* \kappa e^{i(2\theta_3-\theta_2)}\Big)v_\eta v_{\bar{{\eta}}}+2\Re\Big(\lambda^*\kappa e^{i(2\theta_3-\theta_1)}\Big)v_dv_u\nonumber\\
&&\hspace{2.1cm}+\frac{ v_s}{\sqrt{2}}\Re\Big(T_\kappa e^{i3\theta_3}\Big)+\frac{ v_d v_u}{\sqrt{2}v_s} \Re\Big(T_\lambda e^{i(\theta_1+\theta_3)}\Big)+\frac{v_\eta v_{\bar{{\eta}}}}{\sqrt{2}v_s}\Re\Big(T_2 e^{i(\theta_2+\theta_3)}\Big),\nonumber\\
&&{{\cal M}^2_{SS'}}^{(0)}=-\sqrt{2}v_s\Im\Big(T_\kappa e^{i3\theta_3}\Big)+\Im\Big(\lambda_2^* \kappa e^{i(2\theta_3-\theta_2)}\Big)v_\eta v_{\bar{{\eta}}}+\Im\Big(\lambda^*\kappa e^{i(2\theta_3-\theta_1)}\Big)v_dv_u.
\end{eqnarray}
The mass squared matrix of Eq.(\ref{tree-level Higgs}) includes two eigenstates $G_1^0=\frac{1}{v}(-v_d \sigma_d + v_u \sigma_u),\;
G_2^0=\frac{1}{u}(-v_{\eta} \sigma_{\eta} + v_{\bar\eta} \sigma_{\bar\eta})$, which corresponds to the massless Goldstone boson.

The effective potential at the two-loop level can be expressed as $V_{total}=V^0+\Delta V^{(1)}+\Delta V^{(2)}$, where $\Delta V^{(1)}$ is the one-loop effective potential corrections from the top (bottom) and stop (sbottom) particles, and $\Delta V^{(2)}$ is the two-loop effective potential corrections (The detailed calculations of two-loop effective potential corrections are shown in the Appendix B).
\begin{eqnarray}
&&\Delta V^{(1)} = - \frac{3}{32\pi^2} \sum _{q=t,b} \Big [\sum_{k=1,2} m^4_{\tilde q_k} \Big ( \ln\frac{m^2_{\tilde q_k}}{\Lambda^2} - \frac{3}{2}\Big ) - 2{ m}_{q}^4  \Big ( \ln\frac{{ m}_{q}^2}{\Lambda^2} - \frac{3}{2}\Big ) \Big].
\end{eqnarray}
Here, $\Lambda$ represents the renormalization scale in TeV order. ${m}_{q}$ and $ m_{\tilde q}$ correspond to the mass of top (bottom) and stop (sbottom) particles, which can be deduced as:
\begin{eqnarray}
&&{m}_{t(b)}^2 = | Y_{t(b)}|^2 |H_{u(d)}^0|^2,\nonumber\\
&&m^2_{\tilde b_k} = \frac{1}{2} \Big\{\tilde{M}_{Q_3}^2 + \tilde{M}_b^2  + 2 | Y_b |^2 | H_d^0 |^2 - \frac{1}{4} G^2 H^2 + \frac{1}{2} g_B g_{YB} N^2 \nonumber\\
&&\hspace{1.2cm} \pm \sqrt{\Big [ \tilde{M}_{Q_3}^2 - \tilde{M}_b^2  - x_b  H^2 + a_b N^2 \Big ]^2 + 4 | T_b H_d^0 - Y_b \lambda^* H_u^{0\dagger} S^\dagger  |^2} \Big\},\nonumber\\
&&m^2_{\tilde t_k} = \frac{1}{2} \Big\{ \tilde{M}_{Q_3}^2 + \tilde{M}_t^2  + 2 | Y_t |^2 | H_u^0 |^2 + \frac{1}{4} G^2 H^2 - \frac{1}{2} g_B g_{YB} N^2 \nonumber\\
&&\hspace{1.2cm} \pm \sqrt{\Big [ \tilde{M}_{Q_3}^2 - \tilde{M}_t^2  + x_t H^2  + a_t N^2 \Big ]^2 + 4|T_t H_u^0 - Y_t \lambda^* H_d^{0\dagger} S^\dagger  |^2} \Big\},\;(k=1,2)
\end{eqnarray}
with $H^2=\Big ( | H_d^0 |^2 - | H_u^0 |^2 \Big ) $, $N^2=\Big ( |\bar{\eta}  |^2 - |\eta |^2\Big )$, $x_t=\frac{1}{12} \Big ( 3 g_2^2 - 5 g_1^2 - 5 g_{YB}^2 - 2 g_B g_{YB} \Big )$, $a_t = \frac{1}{6} \Big ( 2 g_B^2 + 5 g_B g_{YB}\Big)$, $x_b = \frac{1}{12} \Big ( 3 g_2^2 - g_1^2 - g_{YB}^2 + 2 g_B g_{YB}\Big)$ and $a_b = \frac{1}{6} \Big ( 2 g_B^2 - g_B g_{YB}\Big)$.

Correspondingly, we can deduce the one-loop corrections to the minimization conditions of scalar potential by the formulas $T_{\phi_{d(u,\eta,\bar\eta,s)}}^{(1)} = \left\langle \frac{\partial\Delta V^{(1)}}{\partial{\phi_{d(u,\eta,\bar\eta,s)}}}\right\rangle$ and $T_{\sigma_{d(u,\eta,\bar\eta,s)}}^{(1)} =\left\langle \frac{\partial\Delta V^{(1)}}{\partial{\sigma_{d(u,\eta,\bar\eta,s)}}}\right\rangle$,  whose specific expresses are summarized in the Appendix A.
Then, in the weak basis $\{\phi_d,\phi_u,\phi_\eta,\phi_{\bar{\eta}},\phi_{s},\sigma_d,\sigma_u,\sigma_\eta,\sigma_{\bar{\eta}},\sigma_{s}\}$, the one-loop radiative corrections to the neutral Higgs mass squared matrix is given as:
\begin{eqnarray}
{{{{\cal M}^2_h}^{(1)}}}=\left(
    \begin{array}{cccccc}
       {{\cal M}^2_S}^{(1)} & {{{\cal M}^2_{SY}}}^{(1)}&{{\cal M}^2_{BB'}}^{(1)}  & {{\cal M}^2_{SP}}^{(1)}  & 0_{2\times2}&  {{\cal M}^2_{EE}}^{(1)}  \\
       ({{\cal M}^2_{SY}}^{(1)})^T & {{\cal M}^2_Y}^{(1)}& {{\cal M}^2_{CC'}}^{(1)}  &{{\cal M}^2_{YP}}^{(1)} & 0_{2\times2} &{{\cal M}^2_{FF}}^{(1)} \\
       ({{\cal M}^2_{BB'}}^{(1)})^T & ({{\cal M}^2_{CC'}}^{(1)})^T&{{\cal M}^2_{SS}}^{(1)} &{{\cal M}^2_{EE'}}^{(1)}   &0_{1\times2} &{{\cal M}^2_{SS'}}^{(1)} \\
      ({{\cal M}^2_{SP}}^{(1)})^T &  ({{\cal M}^2_{YP}}^{(1)})^T & ({{\cal M}^2_{EE'}}^{(1)})^T & {{\cal M}^2_{P}}^{(1)} &0_{2\times2}& {{\cal M}^2_{BB}}^{(1)}  \\
         0_{2\times2} & 0_{2\times2} &  0_{2\times2}&  0_{2\times2}&0_{2\times2} &0_{2\times1}  \\
      ({{\cal M}^2_{EE}}^{(1)})^T & ({{\cal M}^2_{FF}}^{(1)})^T &  ({{\cal M}^2_{SS'}}^{(1)})^T& ({{\cal M}^2_{BB}}^{(1)})^T & 0_{1\times2} &{{\cal M}^2_{S'S'}}^{(1)}
    \end{array}
\right),
\end{eqnarray}
where ${{\cal M}^2_S}^{(1)}$, ${{\cal M}^2_Y}^{(1)}$, ${{\cal M}^2_{SS}}^{(1)}$, ${{\cal M}^2_{SY}}^{(1)}$, ${{\cal M}^2_{BB'}}^{(1)}$, ${{\cal M}^2_{CC'}}^{(1)}$, ${{\cal M}^2_P}^{(1)}$, ${{\cal M}^2_{S'S'}}^{(1)}$, ${{\cal M}^2_{BB}}^{(1)}$ and ${{\cal M}^2_{SP}}^{(1)}$, ${{\cal M}^2_{EE}}^{(1)}$, ${{\cal M}^2_{YP}}^{(1)}$, ${{\cal M}^2_{FF}}^{(1)}$, ${{\cal M}^2_{EE'}}^{(1)}$, ${{\cal M}^2_{SS'}}^{(1)}$ denote the one-loop radiation corrections of the CP-even scalar parts, CP-odd scalar parts, the mixing parts of CP-even and CP-odd scalar of the neutral Higgs boson mass squared matrices, respectively. The concrete components are given as follows.
\begin{eqnarray}
&&({{\cal M}^2_S}^{(1)})_{ij}=\frac{3}{16\pi^2}\sum_{q=t,b}\Big\{\sum_{k=1,2}\Big[\Big(\Big\langle\frac{\partial^2 m^2_{\tilde q_k}}{\partial\phi_i \partial\phi_j}\Big\rangle - \frac{\delta_{ij}}{v_i} \Big\langle\frac{\partial  m^2_{\tilde q_k}}{ \partial\phi_j}\Big\rangle\Big)m^2_{\tilde q_k}\Big(\ln\frac{m^2_{\tilde q_k}}{\Lambda^2}-1\Big)\nonumber\\
&&\hspace{1.8cm}+ \Big\langle\frac{\partial  m^2_{\tilde q_k}}{ \partial\phi_i}\Big\rangle \Big\langle\frac{\partial  m^2_{\tilde q_k}}{ \partial\phi_j}\Big\rangle \ln\frac{m^2_{\tilde q_k}}{\Lambda^2}\Big]\
-2\Big[\Big(\Big\langle\frac{\partial^2 m^2_{q}}{\partial\phi_i \partial\phi_j}\Big\rangle - \frac{\delta_{ij}}{v_i} \Big\langle\frac{\partial  m^2_{q}}{ \partial\phi_j}\Big\rangle\Big)m^2_q\Big(\ln\frac{m^2_ q}{\Lambda^2}-1\Big)\nonumber\\
&&\hspace{1.8cm}+\Big\langle\frac{\partial  m^2_{q}}{ \partial\phi_i}\Big\rangle \Big\langle\frac{\partial m^2_{q}}{ \partial\phi_j}\Big\rangle \ln\frac{m^2_q}{\Lambda^2}\Big]\Big\} \quad(i,j=1,2),
\nonumber\\
&&({{\cal M}^2_Y}^{(1)})_{ij}= \frac{3}{16\pi^2} \sum_{q=t,b}\sum_{k=1,2}\Big[\Big(\Big\langle\frac{\partial^2 m^2_{\tilde q_k}}{\partial\phi_i \partial\phi_j}\Big\rangle - \frac{\delta_{ij}}{v_i} \Big\langle\frac{\partial  m^2_{\tilde q_k}}{ \partial\phi_j}\Big\rangle\Big) m^2_{\tilde q_k}\Big(\ln\frac{m^2_{\tilde q_k}}{\Lambda^2}-1\Big)\nonumber\\
&& \hspace{1.9cm}+ \Big\langle\frac{\partial  m^2_{\tilde q_k}}{ \partial\phi_i}\Big\rangle \Big\langle\frac{\partial  m^2_{\tilde q_k}}{ \partial\phi_j}\Big\rangle \ln\frac{m^2_{\tilde q_k}}{\Lambda^2}\Big]  \quad(i,j=3,4),
\nonumber\\
&&({{\cal M}^2_{SS}}^{(1)})_{ij}= ({{\cal M}^2_Y}^{(1)})_{ij} \quad(i=j=5),
\nonumber\\
&&({{\cal M}^2_{SY}}^{(1)})_{ij}= \frac{3}{16\pi^2}\sum_{q=t,b}\sum_{k=1,2}\Big[\Big\langle\frac{\partial^2 m^2_{\tilde q_k}}{\partial\phi_i \partial\phi_j}\Big\rangle m^2_{\tilde q_k}\Big(\ln\frac{m^2_{\tilde q_k}}{\Lambda^2}-1\Big)\nonumber\\
&&\hspace{2.1cm} + \Big\langle\frac{\partial  m^2_{\tilde q_k}}{ \partial\phi_i}\Big\rangle \Big\langle\frac{\partial  m^2_{\tilde q_k}}{ \partial\phi_j}\Big\rangle \ln\frac{m^2_{\tilde q_k}}{\Lambda^2}\Big]  \quad(i=1,2;j=3,4),
\nonumber\\
&&({{\cal M}^2_{BB'}}^{(1)})_{ij}=({{\cal M}^2_{SY}}^{(1)})_{ij} \quad(i=1,2;j=5),\quad({{\cal M}^2_{CC'}}^{(1)})_{ij}= ({{\cal M}^2_{SY}}^{(1)})_{ij}\quad(i=3,4;j=5),\nonumber\\
&&({{\cal M}^2_P}^{(1)})_{ij}=\frac{3}{16\pi^2}\sum_{q=t,b}\Big\{\sum_{k=1,2}\Big[\Big(\Big\langle\frac{\partial^2 m^2_{\tilde q_k}}{\partial\sigma_i \partial\sigma_j}\Big\rangle - \frac{\delta_{ij}}{v_i} \Big\langle\frac{\partial m^2_{\tilde q_k}}{ \partial\sigma_j}\Big\rangle\Big)m^2_{\tilde q_k}\Big(\ln\frac{m^2_{\tilde q_k}}{\Lambda^2}-1\Big)\nonumber\\
&&\hspace{1.9cm}+ \Big\langle\frac{\partial  m^2_{\tilde q_k}}{ \partial\sigma_i}\Big\rangle \Big\langle\frac{\partial  m^2_{\tilde q_k}}{ \partial\sigma_j}\Big\rangle \ln\frac{m^2_{\tilde q_k}}{\Lambda^2}\hspace{-0.1cm}\Big]\hspace{-0.1cm}-\hspace{-0.1cm}2\Big[\Big\langle\frac{\partial^2 m^2_{q}}{\partial\sigma_i \partial\sigma_j}\Big\rangle m^2_q\Big(\hspace{-0.1cm}\ln\frac{m^2_ q}{\Lambda^2}\hspace{-0.1cm}-\hspace{-0.1cm}1\hspace{-0.1cm}\Big)\hspace{-0.1cm}\Big\}\quad(i,j\hspace{-0.1cm}=\hspace{-0.1cm}1,2),
\nonumber\\
&&({{\cal M}^2_{S'S'}}^{(1)})_{ij} =  \frac{3}{16\pi^2}\sum_{q=t,b}\sum_{k=1,2}\Big[\Big(\Big\langle\frac{\partial^2 m^2_{\tilde q_k}}{\partial\sigma_i \partial\sigma_j}\Big\rangle - \frac{\delta_{ij}}{v_i} \Big\langle\frac{\partial m^2_{\tilde q_k}}{ \partial\sigma_j}\Big\rangle\Big)m^2_{\tilde q_k}\Big(\ln\frac{m^2_{\tilde q_k}}{\Lambda^2}-1\Big)\nonumber\\
&&\hspace{2.2cm}+ \Big\langle\frac{\partial  m^2_{\tilde q_k}}{ \partial\sigma_i}\Big\rangle \Big\langle\frac{\partial  m^2_{\tilde q_k}}{ \partial\sigma_j}\Big\rangle \ln\frac{m^2_{\tilde q_k}}{\Lambda^2}\Big] \quad(i=j=5),
\nonumber\\
&&({{\cal M}^2_{BB}}^{(1)})_{ij}=\frac{3}{16 \pi^2} \sum_{q=t,b}\sum_{k=1,2}\Big[\Big\langle\frac{\partial^2 m^2_{\tilde q_k}}{\partial\sigma_i \partial\sigma_j}\Big\rangle m^2_{\tilde q_k}\Big(\ln\frac{m^2_{\tilde q_k}}{\Lambda^2}-1\Big)\nonumber\\
&&\hspace{2.1cm}+\Big\langle\frac{\partial  m^2_{\tilde q_k}}{ \partial\sigma_i}\Big\rangle \Big\langle\frac{\partial  m^2_{\tilde q_k}}{ \partial\sigma_j}\Big\rangle \ln\frac{m^2_{\tilde q_k}}{\Lambda^2}\Big]  \quad(i,=1,2 ,j=5),
\nonumber\\
&&({{\cal M}^2_{SP}}^{(1)})_{ij}=\frac{3}{16 \pi^2} \sum_{q=t,b}\sum_{k=1,2}\Big[\Big(\Big\langle\frac{\partial^2 m^2_{\tilde q_k}}{\partial\phi_i \partial\sigma_j}\Big\rangle- \frac{(1-\delta_{ij})}{v_i} \Big\langle\frac{\partial  m^2_{\tilde q_k}}{ \partial\sigma_j}\Big\rangle\Big)m^2_{\tilde q_k}\Big(\ln\frac{m^2_{\tilde q_k}}{\Lambda^2}-1\Big)\nonumber\\
&&\hspace{2cm}+\Big\langle\frac{\partial  m^2_{\tilde q_k}}{ \partial\phi_i}\Big\rangle \Big\langle\frac{\partial  m^2_{\tilde q_k}}{ \partial\sigma_j}\Big\rangle \ln\frac{m^2_{\tilde q_k}}{\Lambda^2}\Big]  \quad(i,j=1,2),
\nonumber\\
&&({{\cal M}^2_{EE}}^{(1)})_{ij}=({{\cal M}^2_{SP}}^{(1)})_{ij}\quad(i=1,2,j=5),
\nonumber\\
&&({{\cal M}^2_{YP}}^{(1)})_{ij}= \frac{3}{16\pi^2}\sum_{q=t,b}\sum_{k=1,2}\Big[\Big\langle\frac{\partial^2 m^2_{\tilde q_k}}{\partial\phi_i \partial\sigma_j}\Big\rangle m^2_{\tilde q_k}\Big(\ln\frac{m^2_{\tilde q_k}}{\Lambda^2}-1\Big)\nonumber\\
&& \hspace{2.1cm}+ \Big\langle\frac{\partial  m^2_{\tilde q_k}}{ \partial\phi_i}\Big\rangle \Big\langle\frac{\partial  m^2_{\tilde q_k}}{ \partial\sigma_j}\Big\rangle \ln\frac{m^2_{\tilde q_k}}{\Lambda^2}\Big]  \quad(i=3,4;j=1,2),
\nonumber\\
&&({{\cal M}^2_{FF}}^{(1)})_{ij}=({{\cal M}^2_{YP}}^{(1)})_{ij}\quad(i,=3,4,j=5),\quad({{\cal M}^2_{EE'}}^{(1)})_{ij}=({{\cal M}^2_{YP}}^{(1)})_{ij}\quad(i=5,j=1,2),
\nonumber\\
&&({{\cal M}^2_{SS'}}^{(1)}_{ij}) =({{\cal M}^2_{YP}}^{(1)})_{ij} \;\;(i=j=5).
\end{eqnarray}
Here, the concrete expressions of $\frac{\partial m^2_{q_k}}{ \partial\phi_i}, \frac{\partial  m^2_{\tilde q_k}}{ \partial\phi_i}, q_k \in ( t, b), \tilde q_k \in (\tilde t_k, \tilde b_k), \cdot\cdot\cdot$ are discussed specifically in the Appendix C. Hence, the mass squared matrix of the neutral Higgs boson incorporating the tree-level contributions ${{\cal M}^2_h}^{(0)}$, one-loop radiative corrections ${{\cal M}^2_h}^{(1)}$ and two-loop radiative corrections ${{\cal M}^2_h}^{(2)}$ can be expressed as:
\begin{eqnarray}
{{{{\cal M}^2_h}}}=\left(
    \begin{array}{cccccc}
       {{\cal M}^2_S}& ({{\cal M}^2_{SY}})&{{\cal M}^2_{BB'}}  & {{\cal M}^2_{SP}}& {{\cal M}^2_{UL}}&  {{\cal M}^2_{EE}} \\
       {({\cal M}^2_{SY})^T} & {{\cal M}^2_Y}& {{\cal M}^2_{CC'}}  &{{\cal M}^2_{YP}} & 0_{2\times2} &{{\cal M}^2_{FF}} \\
       {({\cal M}^2_{BB'})^T} & {({\cal M}^2_{CC'})^T}&{{\cal M}^2_{SS}} &{{\cal M}^2_{EE'}}  &{{\cal M}^2_{FF'}} &{{\cal M}^2_{SS'}} \\
        {{\cal M}^2_{SP}} &  {({\cal M}^2_{YP})^T} & {({\cal M}^2_{EE'})^T}& {{\cal M}^2_{P}} & {{\cal M}^2_{PR}} & {{\cal M}^2_{BB}} \\
        {({\cal M}^2_{UL})^T} & 0_{2\times2} & {({\cal M}^2_{FF'})^T}&  {({\cal M}^2_{PR})^T}& {{\cal M}^2_{R}} &{{\cal M}^2_{CC}}  \\
      {({\cal M}^2_{EE})^T} & {({\cal M}^2_{FF})^T} &  {({\cal M}^2_{SS'})^T}& {({\cal M}^2_{BB})^T}& {({\cal M}^2_{CC})^T} &{{\cal M}^2_{S'S'}}
    \end{array}
\right)+{{\cal M}_h^2}^{(2)},
\end{eqnarray}
with ${\cal M}^2_S={{\cal M}^2_S}^{(0)}+{{\cal M}^2_S}^{(1)},\;{\cal M}^2_Y={{\cal M}^2_Y}^{(0)}+{{\cal M}^2_Y}^{(1)},{\cal M}^2_{UL}=-{({{\cal M}^2_{YP}}^{(0)})^T},\cdot\cdot\cdot$
Then the mass squared matrix ${\cal M}^2_h$ can be diagonalized by the unitary matrix $Z^H$.
\begin{eqnarray}
m_{h_n}^{diag}=Z^H\cdot{\cal M}^2_{h}\cdot (Z^H)^\dagger.
\end{eqnarray}
Here, $m_{h_n}^{diag}$ is the n-th squared Higgs mass in the mass eigenstate, which includes two massless Goldstone bosons. The lightest nonzero Higgs boson mass corresponds to the mass of 95 GeV scalar excess, and the second-lightest one corresponds to the mass of 125 GeV Higgs boson.
\subsection{The 125 GeV Higgs boson decays}
The signal strengths for the 125 GeV Higgs boson decay channels can be quantified by the following ratios \cite{signal strength ratios}
\begin{eqnarray}
&&\mu_{\gamma\gamma,VV^*}^{ggF}(125)=\frac{\sigma_{NP}(ggF)}{\sigma_{SM}(ggF)}
\frac{Br_{NP}(h^{125}\rightarrow \gamma\gamma,VV^*)}{Br_{SM}(h^{125}\rightarrow \gamma\gamma,VV^*)},(V=Z,W),\nonumber \\&&\mu_{f\bar{f}}^{VBF}(125)=\frac{\sigma_{NP}(VBF)}{\sigma_{SM}(VBF)}
\frac{Br_{NP}(h^{125}\rightarrow f\bar{f})}{Br_{SM}(h^{125}\rightarrow f\bar{f})},(f=b,\tau, c),
\label{mu}
\end{eqnarray}
where $ggF$ and $VBF$ stand for gluon-gluon fusion and vector boson fusion respectively. The Higgs production cross sections can be further simplified as
$\frac{\sigma_{NP}(ggF)}{\sigma_{SM}(ggF)}\approx
\frac{\Gamma_{NP}(h^{125}\rightarrow gg)}{\Gamma_{SM}(h^{125}\rightarrow gg)},\;
\frac{\sigma_{NP}(VBF)}{\sigma_{SM}(VBF)}\approx
\frac{\Gamma_{NP}(h^{125}\rightarrow VV^*)}{\Gamma_{SM}(h^{125}\rightarrow VV^*)}$. Besides, $Br_{NP}(\cdot\cdot\cdot)=\frac{\Gamma_{NP}(\cdot\cdot\cdot)}{\Gamma_{NP}^{h^{125}}}$ and $Br_{SM}(\cdot\cdot\cdot)=\frac{\Gamma_{SM}(\cdot\cdot\cdot)}{\Gamma_{SM}^{h^{125}}}$, $\Gamma_{NP}^{h^{125}}=\sum_{f=b,\tau, c}\Gamma_{NP}(h^{125}\rightarrow f\bar f)+\sum_{V=Z,W}\Gamma_{NP}(h^{125}\rightarrow VV^*)+\Gamma_{NP}(h^{125}\rightarrow gg)+\Gamma_{NP}(h^{125}\rightarrow \gamma\gamma)$ denotes the NP total decay width of 125 GeV Higgs boson, and $\Gamma_{SM}^{h^{125}}$ denotes the SM one. Then we can simplify the signal strengths originating from 125 GeV Higgs boson decay channels.
\begin{eqnarray}
&&\mu_{\gamma\gamma}^{ggF}(125)\approx\frac{\Gamma_{SM}^{h^{125}}}{\Gamma_{NP}^{h^{125}}}\frac{\Gamma_{NP}(h^{125}\rightarrow gg)}{\Gamma_{SM}(h^{125}\rightarrow gg)}
\frac{\Gamma_{NP}(h^{125}\rightarrow \gamma\gamma)}{\Gamma_{SM}(h^{125}\rightarrow\gamma\gamma)},\nonumber \\
&&\mu_{VV^*}^{ggF}(125)\approx\frac{\Gamma_{SM}^{h^{125}}}{\Gamma_{NP}^{h^{125}}}\frac{\Gamma_{NP}(h^{125}\rightarrow gg)}{\Gamma_{SM}(h^{125}\rightarrow gg)}|g_{h^{125}_{S}VV}|^2,(V=Z,W),\nonumber \\
&&\mu_{f\bar{f}}^{VBF}(125)\approx\frac{\Gamma_{SM}^{h^{125}}}{\Gamma_{NP}^{h^{125}}}|g_{h^{125}_{S}VV}|^2|g_{h^{125}ff}|^2,(f=b,\tau, c).
\end{eqnarray}
Here, $\frac{\Gamma_{NP}(h^{125}\rightarrow VV^*)}{\Gamma_{SM}(h^{125}\rightarrow VV^*)}=|g_{h^{125}_{S}VV}|^2$ and $\frac{\Gamma_{NP}(h^{125}\rightarrow f\bar{f})}{\Gamma_{SM}(h^{125}\rightarrow f\bar{f})}=|g_{h^{125}ff}|^2$ with $|g_{h^{125}ff}|^2= |g_{h^{125}_{S}ff}|^2+|g_{h^{125}_{A}ff}|^2$. $S$ representing the CP-even scalar Higgs component and $A$ representing the CP-odd scalar Higgs component. The concrete expressions of $g^{B-L}_{h^{125}_{S,A}ff}$ and the following $g_{{h^{125}_{A}\tilde{f}\tilde{f}}}^{B-L}$ $g^{B-L}_{{h^{125}_{S,A}\chi_i^+\chi_i^-}},\cdot\cdot\cdot$ are collected in the Appendix D.

The gluon fusion is the primary way to produce the Higgs boson at the LHC. In the NB-LSSM, the decay width of $ h^0\rightarrow gg$ with one-loop contributions is shown as\cite{CPHiggs dacay,h2glon2gamma1,h2glon2gamma2,h2glon2gamma3}:
\begin{eqnarray}
&&\Gamma_{{NP}}(h^{125}\rightarrow gg)={G_{F}\alpha_s^2m_{{h^{125}}}^3\over64\sqrt{2}\pi^3}
\Big|\sum\limits_{q=u,d}g_{{h^{125}_{S}qq}} A_{1/2}(x_q)
+\sum\limits_{\tilde q=\tilde U,\tilde D}g_{{h^{125}_{S}\tilde{q}\tilde{q}}}{m_{{\rm Z}}^2\over m_{{\tilde q}}^2}A_{0}(x_{{\tilde{q}}})\Big|^2 \nonumber\\
&&\hspace{3.1cm}+{G_{F}\alpha_s^2m_{{h^{125}}}^3\over64\sqrt{2}\pi^3}\Big|\sum\limits_{q=u,d}g_{{h^{125}_{A}qq}} A_{2} (x_q)\Big|^2,\label{hgg}
\end{eqnarray}
with $x_a=m_{{h^{125}}}^2/(4m_a^2)$, $q$ and $\tilde{q}$ correspond to NB-LSSM quarks and squarks respectively. The form factors $A_{1/2}(x)$, $A_{2}(x)$, $A_0(x)$ and the following $A_1(x)$, $F(x)$ are defined in the Appendix E.
The decay width of $h^{125}\rightarrow \gamma\gamma$ with one-loop corrections can be expressed as\cite{CPHiggs dacay,h2glon2gamma1,h2glon2gamma2,h2glon2gamma3}:
\begin{eqnarray}
&&\Gamma_{{NP}}(h^{125}\rightarrow\gamma\gamma)\hspace{-0.1cm}=\hspace{-0.1cm}{G_{F}\alpha^2m_{{h^{125}}}^3\over128\sqrt{2}\pi^3}
\Big|\sum\limits_{f=u,d,l} N_c Q_{f}^2 g_{{h^{125}_{S}ff}} A_{1/2}(x_f)+\sum\limits_{\tilde f=\tilde U,\tilde D,\tilde L}N_cQ_{f}^2g_{{h^{125}_{S}\tilde{f}\tilde{f}}}{m_{ Z}^2\over m_{{\tilde f}}^2} A_{0}(x_{{\tilde{f}}})\nonumber\\
&&\hspace{3.cm}+g_{{h^{125}_{S}H^+H^-}}{m_{{\rm Z}}^2\over m_{{H^\pm}}^2}A_0(x_{{H^\pm}})\hspace{-0.1cm}+\hspace{-0.1cm}g_{{h^{125}_{S}WW}}A_1(x_{{\rm W}})\hspace{-0.1cm}+\hspace{-0.1cm}\sum\limits_{i=1}^2\hspace{-0.1cm}{m_{{\rm W}}\over m_{{\chi_i^{\pm}}}}g_{{h^{125}_{S}\chi_i^+\chi_i^-}} A_{1/2}(x_{{\chi_i^{\pm}}})\Big|^2\nonumber\\
&&\hspace{3.cm}+{G_{F}\alpha^2m_{{h^{125}}}^3\over128\sqrt{2}\pi^3}\Big|\sum\limits_{f=u,d,l} N_c Q_{f}^2 g_{{h^{125}_{A}ff}} A_{2}(x_f)+\sum\limits_{i=1}^2{m_{{\rm W}}\over m_{{\chi_i^{\pm}}}}g_{{h^{125}_{A}\chi_i^+\chi_i^-}} A_{2}(x_{{\chi_i^{\pm}}})\Big|^2.
\label{hpp}
\end{eqnarray}
The decay widths for $h^0\rightarrow ZZ, WW^*$ can be shown as\cite{h2W2Z}: 
\begin{eqnarray}
&&\Gamma(h^{125}\rightarrow WW^*)={3e^4m_{{h^{125}}}\over512\pi^3s_{ W}^4}|g_{h^{125}_{S}WW}|^2
F({m_{_{\rm W}}\over m_{h^{125}}}),\;\nonumber\\
&&\Gamma(h^{125}\rightarrow ZZ)={e^4m_{{h^{125}}}\over2048\pi^3s_{W}^4c_{W}^4}|g_{h^{125}_{S}ZZ}|^2
\Big(7-{40\over3}s_{W}^2+{160\over9}s_{W}^4\Big)F({m_{Z}\over m_{_{h^{125}}}}).
\end{eqnarray}
The decay width of $h^{125}\rightarrow f\bar{f}$ can be deduced with the Born approximation\cite{CPHiggs dacay}: 
\begin{eqnarray}
\Gamma(h^{125}\rightarrow f\bar{f})=N_c{G_Fm_f^2m_{{h^{125}}}\over4\sqrt{2}\pi}|g_{h^{125}ff}|^2
(1-{4m_f^2\over m_{h^{125}}^2})^{3/2}.\label{hff}
\end{eqnarray}

\subsection{The 95 GeV scalar excess signals }
The signal strengths for the 95 GeV scalar excess can be quantified by the following ratios:
\begin{eqnarray}
&&\mu_{\gamma\gamma}(95)=\frac{\sigma_{NP}(gg\rightarrow h^{95})}{\sigma_{SM}(gg\rightarrow h^{95})}
\frac{Br_{NP}(h^{95}\rightarrow \gamma\gamma)}{Br_{SM}(h^{95}\rightarrow \gamma\gamma)}\nonumber \\&&\hspace{1.5cm}\approx\frac{\Gamma_{NP}(h^{95}\rightarrow gg)}{\Gamma_{SM}(h^{95}\rightarrow gg)}
\frac{\Gamma_{NP}(h^{95}\rightarrow \gamma\gamma)}{\Gamma_{SM}(h^{95}\rightarrow\gamma\gamma)}\times\frac{\Gamma_{SM}^{h^{95}}}{\Gamma_{NP}^{h^{95}}},\nonumber \\&&\mu_{b\bar{b}}(95)=\frac{\sigma_{NP}(Z^*\rightarrow Zh^{95})}{\sigma_{SM}(Z^*\rightarrow Zh^{95})}
\frac{Br_{NP}(h^{95}\rightarrow b\bar{b})}{Br_{SM}(h^{95}\rightarrow b\bar{b})}\nonumber \\
&&\hspace{1.5cm}\approx|g_{h^{95}_{S}ZZ}|^2\times\frac{\Gamma_{NP}(h^{95}\rightarrow b\bar{b})}{\Gamma_{SM}(h^{95}\rightarrow b\bar{b})}\times\frac{\Gamma_{SM}^{h^{95}}}{\Gamma_{NP}^{h^{95}}},
\end{eqnarray}
here, $\Gamma_{NP}^{h^{95}}\approx \sum_{f=b,\tau, c}\Gamma_{NP}(h^{95}\rightarrow f\bar f)+\Gamma_{NP}(h^{95}\rightarrow gg)$ denotes the NP total decay width of 95 GeV scalar excess, and $\Gamma_{SM}^{h^{95}}$ denotes the SM one. The functions of $\Gamma_{NP(SM)}(h^{95}\rightarrow gg,\gamma\gamma, f\bar f)$ are similar as Eqs.(\ref{hgg}), (\ref{hpp}), (\ref{hff}) with $h^{125}$ changing to $h^{95}$.
\section{Numerical analyses}
Taking the lightest Higgs boson mass at around 95 GeV and the second-lightest Higgs boson as the 125 GeV SM-like Higgs boson, we discuss the numerical results of these Higgs boson masses and the corresponding signal strengths in the NB-LSSM. To investigate the numerical evaluations, we consider some experimental constraints:

1. Ref.\cite{Zpupper} indicates that the $Z'$ boson mass satisfies $M'_Z\geq 5.15~ {\rm TeV}$ with $95\%$ confidence level (CL). An upper bound of the ratio between the $Z'$ boson mass and its gauge coupling is given by Refs.\cite{Zpupper1,Zpupper2} $\frac{M'_Z}{g_B}\geq 6~{\rm TeV}$ with $99\%$ CL, so $g_B$ is restricted in the region of $0 < g_B \leq0.85$.

2. The large $\tan\beta$ has been excluded by the $\bar{B}\rightarrow X_s\gamma$ experiment\cite{BSgamma1,BSgamma2}.

3. The LHC experimental data constrains $\tan\beta'< 1.5$\cite{tanB}.

4. The mass constraints of some supersymmetric particles are considered: The slepton mass is greater than 0.7 TeV, the neutralino mass is larger than 0.5 TeV, the chargino mass is greater than 0.9 TeV\cite{PDG2024} and the charged Higgs mass is greater than 0.6 TeV\cite{ATLAS2024hya}. To coincide with the constraints from the direct searches of squarks and gluinos at the LHC\cite{squarks1,squarks2} and the observed Higgs signal\cite{squarks3}, we take the masses of squarks and gluinos larger than 2 TeV.

5. The second-lightest Higgs boson is aligned to the SM-like Higgs boson, whose mass and decays $(h^{125} \rightarrow \gamma\gamma, ZZ, WW^*, b \bar b, \tau \bar\tau)$ are all constrained within $3\sigma$ experimental limits. The lightest Higgs boson is used to simulate a hypothetical scenario with mass around 95 GeV, and thereby the mass of the lightest Higgs boson is limited within the region of 93-97 GeV, and the corresponding decays $h^{95} \rightarrow \gamma\gamma, b \bar b$ both satisfy $3\sigma$ experimental limits.

6. We consider the experimental constraints of the rare decays $h^{125}\rightarrow l_i l_j$ and $l_j\rightarrow l_i \gamma$. These processes influenced by some new particles and interactions beyond SM have been discussed specifically in our previous work\cite{LFV Higgs decay}, and the corresponding parameter spaces coinciding with the current experimental constraints are also suitable in this work.

7. We consider the experimental constraints of the lepton electric dipole moment (EDM). In our previous work, we have already studied the two-loop corrections of lepton EDM in a supersymmetric model\cite{EDM}. We find that the parameters from the new particles such as chargino, neutralino, slepton can significantly affect the EDM, but seldom influence the Higgs mass and decays. Hence, we can coincide with the experimental constraints of EDM in this work by some suitable parameter values from the new particles such as chargino, neutralino, slepton. The specific research of lepton EDM in the NB-LSSM will be discussed in our latter work.

In order to study how generic the Higgs boson mass of 95 GeV is, we consider the $\chi^2$ analyses for the corresponding theoretical and experimental data of Higgs boson, in which the theoretical values obtained for our model $\mu_{\xi}^{theo}$ are confronted with the experimental measurements $\mu_{\xi}^{exp}$, $\delta_\xi$ represents the statistic and systematic errors.
\begin{eqnarray}
&&\chi^2=\sum_\xi(\frac{\mu_{\xi}^{theo}-\mu_{\xi}^{exp}}{\delta_\xi})^2
\nonumber\\
 &&\hspace{0.6cm}=(\frac{m_{h^{125}}-m_{h^{125}}^{exp}}{\delta_{m_{h}^{125}}})^2
+(\frac{\mu_{\gamma\gamma}(125)-\mu_{\gamma\gamma}^{exp}(125)}{\delta_{\mu_{\gamma\gamma}^{125}}})^2
+(\frac{\mu_{WW^*}(125)-\mu_{WW^*}^{exp}(125)}{\delta_{\mu_{WW^*}}^{125}})^2
\nonumber\\
 &&\hspace{0.6cm}+(\frac{\mu_{ZZ}(125)-\mu_{ZZ}^{exp}(125)}{\delta_{\mu_{ZZ}^{125}}})^2
+(\frac{\mu_{b\bar{b}}(125)-\mu_{b\bar{b}}^{exp}(125)}{\delta_{\mu_{b\bar{b}}^{125}}})^2
+(\frac{\mu_{\tau\bar{\tau}}(125)-\mu_{\tau\bar{\tau}}^{exp}(125)}{\delta_{\mu_{\tau\bar{\tau}}^{125}}})^2
\nonumber\\
 &&\hspace{0.6cm}+(\frac{\mu_{\gamma\gamma}(95)-\mu_{\gamma\gamma}^{exp}(95)}{\delta_{\mu_{\gamma\gamma}^{95}}})^2
+(\frac{(\mu_{b\bar{b}}(95)-\mu_{b\bar{b}}^{exp}(95)}{\delta_{\mu_{b\bar{b}}^{95}}})^2.
\end{eqnarray}

\begin{figure}
\centering
\includegraphics[width=0.325\textwidth]{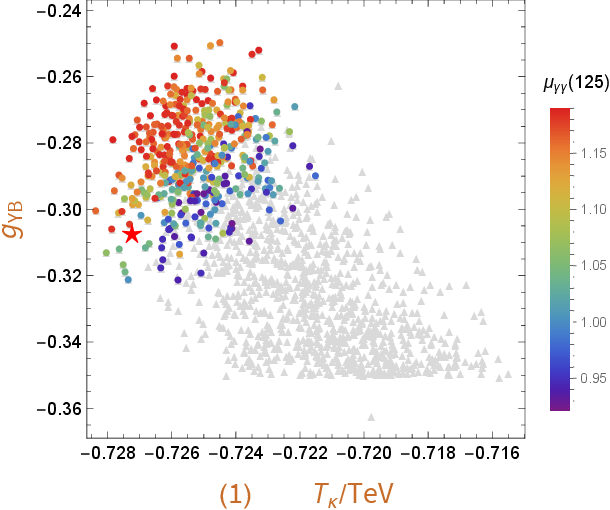}
\hspace{0.1cm}\includegraphics[width=0.325\textwidth]{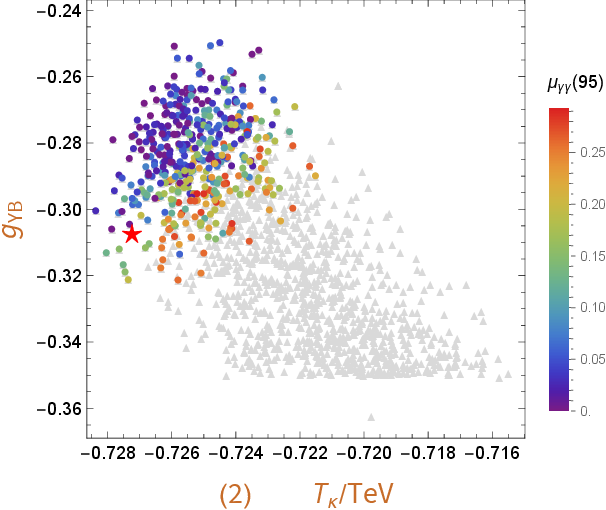}
\hspace{0.1cm}\includegraphics[width=0.32\textwidth]{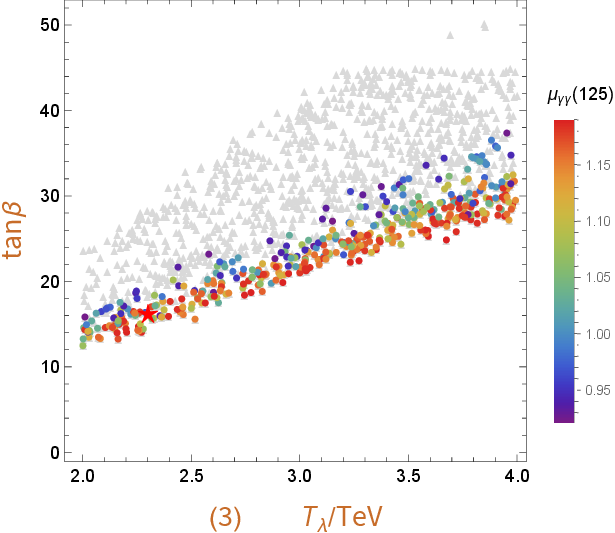}\\
\includegraphics[width=0.32\textwidth]{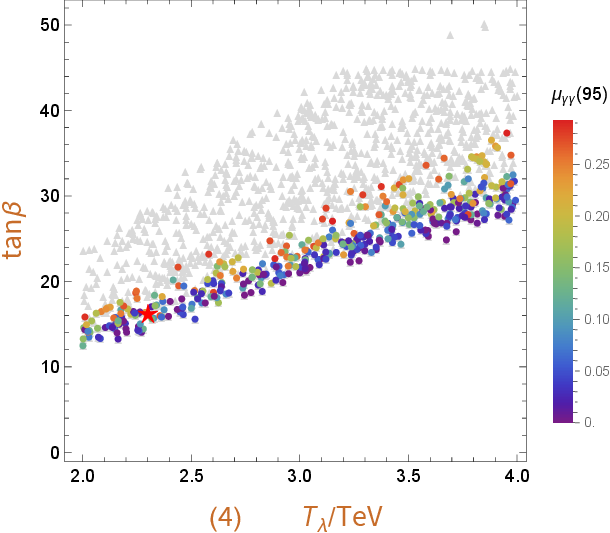}
\hspace{0.1cm}\includegraphics[width=0.325\textwidth]{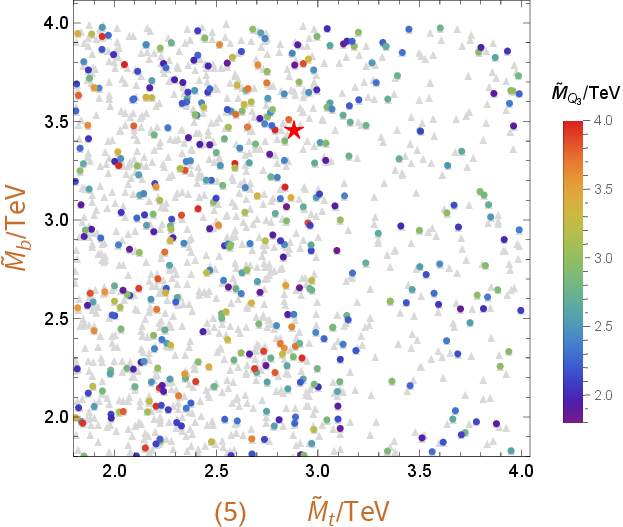}
\hspace{0.1cm}\includegraphics[width=0.32\textwidth]{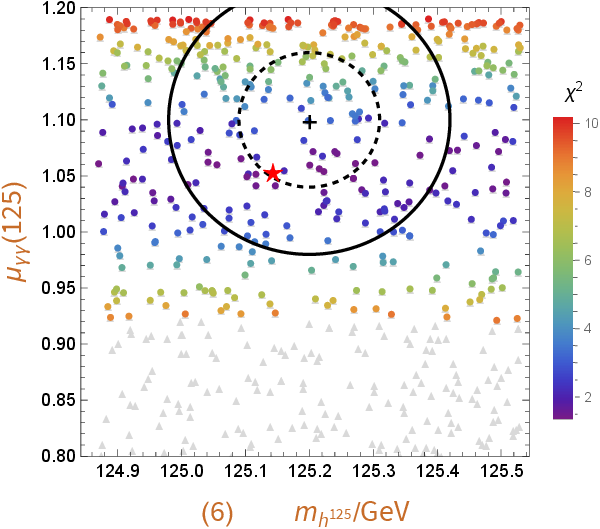}\\
\includegraphics[width=0.325\textwidth]{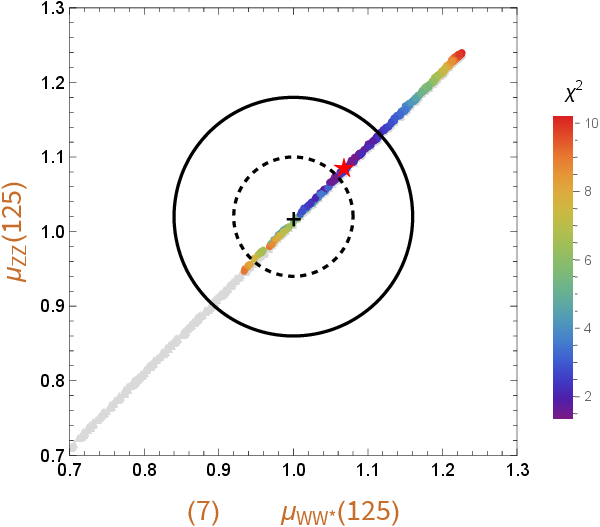}
\hspace{0.1cm}\includegraphics[width=0.32\textwidth]{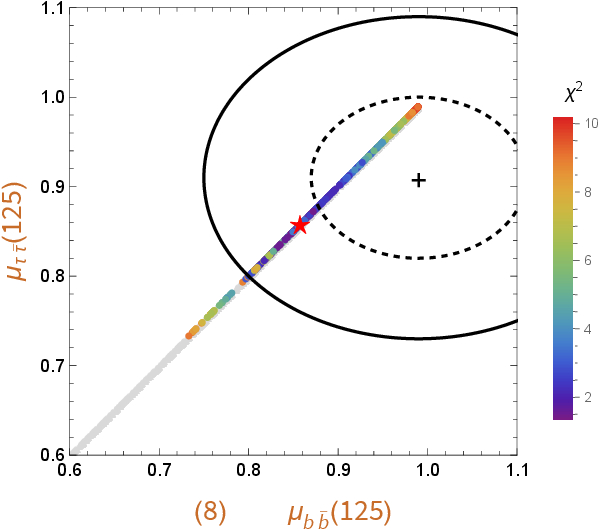}
\hspace{0.1cm}\includegraphics[width=0.32\textwidth]{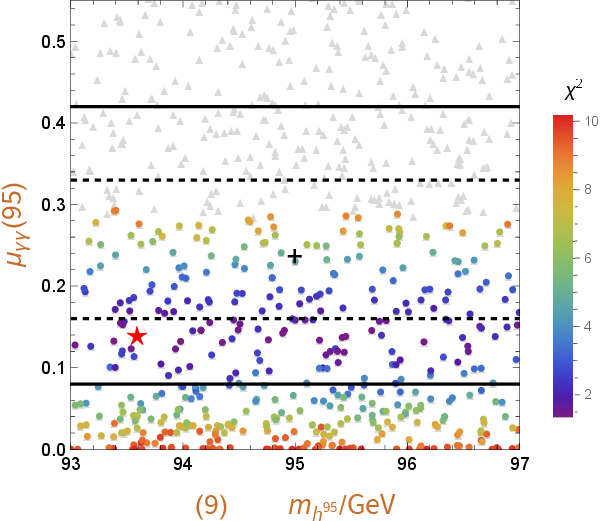}\\
\includegraphics[width=0.32\textwidth]{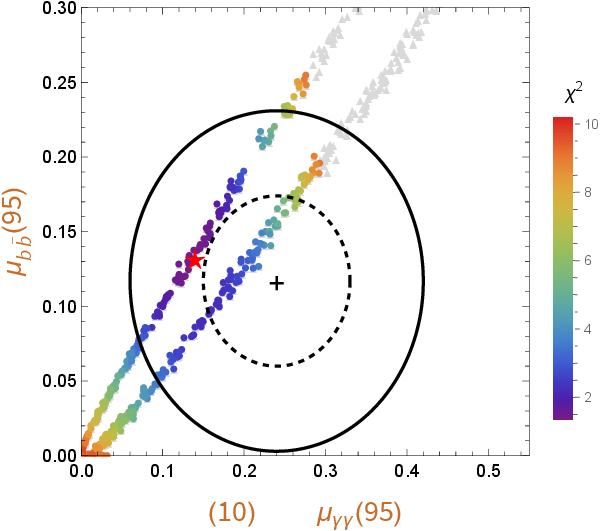}
\hspace{0.1cm}\includegraphics[width=0.32\textwidth]{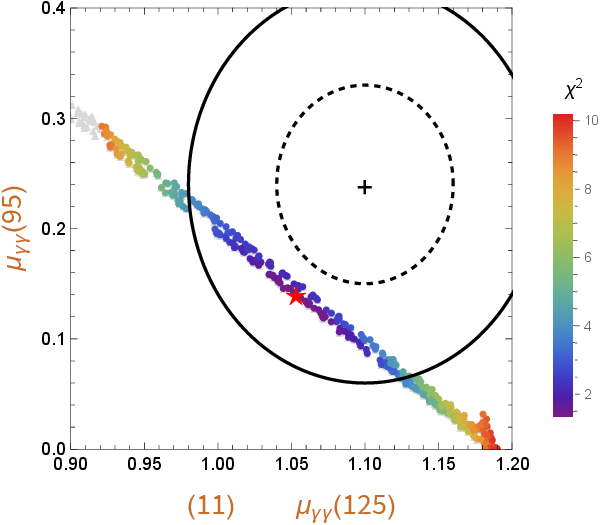}
\hspace{0.1cm}\includegraphics[width=0.325\textwidth]{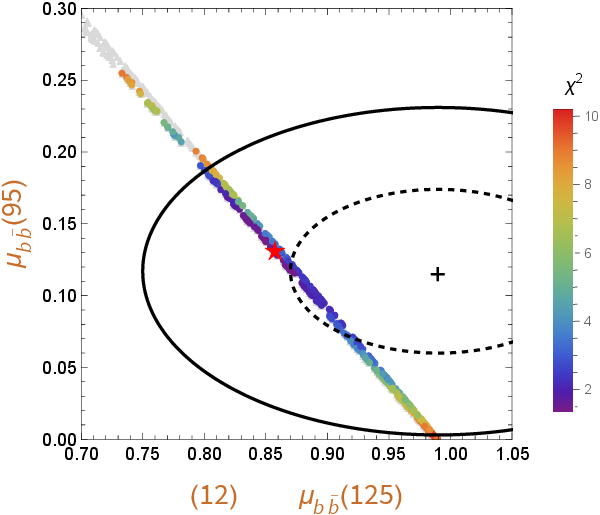}
\caption[]{(1)-(5): The scatter points for parameters $T_{\kappa},g_{YB},T_{\lambda},\tan\beta, \tilde{M}_t,\tilde{M}_b,\tilde{M}_{Q_3}$ and signal strengths of $\mu_{\gamma\gamma}(125)$ and $\mu_{\gamma\gamma}(95)$. (6)-(12): The scatter points for Higgs masses and signal strengths.}
\label{fig111}
\end{figure}

We select 9 random variables in the suitable regions (here we do not consider the CP violation phases): $\tan\beta\in(15,55)$, $g_{YB}\in(-0.45,-0.05)$,
$T_\kappa\in(-2,-0.1)~\mathrm{TeV}$, $T_\lambda\in(2,4)~\mathrm{TeV}$, $\tilde{M}_{Q_3},\tilde{M}_{t},\tilde{M}_{b} \in(1.8,4)~\mathrm{TeV}$ and $A_{t},A_{b} \in(0.3,2)~\mathrm{TeV}$. In FIG.\ref{fig111}, the gray triangle scatter points satisfy $3\sigma$ constraints of SM-like Higgs detections for the second-lightest Higgs boson, and the lightest Higgs boson mass falls within the range of 93 GeV $\sim$ 97 GeV. Then we further consider that the Higgs signal strengths $\mu_{\gamma\gamma,\; WW^*,\; ZZ,\; b\bar{b},\; \tau\bar{\tau}}(125)$ and $\mu_{\gamma\gamma,\; b\bar{b}}(95)$ all satisfy $3\sigma$ experimental limits, the reasonable parameter space is selected to rainbow scatter points, which also satisfy $95\%$ CL constraint in the domain $\chi^2\leq16.92$ with 9 freedom parameters. The red star represents the best-fit point ($\chi^2_{best}=1.336$). "+" is confronted with the experimental measured central value, and black dashed (real) line is the corresponding $1\sigma$ ($2\sigma$) uncertainty.

FIG.\ref{fig111}(1)-(5) show that the parameter values are $g_{YB}\simeq -0.31$,  $\tan\beta\simeq16$, $T_{\kappa}\simeq-0.727~\mathrm{TeV}$, $T_{\lambda}\simeq2.3~\mathrm{TeV}$, $\tilde{M}_{Q_3}\simeq1.9~\mathrm{TeV}$, $\tilde{M}_{t}\simeq2.9~\mathrm{TeV}$ and $\tilde{M}_{b}\simeq3.4~\mathrm{TeV}$ corresponding to the best-fit point $\chi^2_{best}=1.336$. As the sensitive parameters, $g_{YB}$ and $T_{\kappa}$ are limited within the small regions $-0.32\leq g_{YB}\leq-0.25$ and $-0.729~\mathrm{TeV}\leq T_{\kappa}\leq-0.722~\mathrm{TeV}$ after considering all constraints, while the larger $\mu_{\gamma\gamma}(125)$ (smaller $\mu_{\gamma\gamma}(95)$) will be obtained by pushing $g_{YB}$ towards -0.25. In FIG.\ref{fig111}(3) and (4), the parameter $\tan\beta$ is confined to the range of 11 to 37.5 by the experimental limits, and gradually increases as the value of $T_{\lambda}$ increases. When $T_{\lambda}$ takes a certain value, the larger value of $\tan\beta$ will ensure Higgs signal strength $\mu_{\gamma\gamma}(125)$ ($\mu_{\gamma\gamma}(95)$) smaller (larger). In contrast, when $\tan\beta$ takes a certain value, the larger value of $T_{\lambda}$ may ensure the Higgs signal strength $\mu_{\gamma\gamma}(125)$ ($\mu_{\gamma\gamma}(95)$) larger (smaller). FIG.\ref{fig111}(5) indicates that the scatter points of $\tilde{M}_{Q_3}$ and $\tilde{M}_{t}$ between 1.8 and 3 TeV are significantly denser than those in other ranges, and there are almost no points for $\tilde{M}_{Q_3}\geq3$ TeV as $\tilde{M}_{t}\geq3$ TeV. Furthermore, $\tilde{M}_{b}$ has little influence on the numerical results and can be used throughout the selected parameter space.

In addition, we study the scatter points of Higgs mass and signal strengths in FIG.\ref{fig111}(6)-(12). FIG.\ref{fig111}(6) points out that the largest value of $\mu_{\gamma\gamma}(125)$ tends to 1.2, which can cause a deviation about 20\% from the SM and prove the considerable NP contributions from the NB-LSSM. One can observe that the values of $\mu(h^{125})_{ZZ}$ ($\mu(h^{125})_{\tau\bar{\tau}}$, $\mu(h^{95})_{b\bar{b}}$) overall increase with the increasing value of $\mu(h^{125})_{WW^*}$ ($\mu(h^{125})_{b\bar{b}}$, $\mu_{\gamma\gamma}(95)$) in FIG.\ref{fig111}(7), (8), (10), while the values of $\mu_{\gamma\gamma}(125)$ ( $\mu(h^{125})_{b\bar{b}}$) decrease as the values of $\mu_{\gamma\gamma}(95)$ ($\mu(h^{95})_{b\bar{b}})$ enlarge. And the Higgs mass and the concrete signal strengths tend to $m_{h^{125}}\simeq125.14$ GeV, $m_{h^{95}}\simeq93.6$ GeV, $\mu_{\gamma\gamma}(125)\simeq1.05, \mu_{WW^*}(125)\simeq1.07, \mu_{ZZ}(125)\simeq1.08, \mu_{b\bar{b}}(125)\simeq0.86, \mu_{\tau\bar{\tau}}(125)\simeq0.86$ and $\mu_{\gamma\gamma}(95)\simeq0.14, \mu_{b\bar{b}}(95)\simeq0.13$ corresponding to the best-fit point $\chi^2_{best}=1.336$. Besides, there are parameter points within the dashed lines ($1\sigma$ uncertainty intervals), which means that the parameter space we selected under the NB-LSSM can fit the current Higgs experimental observations (the LHC detections for SM-like Higgs boson mass and decays, the LHC diphoton 95 GeV excess and the $b\bar{b}$ 95 GeV excess observed at LEP) well. Therefore, we have a relatively high possibility of simultaneously fitting the 95 GeV excess and the 125 GeV SM-like Higgs boson in the NB-LSSM.
\begin{figure}[t]
\centering
\includegraphics[width=0.4\textwidth]{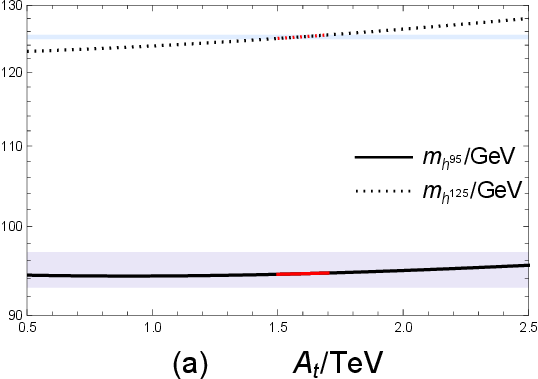}
\hspace{0.5cm}\includegraphics[width=0.4\textwidth]{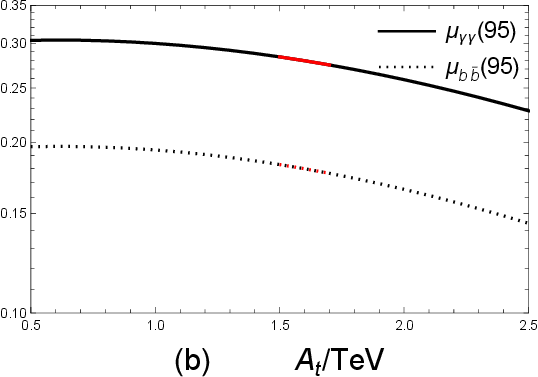}\\
\includegraphics[width=0.3\textwidth]{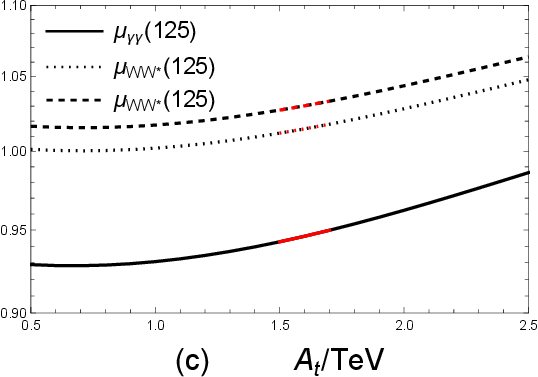}
\hspace{0.5cm}\includegraphics[width=0.3\textwidth]{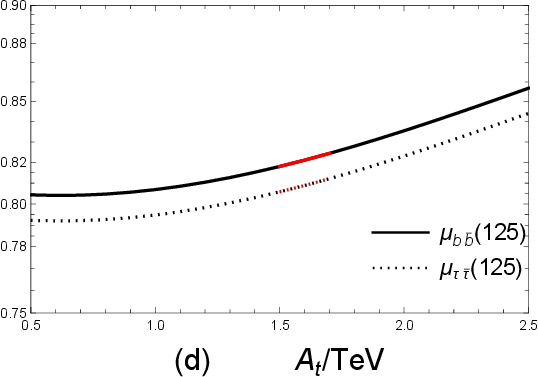}
\hspace{0.5cm}\includegraphics[width=0.3\textwidth]{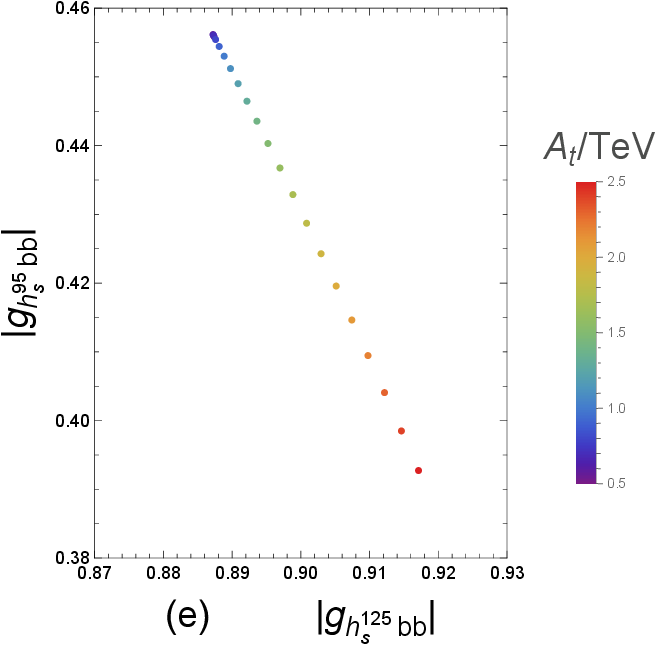}
\caption[]{The Higgs boson masses and signal strengths change with parameter $A_t$, where the red lines meet the range of (93 GeV $\sim$ 97 GeV) and $3\sigma$ experimental limit of SM-like Higgs boson mass. The red lines in following figures satisfy the same constraints as here.}
\label{fig1}
\end{figure}
\begin{figure}[t]
\centering
\includegraphics[width=0.4\textwidth]{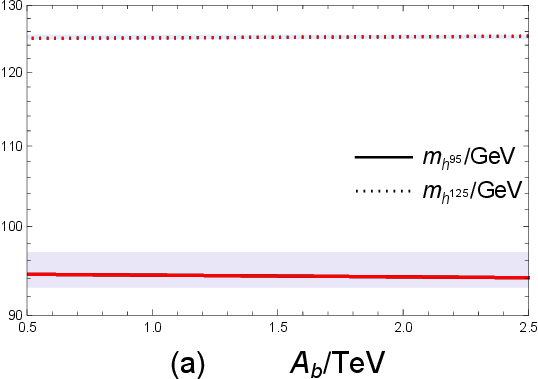}
\hspace{0.5cm}\includegraphics[width=0.4\textwidth]{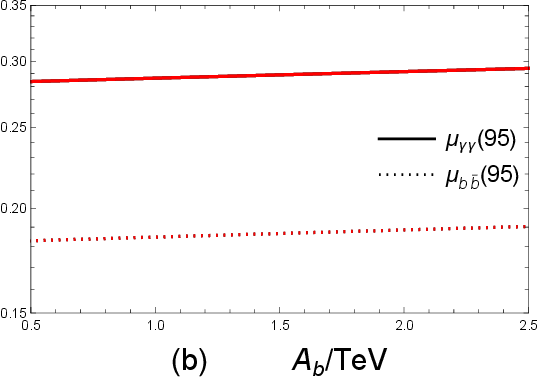}\\
\includegraphics[width=0.4\textwidth]{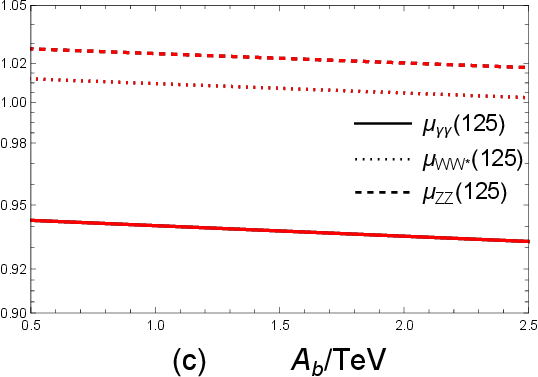}
\hspace{0.5cm}\includegraphics[width=0.4\textwidth]{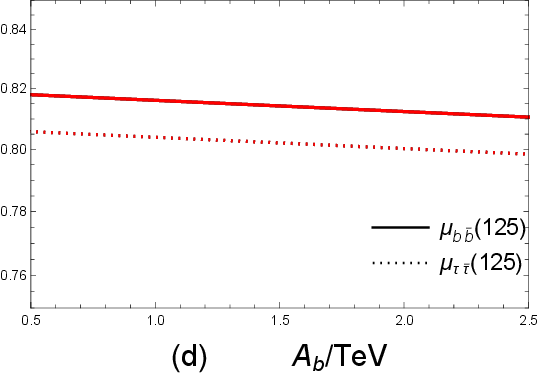}
\caption[]{The Higgs boson masses and signal strengths change with parameter $A_b$.}
\label{fig2}
\end{figure}

Then, we further investigate the Higgs mass and corresponding signal strengths within the CP-conserved NB-LSSM (Here we take the definition $T_{t}=A_{t}=|A_{t}|e^{i\theta_4},T_{b}=A_{b}=|A_{b}|e^{i\theta_5}, T_{\kappa}=|T_{\kappa}|e^{i\theta_6},T_{\lambda}=|T_{\lambda}|e^{i\theta_7},
T_{\lambda_2}=|T_{\lambda_2}|e^{i\theta_8},T_{e}=A_{e}=|A_{e}|e^{i\theta_9}$ with the CP-violating phases $\theta_i=0$). In order to discuss the effect of parameters on numerical results, we select some suitable parameter spaces.
\begin{eqnarray}
&& |A_{t}|=1.7~\mathrm{TeV},\;|A_{b}|=0.5~\mathrm{TeV},\; |A_{e}|=0.3~\mathrm{TeV},\; |M_{2}|=0.6~\mathrm{TeV},\;\Lambda=1~\mathrm{TeV},\nonumber\\
&&|T_{\kappa}|=-0.725~\mathrm{TeV},\;|T_{\lambda}|=2.6~\mathrm{TeV},\;|T_{\lambda_2}|=0.04~\mathrm{TeV},\;
\tilde{M}_{t}=2.7~\mathrm{TeV},\nonumber\\
&&{M}_{\tilde g}=\tilde{M}_{Q_3}=2.5~\mathrm{TeV},\;\tilde{M}_{b}=2.3~\mathrm{TeV},\;M_{EE}=M_{LL}=v_S=1.5~\mathrm{TeV}, \nonumber\\
&&g_{B}=\kappa=0.3,\; \lambda=0.5,\;\lambda_2=-0.4,\;g_{YB}=-0.3,\; \tan \beta=19,\; \tan \beta^{\prime}=1.4.
\end{eqnarray}

In FIG.\ref{fig1}, we study the Higgs mass and corresponding signal strengths changing with parameter $A_t$. Parameter $A_t$ is the non-diagonal element of stop quark matrix, which affects the Higgs masses by the effective potential corrections related to stop quarks. In FIG.\ref{fig1}(a), the gray area represents the lightest Higgs boson falling within the range of 93 GeV $\sim$ 97 GeV, while the blue area represents the second-lightest Higgs boson falling within the $3\sigma$ experimental error range (124.87 GeV $\sim$ 125.53 GeV). We find that when parameter $A_t$ changes from $0.5~\mathrm{TeV}$ to $2.5~\mathrm{TeV}$, the mass of the lightest Higgs boson which satisfies the constraints (93 GeV $\sim$ 97 GeV) increases slowly. Meanwhile, the parameter $A_t$ is further limited in a small region $1.5\sim1.7~\mathrm{TeV}$ with the $3\sigma$ experimental limit of the second-lightest Higgs boson mass. In FIG.\ref{fig1}(b)-(d), the signal strengths $\mu_{\gamma\gamma}(125)$, $\mu_{WW^*}(125)$, $\mu_{ZZ}(125)$, $\mu_{b\bar{b}}(125)$ and $\mu_{\tau\bar{\tau}}(125)$ all enlarge with the increasing $A_t$, and the signal strengths $\mu_{\gamma\gamma}(95)$ and $\mu_{b\bar{b}}(95)$ both decrease with the increasing $A_t$. Under the experimental limitations of Higgs mass, $\mu_{\gamma\gamma}(125)\simeq0.95$, $\mu_{WW^*}(125)\simeq1.01$, $\mu_{ZZ}(125)\simeq1.03$, $\mu_{bb}(125)\simeq0.82$, $\mu_{\tau\bar{\tau}}(125)\simeq0.8$, $\mu_{\gamma\gamma}(95)\simeq0.28$ and $\mu_{b\bar{b}}(95)\simeq0.18$, which all fit well with the current experiments of Higgs decays. The influence from parameter $A_b$ is figured in FIG.\ref{fig2}, which indicates that parameter $A_b$ possesses opposite effects on Higgs masses and decays comparing the trends of FIG.\ref{fig1}. Whilst $A_b$ is a non-sensitive parameter as all the values of $A_b$ gently affect Higgs masses and decays.

The curves in FIG.\ref{fig1}(b) and FIG.\ref{fig1}(c),(d) show opposite trends of changes, and this feature is present in FIG.\ref{fig111}(11),(12) and all the images below. To understand more clearly the changes in signal strengths at around 95 GeV and 125 GeV, we choose the signal strengthes $\mu_{b\bar{b}}(125)$ and $\mu_{b\bar{b}}(95)$ for example, figure the absolute values of the related couplings $g_{h^{125}_sbb}$ and $g_{h^{95}_sbb}$ in FIG.\ref{fig1}(e), and discuss the influence of parameter $A_t$ on the couplings $|g_{h^{125}_sbb}|$ and $|g_{h^{95}_sbb}|$. The figure shows that as the parameter $A_t$ increases, $g_{h^{95}_sbb} $ gradually decreases, while $g_{h^{125}_sbb}$ gradually increases, which is in perfect correspondence with the curve distributions in FIG.\ref{fig1}(b) and FIG.\ref{fig1}(c),(d). So the reason why the signal strengths $\mu_{b\bar{b}}(125)$ and $\mu_{b\bar{b}}(95)$ rendering instead changes may come from the relevant Higgs couplings.

\begin{figure}[t]
\centering
\includegraphics[width=0.4\textwidth]{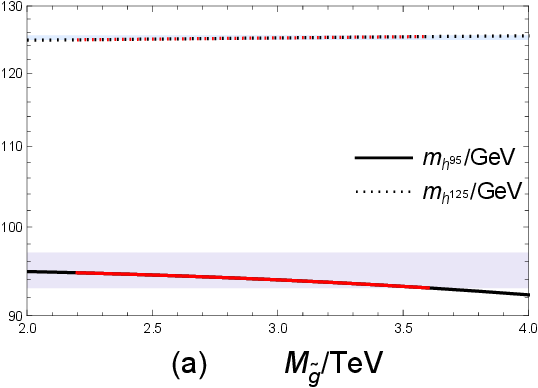}
\hspace{0.5cm}\includegraphics[width=0.4\textwidth]{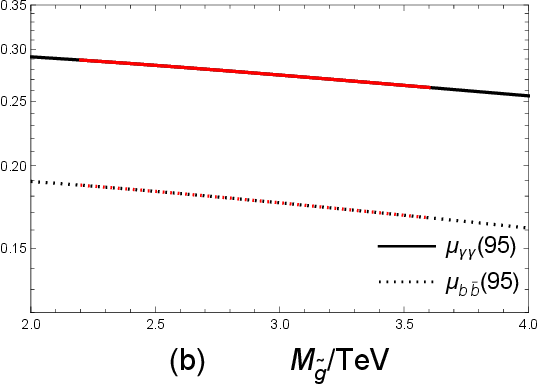}\\
\includegraphics[width=0.4\textwidth]{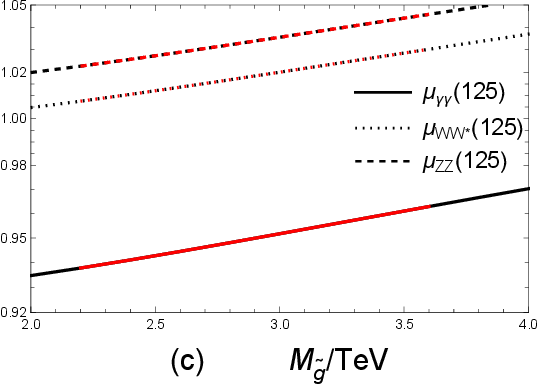}
\hspace{0.5cm}\includegraphics[width=0.4\textwidth]{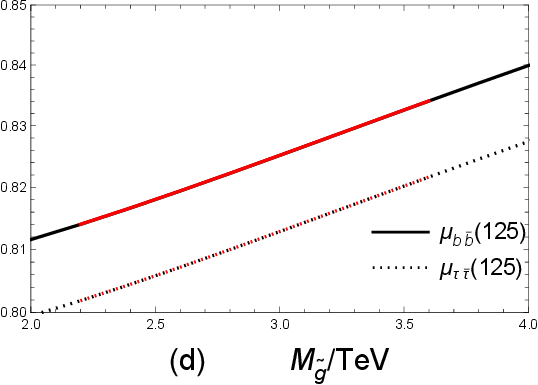}
\caption[]{The Higgs boson masses and corresponding signal strengths change with parameter $M_{\tilde{g}}$.}
\label{fig3}
\end{figure}
\begin{figure}[t]
\centering
\includegraphics[width=0.4\textwidth]{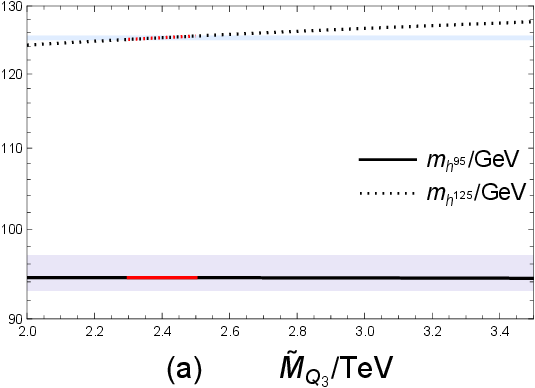}
\hspace{0.5cm}\includegraphics[width=0.4\textwidth]{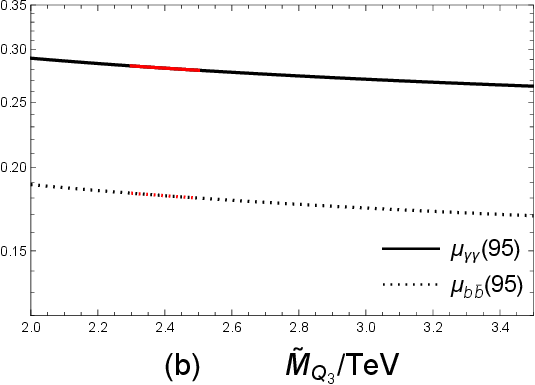}\\
\includegraphics[width=0.4\textwidth]{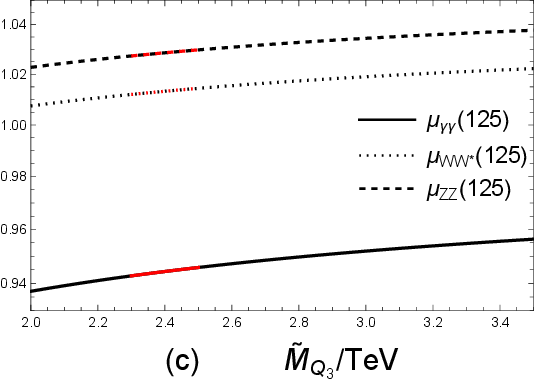}
\hspace{0.5cm}\includegraphics[width=0.4\textwidth]{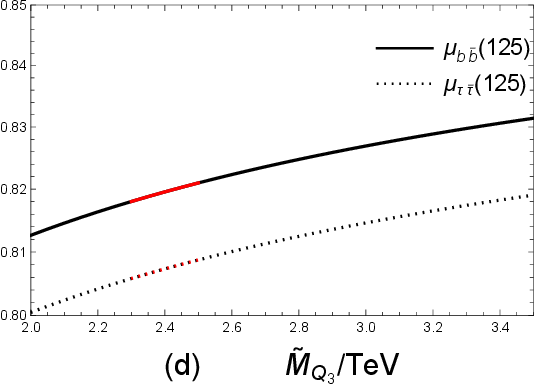}
\caption[]{The Higgs boson masses and corresponding signal strengths change with parameter $\tilde{M}_{Q_3}$.}
\label{fig4}
\end{figure}
\begin{figure}[t]
\centering
\includegraphics[width=0.4\textwidth]{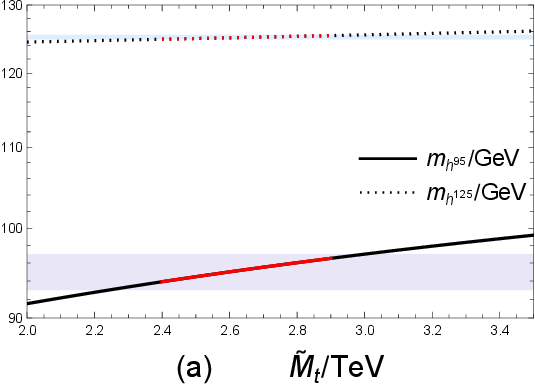}
\hspace{0.5cm}\includegraphics[width=0.4\textwidth]{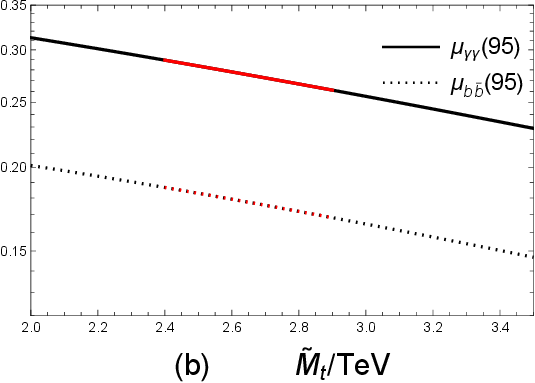}\\
\includegraphics[width=0.4\textwidth]{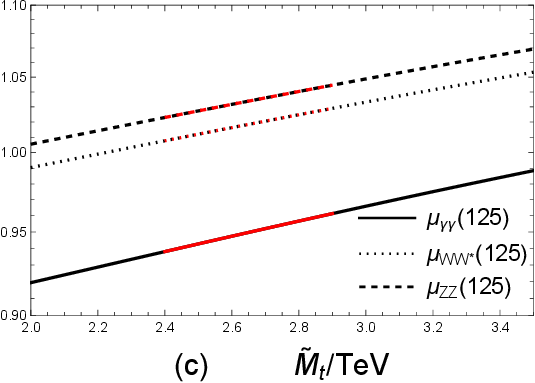}
\hspace{0.5cm}\includegraphics[width=0.4\textwidth]{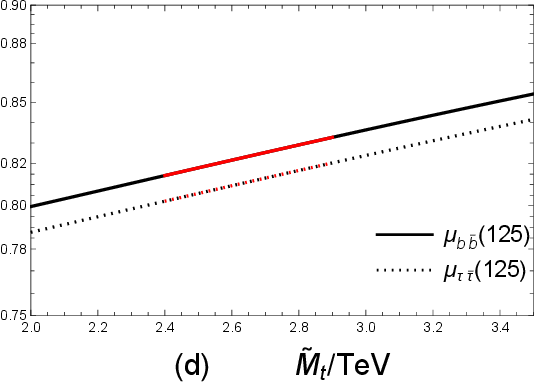}
\caption[]{The Higgs boson masses and corresponding signal strengths change with parameter $\tilde{M}_{t}$.}
\label{fig5}
\end{figure}

The gluinos can affect Higgs masses through two-loop effective potential. As the diagonal element of stop quark matrix, parameters $\tilde{M}_{Q_3}$ and $\tilde M_{{t}}$ both correct the Higgs masses and signal strengths by stop particle. In this part, we study how parameters ${M}_{\tilde g}$, $\tilde{M}_{Q_3}$ and $\tilde{M}_{t}$ influence the numerical results in FIG.\ref{fig3}-FIG.\ref{fig5}. It is easy to see that experimental constraint of the SM-like Higgs boson is strict, and the lightest and second-lightest Higgs boson mass can be both within the experimental restrictions as $2.2~{\rm TeV}<\tilde{M}_{Q_3}<3.6~{\rm TeV}$, $2.3~{\rm TeV}<\tilde{M}_{Q_3}<2.5~{\rm TeV}$ and ${2.4~\rm TeV}<\tilde{M}_{t}<2.9~{\rm TeV}$. The reasonable regions of these parameters also reveal that $\tilde{M}_{Q_3}$ and $\tilde{M}_{t}$ are more sensitive parameters than ${M}_{\tilde g}$. $\mu_{\gamma\gamma}(95)$ and $\mu_{b\bar{b}}(95)$ both decrease with the enlarging ${M}_{\tilde g}$, $\tilde{M}_{Q_3}$ or $\tilde{M}_{t}$, while the signal strengths of second-lightest Higgs boson all increase with the enlarging ${M}_{\tilde g}$, $\tilde{M}_{Q_3}$ or $\tilde{M}_{t}$. All the signal strengths corresponding to the suitable parameters ${M}_{\tilde g}$, $\tilde{M}_{Q_3}$ and $\tilde{M}_{t}$ can satisfy the current experimental constraints of Higgs decays well.

\begin{figure}[t]
\centering
\includegraphics[width=0.4\textwidth]{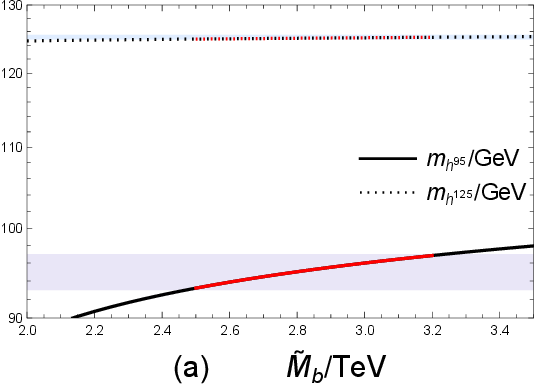}
\hspace{0.5cm}\includegraphics[width=0.4\textwidth]{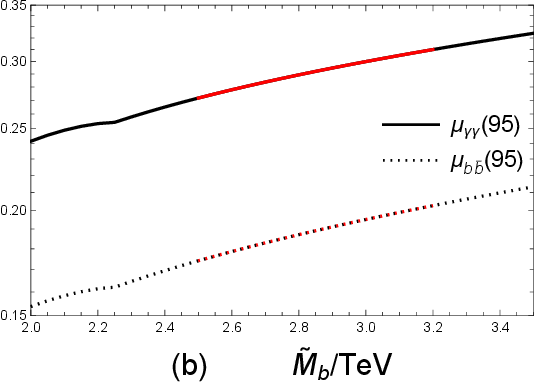}\\
\includegraphics[width=0.4\textwidth]{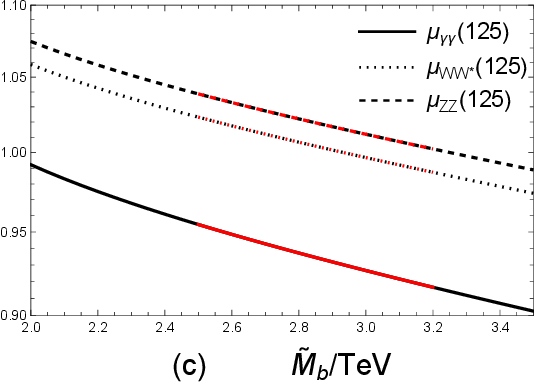}
\hspace{0.5cm}\includegraphics[width=0.4\textwidth]{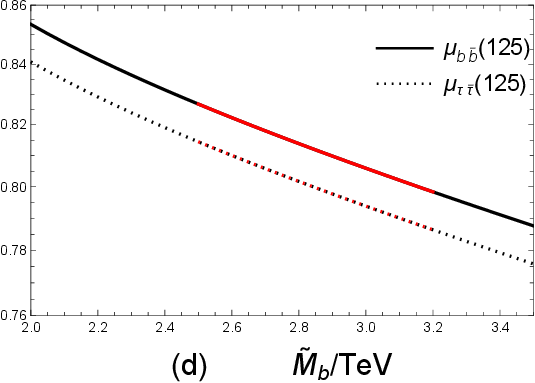}
\caption[]{The Higgs boson masses and corresponding signal strengths change with parameter $\tilde{M}_{b}$.}
\label{fig6}
\end{figure}
We then study the Higgs masses and corresponding signal strengths changing with parameters $\tilde{M}_{b}$ in FIG.\ref{fig6}. As the diagonal element of sbottom quark matrix, parameters $\tilde M_{{b}}$ correct the Higgs masses and signal strengths by sbottom quarks. The experimental constraints from the SM-like Higgs boson mass are not strict on the parameter $\tilde M_{{b}}$, but it is constrained strictly in the region of 2.5 $\sim$ 3.2 TeV after considering the constraints of lightest Higgs mass. Besides, we can see that the signal strengths fitted the experimental constraints well in FIG.\ref{fig6}(b) pronounce upward trend and the ones in FIG.\ref{fig6}(c)-(d) pronounce downward trend with the increasing $\tilde M_{{b}}$. Above analyses indicate that $\tilde M_{{b}}$ is a sensitive parameter to the Higgs masses and signal strengths.

\begin{figure}[t]
\centering
\includegraphics[width=0.4\textwidth]{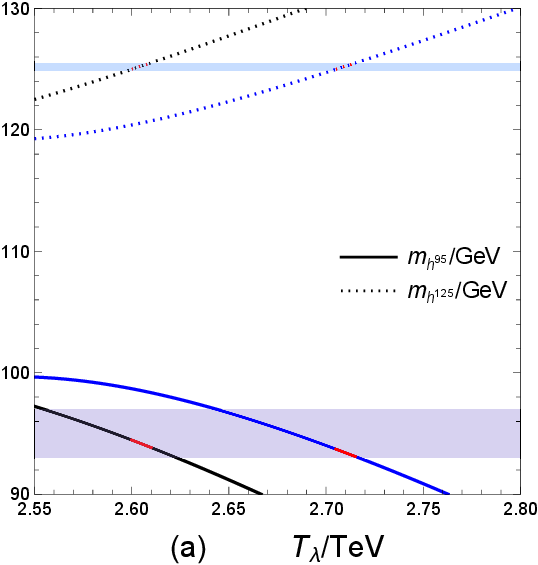}
\hspace{0.5cm}\includegraphics[width=0.4\textwidth]{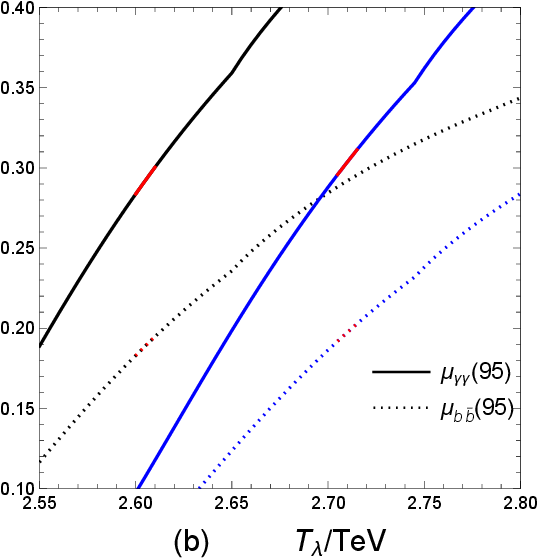}\\
\includegraphics[width=0.4\textwidth]{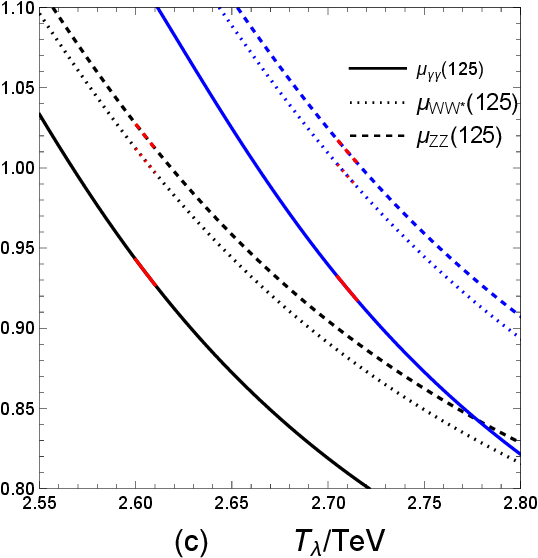}
\hspace{0.5cm}\includegraphics[width=0.4\textwidth]{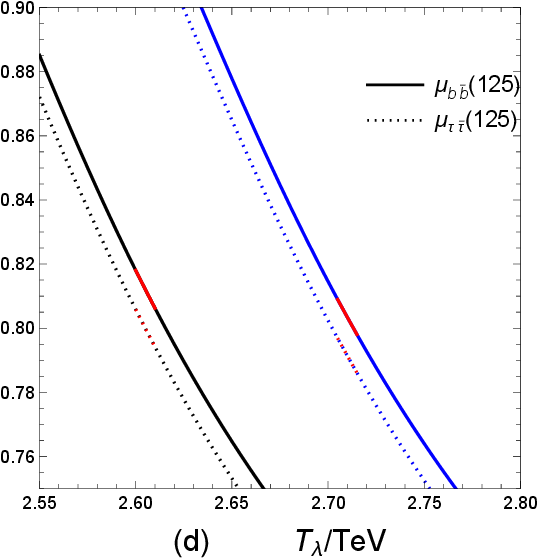}
\caption[]{The Higgs boson masses and corresponding signal strengths change with parameter $T_\lambda$.}
\label{fig7}
\end{figure}
In FIG.\ref{fig7}, we discuss the numerical results influenced by the soft trilinear Higgs mixing mass $T_\lambda$, which is a new introduced parameters in the NB-LSSM. FIG.\ref{fig7}(a) shows the parameter $T_\lambda$ is obviously constrained around $2.6~{\rm TeV}$ as $\tan\beta=19$ (the black line) and $T_\lambda\simeq2.7~{\rm TeV}$ as $\tan\beta=20$ (the blue line) by the limitations of the lightest and second-lightest Higgs boson masses, thereby the values of $T_\lambda$ and $\tan\beta$ are determined by each other if we hope to satisfy $3\sigma$ constraints of SM-like Higgs mass detections for the second-lightest Higgs boson and the range of 93 GeV $\sim$ 97 GeV for the lightest Higgs boson mass spontaneously. Furthermore, the signal strengths for the 95 GeV scalar excess all increase quickly and the ones for the SM-like Higgs boson all decrease quickly with the increasing $T_\lambda$. So the Higgs masses and signal strengths are highly influenced by parameters $T_\lambda$ and $\tan\beta$.

\begin{figure}[t]
\centering
\includegraphics[width=0.4\textwidth]{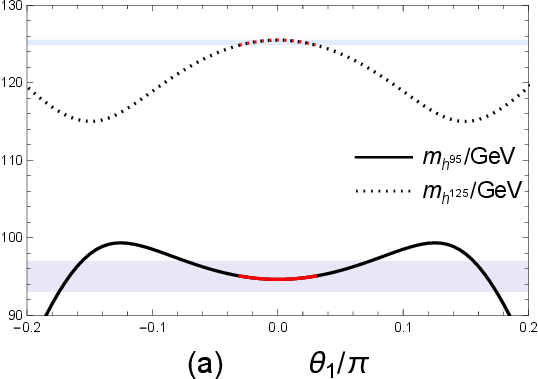}
\hspace{0.5cm}\includegraphics[width=0.4\textwidth]{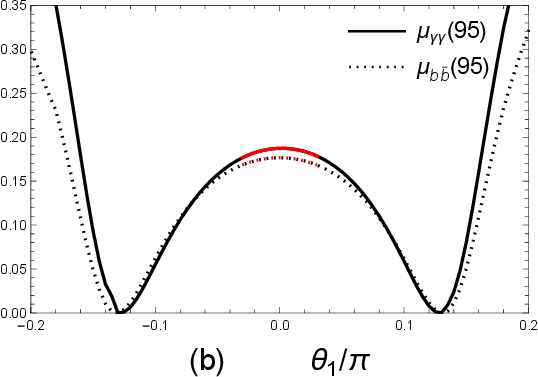}\\
\includegraphics[width=0.4\textwidth]{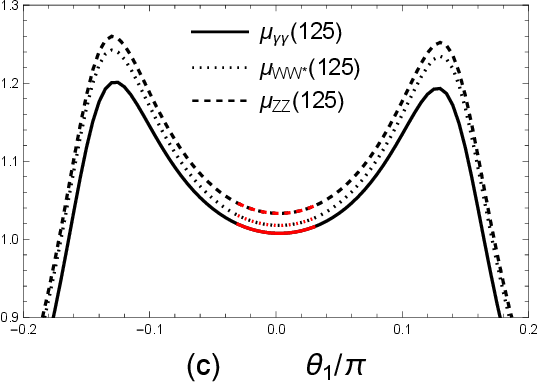}
\hspace{0.5cm}\includegraphics[width=0.4\textwidth]{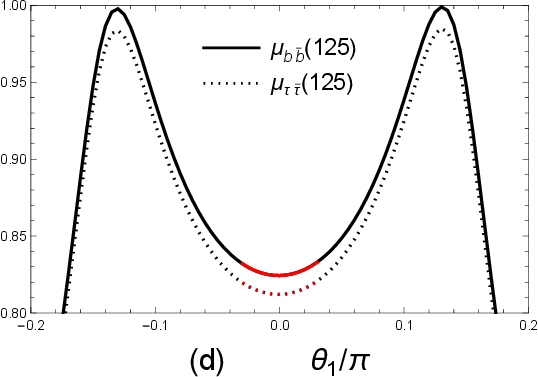}
\caption[]{The Higgs boson mass and signal strengths vary with CP phase $\theta_1$.}
\label{fig8}
\end{figure}
In the following part, we study the contributions from some CP phases. We first study the numerical results changing with CP phase $\theta_1$ when other CP phases equal zero. As the relative phase of $v_u$ and $v_d$, $\theta_1$ is limited in a small region around $|0\sim0.08\pi|$ and $|0.155\pi\sim0.175\pi|$ by the experimental limitation of the lightest Higgs boson mass, and further constrained around $-0.03\pi\sim0.03\pi$ by the experimental constraint of the SM-like Higgs boson mass, and the corresponding signal strengths then all fit the current experimental limits. So CP phase $\theta_1$ has a sensitive effect on the Higgs boson masses. In spite of the signal strengths of lightest (second-lightest) Higgs boson first decrease (increase) and then increase (decease) quickly with the enlarging absolute value $|\theta_1|$, the Higgs signal strengths all change slightly when satisfying the constraints of Higgs boson masses, which indicates that the reasonable CP phase $\theta_1$ may lead to a tiny fluctuation on the Higgs signal strengths.

\begin{figure}[t]
\centering
\includegraphics[width=0.4\textwidth]{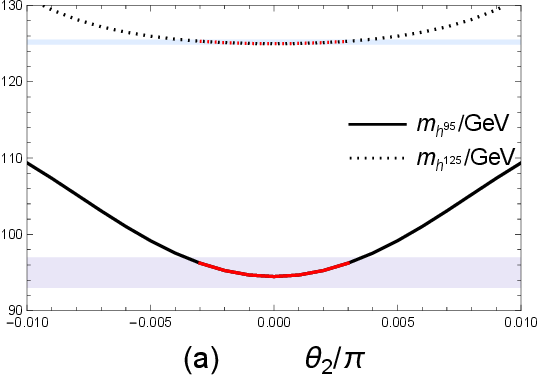}
\hspace{0.5cm}\includegraphics[width=0.4\textwidth]{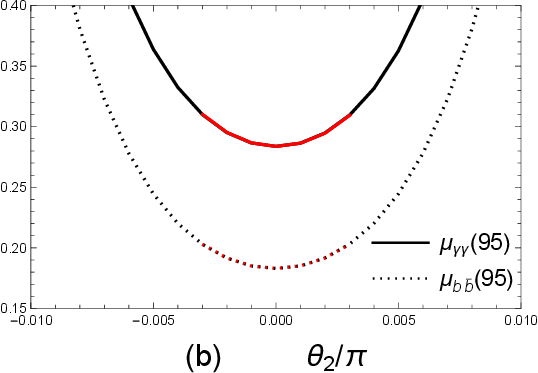}\\
\includegraphics[width=0.4\textwidth]{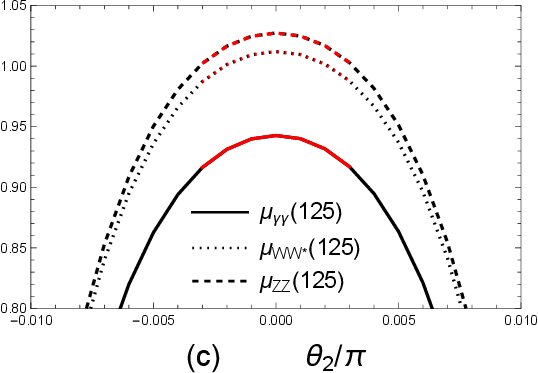}
\hspace{0.5cm}\includegraphics[width=0.4\textwidth]{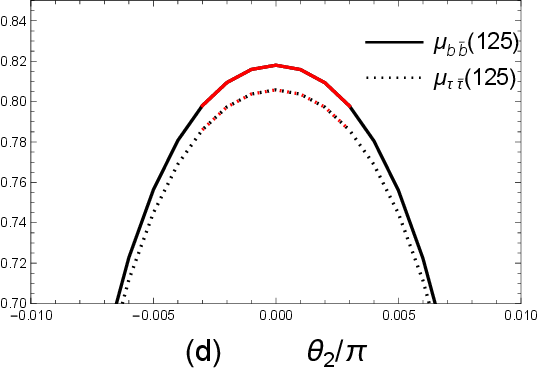}
\caption[]{The Higgs boson masses and signal strengths vary with CP phase $\theta_2$.}
\label{fig9}
\end{figure}
\begin{figure}[t]
\centering
\includegraphics[width=0.4\textwidth]{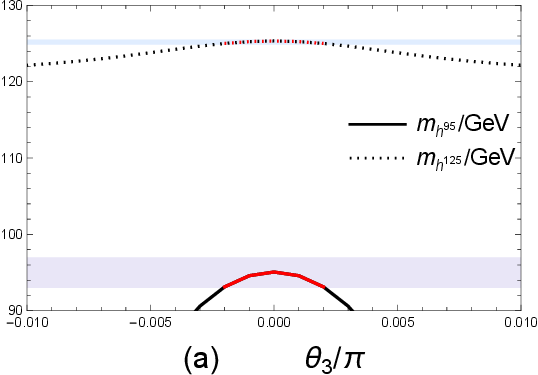}
\hspace{0.5cm}\includegraphics[width=0.4\textwidth]{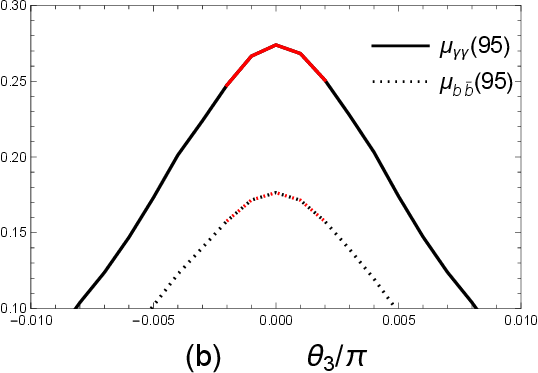}\\
\includegraphics[width=0.4\textwidth]{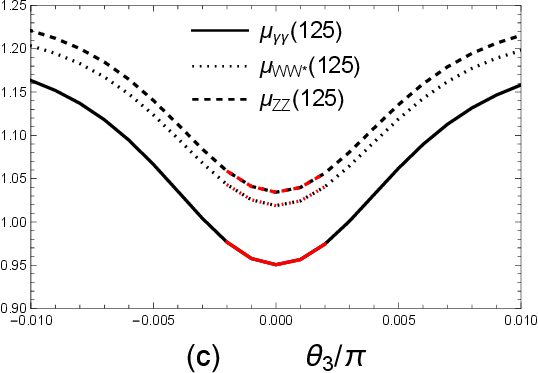}
\hspace{0.5cm}\includegraphics[width=0.4\textwidth]{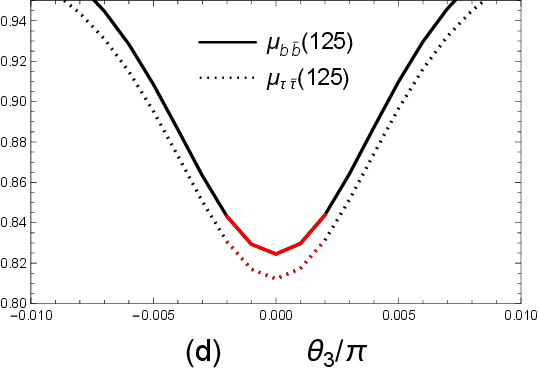}
\caption[]{The Higgs boson masses and signal strengths vary with CP phase $\theta_3$.}
\label{fig9-1}
\end{figure}
In FIG.\ref{fig9} (FIG.\ref{fig9-1}), we study the numerical results changing with CP phase $\theta_2$ ($\theta_3$) when other CP phases equal zero. As the relative phase of $v_\eta$ and $v_{\bar{\eta}}$, the two Higgs boson masses both increase rapidly with the increasing absolute $\theta_2$, so $\theta_2$ as a sensitive parameter is limited in a small region around $-0.003\pi\sim0.003\pi$ by the experimental constraint of Higgs boson masses. On the contrary, the two Higgs boson masses both decrease quickly with the increasing absolute $\theta_3$, and thereby $\theta_3$ is limited in a small region around $-0.002\pi\sim0.002\pi$ by the experimental constraints of Higgs boson masses, hence $\theta_3$ is a sensitive parameter as well. Under the aforementioned constraints, the Higgs signal strengths are all limited in small regions and fit the current experimental limits well.

\begin{figure}[t]
\centering
\includegraphics[width=0.4\textwidth]{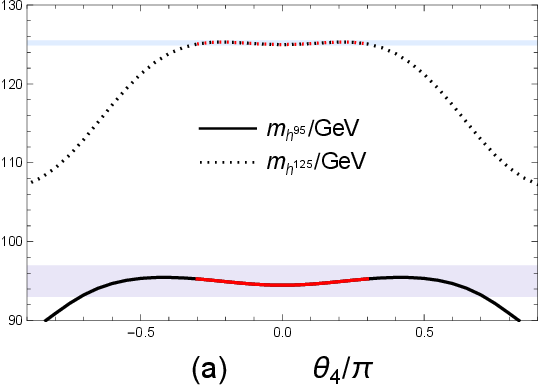}
\hspace{0.5cm}\includegraphics[width=0.4\textwidth]{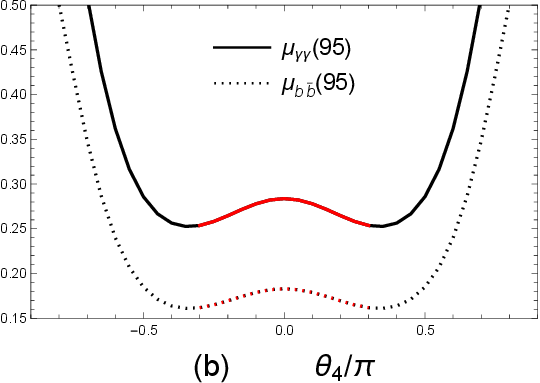}\\
\includegraphics[width=0.4\textwidth]{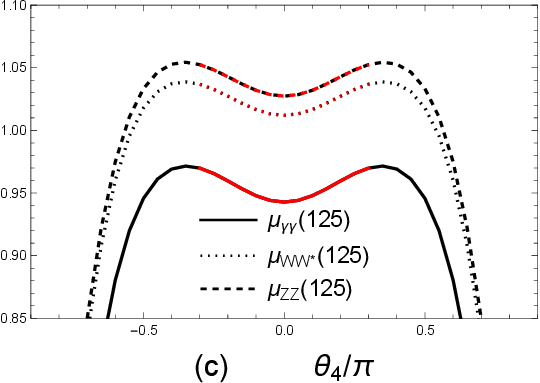}
\hspace{0.5cm}\includegraphics[width=0.4\textwidth]{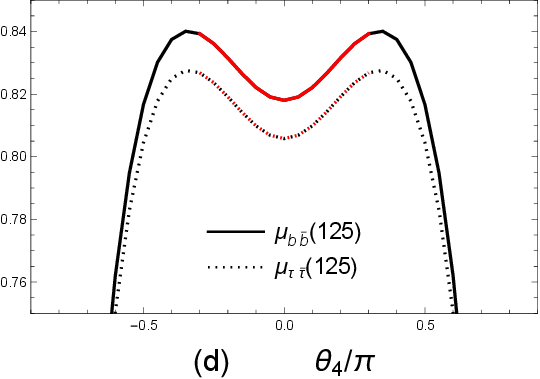}
\caption[]{The Higgs boson masses and signal strengths vary with CP phase $\theta_4$.}
\label{fig10}
\end{figure}
\begin{figure}[t]
\centering
\includegraphics[width=0.4\textwidth]{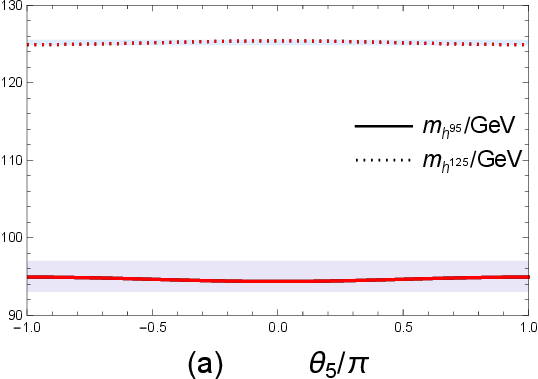}
\hspace{0.5cm}\includegraphics[width=0.4\textwidth]{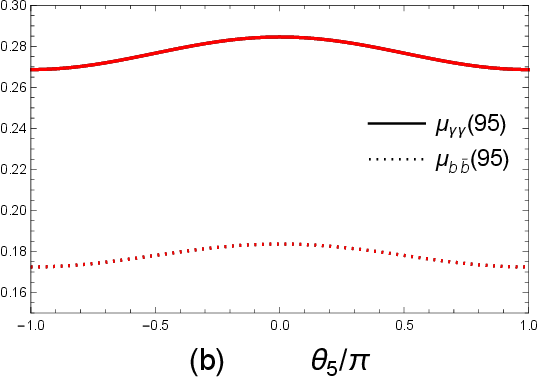}\\
\includegraphics[width=0.4\textwidth]{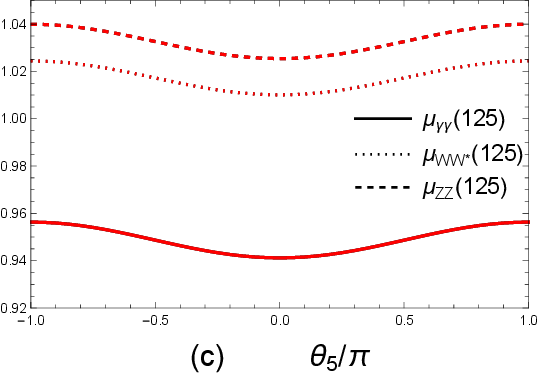}
\hspace{0.5cm}\includegraphics[width=0.4\textwidth]{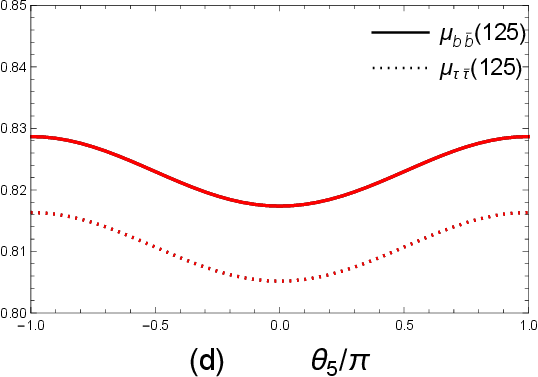}
\caption[]{The Higgs boson masses and signal strengths vary with CP phase $\theta_5$.}
\label{fig11}
\end{figure}
In FIG.\ref{fig10} and FIG.\ref{fig11}, we consider the numerical results fluctuated with the CP phases $\theta_4$ and $\theta_5$, which relate to the non-diagonal elements ($A_t$ and $A_b$) of stop and sbottom quark matrices. With the $3\sigma$ experimental limitation of second-lightest Higgs boson mass, the CP phase $\theta_4$ is restricted in the region of $-0.3\pi\sim0.3\pi$. Furthermore, the signal strengths of lightest (second-lightest) Higgs boson with slight changes can be smaller (larger) as $\theta_4$ trends to $|0.3\pi|$, which means that the numerical results can be influenced by the $\theta_4$ in a certain degree. However, FIG.\ref{fig11} addresses that $\theta_5$ can take the values from $-\pi$ to $\pi$ and has a slight influence on both the Higgs masses and the corresponding signal strengths, so the contributions from $\theta_5$ can be ignored.

\begin{figure}[t]
\centering
\includegraphics[width=0.4\textwidth]{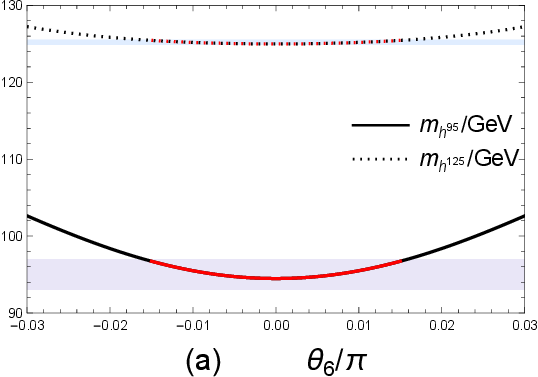}
\hspace{0.5cm}\includegraphics[width=0.4\textwidth]{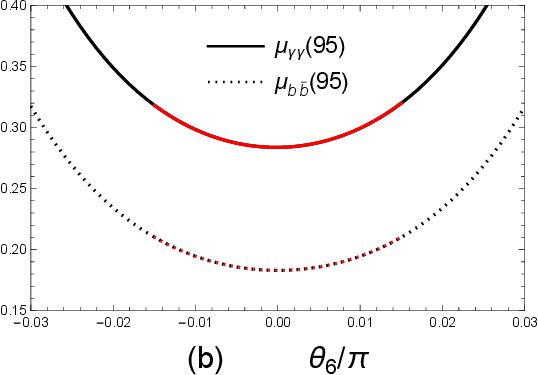}\\
\includegraphics[width=0.4\textwidth]{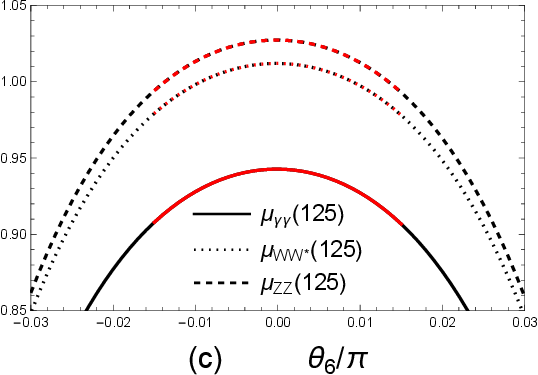}
\hspace{0.5cm}\includegraphics[width=0.4\textwidth]{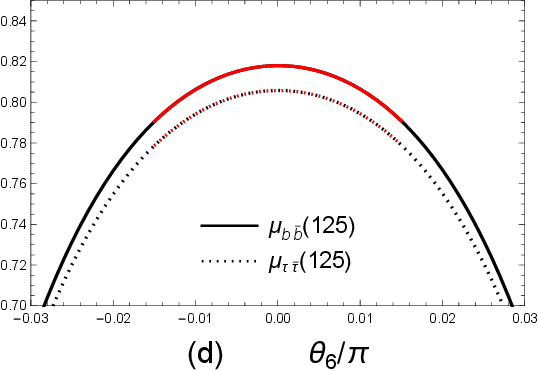}
\caption[]{The Higgs boson masses and signal strengths vary with CP phase $\theta_6$.}
\label{fig12}
\end{figure}
\begin{figure}[t]
\centering
\includegraphics[width=0.4\textwidth]{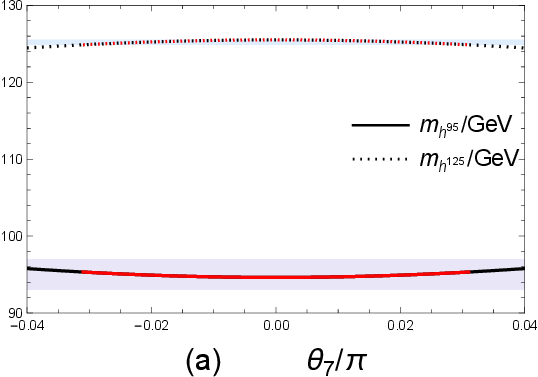}
\hspace{0.5cm}\includegraphics[width=0.4\textwidth]{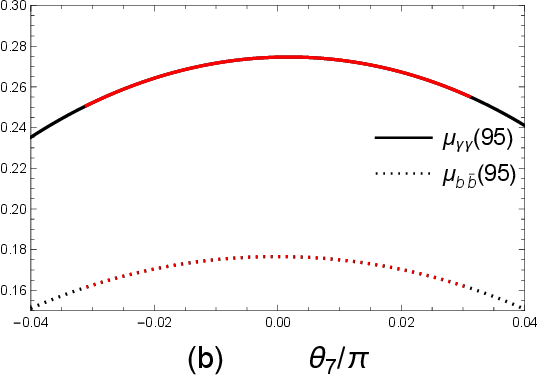}\\
\includegraphics[width=0.4\textwidth]{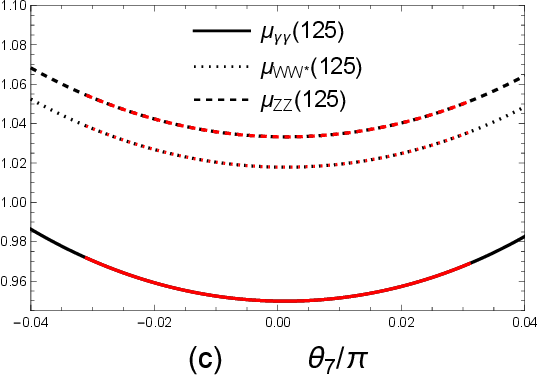}
\hspace{0.5cm}\includegraphics[width=0.4\textwidth]{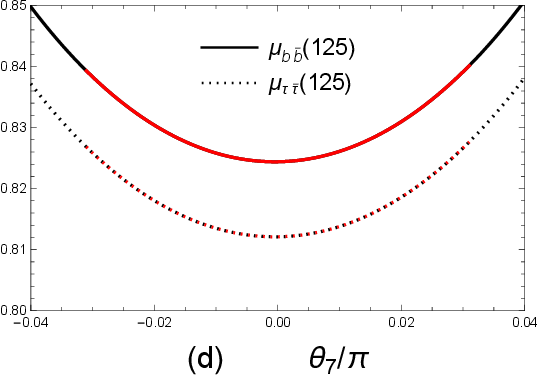}
\caption[]{The Higgs boson masses and signal strengths vary with CP phase $\theta_7$.}
\label{fig13}
\end{figure}
\begin{figure}[t]
\centering
\includegraphics[width=0.4\textwidth]{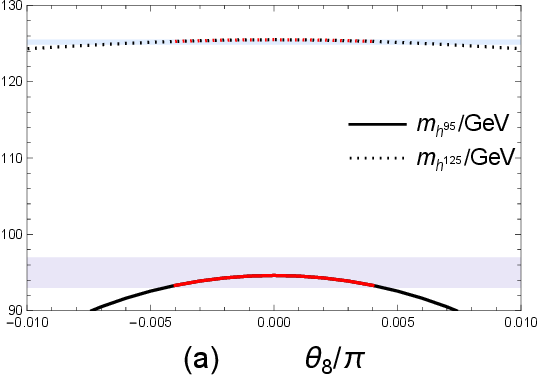}
\hspace{0.5cm}\includegraphics[width=0.4\textwidth]{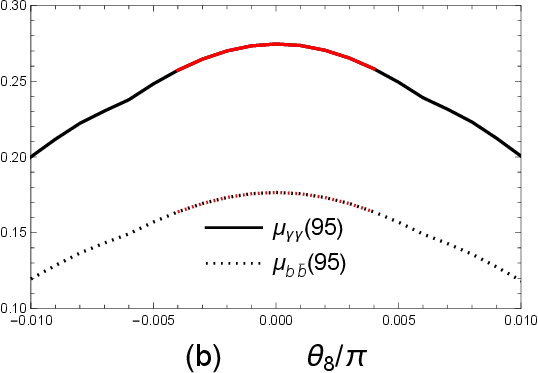}\\
\includegraphics[width=0.4\textwidth]{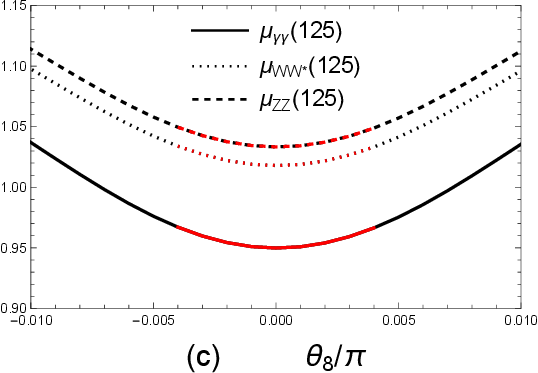}
\hspace{0.5cm}\includegraphics[width=0.4\textwidth]{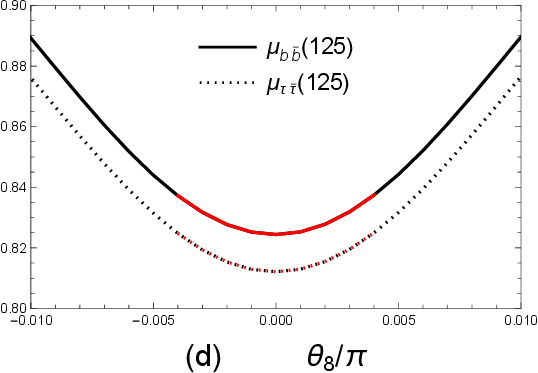}
\caption[]{The Higgs boson masses and signal strengthsvary with CP phase $\theta_8$.}
\label{fig14}
\end{figure}
Last but not least, we study the contributions from CP phases $\theta_{6,7,8}$ (which are related with the soft trilinear Higgs mixing masses $T_{\kappa},T_{\lambda},T_{\lambda_2}$ newly introduced in NB-LSSM) in FIG.\ref{fig12},\ref{fig13},\ref{fig14}. We can easily find that the experimental data of lightest and second-lightest Higgs boson mass have strict limitations on these three CP phases: $\theta_6\in(-0.015\pi,0.015\pi)$, $\theta_7\in(-0.03\pi,0.03\pi)$ and $\theta_8\in(-0.004\pi,0.004\pi)$. We can also see that CP phases $\theta_{6,7,8}$ have remarkable influences on all signal strengths, but their large variation regions are excluded by present experimental constraints of Higgs boson masses. Additionally, the signal strengths varying with the suitable values of CP phases $\theta_{6,7,8}$ all satisfy the experimental limits. Therefore, the CP phases $\theta_{6,7,8}$ all possess obvious influences on the Higgs masses and signal strengths.
\section{Discussion and conclusion}
In this paper, we introduce the CP violation in the NB-LSSM, which leads to the mixing of CP-even and CP-odd Higgs sectors. The mass squared matrix Higgs boson is modified by the two-loop effective potential corrections taken into account the CP violation. The mixing from the CP violation is contributed to produce the lightest Higgs boson mass around 95 GeV, which can explain the diphoton, $b\bar{b}$ excesses events. And the second-lightest Higgs boson mass around 125 GeV is the so called SM-like Higgs boson.

The reasonable parameter space is selected to scatter points after considering the $\chi^2$ analyses for the corresponding theoretical and experimental data of Higgs boson. There are parameter points within $1\sigma$ uncertainty intervals, which means that we have a relatively high possibility of simultaneously fitting the 95 GeV excess and the 125 GeV SM-like Higgs boson in the NB-LSSM. Considering the lightest Higgs boson mass in the range of 93 GeV $\sim$ 97 GeV and the second-lightest Higgs boson mass within $3\sigma$ experimental limitation, it is found that the parameters $A_t$, ${M}_{\tilde g}$, $\tilde{M}_{Q_3}$, $\tilde{M}_{t}$, $\tilde{M}_{b}$, $\tan\beta$, $T_\lambda$, $g_{YB}$ and $T_\kappa$ are constrained strictly in suitable ranges, which are sensitive to the numerical results. With the consideration of the CP-violating phases $\theta_{1,2,3,4,6,7,8}$, the Higgs masses and signal strengths are also influenced in a certain degree. And all of these parameters can affect the distributions of Higgs signal strengthes by the correlated Higgs couplings. The signal strengths of SM-like Higgs decay channels can meet well with the present experimental data, and the specific Higgs states in the NB-LSSM can also account for the diphoton, $b\bar{b}$ excesses at around 95 GeV simultaneously. With the improvement of experimental precise measurements, we hope more fundamental scalar particles can be detected in the near future.

\begin{acknowledgments}
\indent
This work is supported by the Major Project of National Natural Science Foundation of China (NNSFC) (No. 12235008), the National Natural Science Foundation of China (NNSFC) (No. 12075074, No. 12075073), the Natural Science Foundation of Hebei province(No.A2022201022, No. A2023201041, No. A2022201017), the Natural Science Foundation of Hebei Education Department(No. QN2022173), the Project of the China Scholarship Council (CSC) (No. 202408130113). This work is also supported by Funda\c{c}\~{a}o para a Ci\^{e}ncia e a Tecnologia (FCT, Portugal) through the projects CFTP-FCT Unit UIDB/00777/2020, UIDP/00777/2020 and UID/00777/2025 (https://doi.org/10.54499/UID/00777/2025), which are partially funded through POCTI (FEDER), COMPETE, QREN and EU.
\end{acknowledgments}

\appendix
\section{the tadpole contributions }
Correspondingly, the tadpole contributions at the tree-level and one-loop level are given as:
\begin{eqnarray}
&&T_{\phi_{d(u)}}^{(0)} =\left\langle \frac{\partial{V^0}}{\partial{\phi_{d(u)}}}\right\rangle=v_{d(u)}\Big[\frac{1}{8}G^2\Big(v_{d(u)}^2-v_{u(d)}^2\Big)+\frac{1}{4}g_B g_{YB}\Big(v_{\eta(\bar{\eta})}^2-v_{{\bar{\eta}}(\eta)}^2\Big)+m_{H_{d(u)}}^2\Big]\nonumber\\
&&\hspace{3.4cm}+\frac{1}{2} |\lambda|^2 v_{d(u)}\Big(v_s^2+v_{u(d)}^2\Big)-\frac{1}{\sqrt{2}} \Re\Big(T_\lambda e^{i(\theta_1+\theta_3)}\Big)v_{u(d)} v_s\nonumber\\&&\hspace{3.4cm}+\frac{1}{2}\Re\Big(\lambda_2^*\lambda e^{i(\theta_1-\theta_2)}\Big)v_{u(d)}v_\eta v_{\bar{\eta}}-\frac{1}{2}\Re\Big(\lambda^* \kappa e^{i(2\theta_3-\theta_1)}\Big)v_{u(d)} v_s^2,
\nonumber\\
&&T_{\phi_{\eta(\bar\eta)}}^{(0)} =\left\langle \frac{\partial{V^0}}{\partial{\phi_{\eta(\bar\eta)}}}\right\rangle=v_{\eta(\bar\eta)}\Big[\frac{1}{2}g_B^2\Big(v_{\eta(\bar\eta)}^2-v_{{\bar\eta}(\eta)}^2\Big)+\frac{1}{4}g_B g_{YB}\Big(v_{d(u)}^2-v_{u(d)}^2\Big)+m_{{\eta}(\bar\eta)}^2\Big]\nonumber\\
&&\hspace{3.4cm}+\frac{1}{2} |\lambda_2|^2 v_{\eta(\bar\eta)}\Big(v_s^2+v_{\bar\eta(\eta)}^2\Big)-\frac{1}{\sqrt{2}} \Re\Big(T_2 e^{i(\theta_2+\theta_3)}\Big)v_{\bar\eta(\eta)} v_s\nonumber\\&&\hspace{3.4cm}+\frac{1}{2}\Re\Big(\lambda_2^*\lambda e^{i(\theta_1-\theta_2)}\Big)v_{\bar\eta(\eta)} v_d v_u-\frac{1}{2}\Re\Big(\lambda_2^* \kappa e^{i(2\theta_3-\theta_2)}\Big)v_{\bar\eta(\eta)} v_s^2,
\nonumber\\
&&T_{\sigma_{d(u)}}^{(0)} = \left\langle \frac{\partial{V^0}}{\partial{\sigma_{d(u)}}}\right\rangle =\frac{1}{\sqrt{2}}\Im\Big(T_\lambda e^{i(\theta_1+\theta_3)}\Big)v_{u(d)}v _s-\frac{1}{2}\Im\Big(\lambda_2^*\lambda e^{i(\theta_1-\theta_2)}\Big)v_{u(d)}v_{\eta}v_{\bar\eta}\nonumber\\
&&\hspace{3.4cm}-\frac{1}{2}\Im\Big(\lambda^* \kappa e^{i(2\theta_3-\theta_1)}\Big)v_{u(d)} v_s^2,
\nonumber\\
&&T_{\sigma_{\eta (\bar\eta)}}^{(0)} =\left\langle \frac{\partial{V^0}}{\partial{\sigma_{\eta(\bar\eta)}}}\right\rangle = \frac{1}{\sqrt{2}}\Im\Big(T_2 e^{i(\theta_2+\theta_3)}\Big)v_{\bar\eta(\eta)} v_s- \frac{1}{2} \Im \Big(\lambda_2^* \lambda e^{i(\theta_1-\theta_2)}\Big)v_{\bar\eta(\eta)} v_d v_u\nonumber\\
&&\hspace{3.4cm}-\frac{1}{2}\Im \Big(\lambda_2^* \kappa e^{i(2\theta_3-\theta_2)}\Big)v_{\bar\eta(\eta)} v_s^2,
\nonumber\\
&&T_{\phi_{s}}^{(0)}=\left\langle \frac{\partial{V^0}}{\partial{\phi_s}}\right\rangle =\frac{1}{2} |\lambda|^2 v_s\Big(v_u^2+v_d^2\Big)+\frac{1}{2}|\lambda_2|^2 v_s\Big(v_{\eta}^2+v_{{\bar{\eta}}}^2\Big)+|\kappa|^2 v_s^3+m_s^2 v_s\nonumber\\
&&\hspace{2.8cm}+\frac{1}{\sqrt{2}}\Re\Big(T_\kappa e^{i3\theta_3}\Big) v_s^2-\frac{1}{\sqrt{2}} \Re\Big(T_\lambda e^{i(\theta_1+\theta_3)}\Big)v_d v_u-\frac{1}{\sqrt{2}}\Re\Big(T_2 e^{i(\theta_2+\theta_3)}\Big)v_\eta v_{\bar{{\eta}}}\nonumber\\
&&\hspace{2.8cm}-\Re\Big(\lambda_2^* \kappa e^{i(2\theta_3-\theta_2)}\Big)v_s v_\eta v_{\bar{{\eta}}}-\Re\Big(\lambda^*\kappa e^{i(2\theta_3-\theta_1)}\Big)v_s v_d v_u,
\nonumber\\
&&T_{\sigma_{s}}^{(0)}=\left\langle \frac{\partial{V^0}}{\partial{\sigma_s}}\right\rangle =\frac{1}{\sqrt{2}} \Im \Big(T_\kappa e^{i3\theta_3}\Big) v_s^2+\frac{1}{\sqrt{2}} \Im\Big(T_\lambda e^{i(\theta_1+\theta_3)}\Big)v_d v_u+\frac{1}{\sqrt{2}}\Big(T_2 e^{i(\theta_2+\theta_3)}\Big)v_\eta v_{\bar{\eta}}\nonumber\\
&&\hspace{2.8cm}+\Im\Big(\lambda_2^* \kappa  e^{i(2\theta_3-\theta_2)}\Big) v_s v_\eta v_
{\bar{\eta}} +\Im\Big(\lambda^* \kappa  e^{i(2\theta_3-\theta_1)}\Big) v_s v_d v_u,
\nonumber\\
&&T_{\phi_{d(u)}}^{(1)} = \left\langle \frac{\partial \Delta V^{(1)}}{\partial{\phi_{d(u)}}}\right\rangle=-\frac{3}{16\pi^2}\sum_{q=t,b}\Big[\sum_{k=1,2} 2\left\langle\frac{\partial{{m}_q}^2}{\partial{\phi_{d(u)}}}\right\rangle m_q^2\Big(\ln {\frac{m_q^2}{\Lambda^2}-1}\Big)\nonumber\\
&&\hspace{3.8cm}-\left\langle\frac{\partial{m}_{\tilde q_k}^2}{\partial\phi_{d(u)}}\right\rangle m_{\tilde{q}_k}^2 \Big(\ln {\frac{m_{\tilde q_k}^2}{\Lambda^2}-1}\Big)\Big],\nonumber\\
&&T_{\phi_{\eta(\bar\eta,s)}}^{(1)} = \left\langle \frac{\partial \Delta V^{(1)}}{\partial{\phi_{\eta(\bar\eta,s)}}}\right\rangle = \frac{3}{16\pi^2}\sum_{q=t,b}\sum_{k=1,2}\left\langle \frac{\partial{{m}}_{\tilde q_k}^2}{\partial\phi_{{\eta}(\bar\eta,s)}}\right\rangle m_{\tilde{q}_k}^2 \Big(\ln {\frac{m_{\tilde q_k}^2}{\Lambda^2}-1}\Big),
\nonumber\\
&&T_{\sigma_{d(u,s)}}^{(1)} =\left\langle \frac{\partial \Delta V^{(1)} }{\partial{\sigma_{d(u,s)}}}\right\rangle
=\frac{3}{16\pi^2}\sum_{q=t,b}\sum_{k=1,2} \left\langle \frac{\partial{{m}}_{\tilde q_k}^2}{\partial\sigma_{d(u,s)}}\right\rangle m_{\tilde{q}_k}^2 \Big(\ln {\frac{m_{\tilde q_k}^2}{\Lambda^2}-1}\Big),
\nonumber\\
&&T_{\sigma_{\eta (\bar\eta)}}^{(1)} = \left\langle \frac{\partial\Delta V^{(1)}}{\partial{\sigma_{\eta(\bar\eta)}}}\right\rangle = 0,
\end{eqnarray}
where the concrete expressions of the derivatives $\left\langle\partial{{m}_{q_k}^2}/\partial\phi_{d(u)}\right\rangle$, $\left\langle\partial{{m}_{\tilde q_k}^2}/\partial\phi_{d(u,\eta,\bar\eta,s)}\right\rangle$ and $\left\langle\ \partial {{m}_{\tilde q_k}^2}/\partial{\sigma_{d(u,s)}}\right\rangle$ are given in the Appendix C.
\section{The two-loop corrections}
In Refs.\cite{2deltaV1,2deltaV2,2deltaV3,2deltaV4}, the authors researched the two-loop corrections to the Higgs boson mass matrix by the effective potential approach in some supersymmetric models. We follow their method and also discuss the two-loop corrections involving top, stop, bottom, sbottom and gluinos to the Higgs mass matrix in the NB-LSSM. The two-loop effective potential can be shown as:
\begin{eqnarray}
&&\Delta V^{(2)}=  \frac{\alpha_{s}}{16 \pi^{3}}\sum_{q=t,b}\{2 J(m_{q}^{2}, m_{q}^{2})-4 m_{q}^{2} I(m_{q}^{2}, m_{q}^{2}, 0)+[2 m_{\tilde{q}_{1}}^{2} I(m_{\tilde{q}_{1}}^{2}, m_{\tilde{q}_{1}}^{2}, 0)\nonumber\\
&&+2 L(m_{\tilde{q}_{1}}^{2}, M_{\tilde{g}}^{2}, m_{q}^{2})  -4 m_{q} M_{\tilde{g}} S_{2 \bar{\theta}_q} I(m_{\tilde{q}_{1}}^{2}, M_{\tilde{g}}^{2}, m_{q}^{2})+\frac{1}{2}(1+2 C_{2 \bar{\theta}_q}^{2}) J(m_{\tilde{q}_{1}}^{2}, m_{\tilde{q}_{1}}^{2})\nonumber\\
&&+\frac{S_{2 \bar{\theta}_q}^{2}}{2} J(m_{\tilde{q}_{1}}^{2}, m_{\tilde{q}_{2}}^{2})+(m_{\tilde{q}_{1}} \leftrightarrow m_{\tilde{q}_{2}}, S_{2 \bar{\theta}_q} \leftrightarrow-S_{2 \bar{\theta}_q})]\},
\end{eqnarray}
where $M_{\tilde{g}}$ is the mass of gluinos and $S_{2 \bar{\theta}_q}=\frac{2|m_{\tilde{q}_{12}}|}{m_{\tilde{q}_{1}}-m_{\tilde{q}_{2}}}$ with $m_{\tilde{b}_{12}}=T_b H_d^0 - Y_b \lambda^* H_u^{0\dagger} S^\dagger$ and $m_{\tilde{t}_{12}}=T_t H_u^0 - Y_t \lambda^* H_d^{0\dagger} S^\dagger$. The function of $I$, $J$ and $L$ can be found in the Appendix of Refs.\cite{2deltaV1,2deltaV2,2deltaV3,2deltaV4}. The corrections from the effective potential $\Delta V^{(2)}$ to the Higgs boson mass in the NB-LSSM can be written as
\begin{eqnarray}
&&{\cal M}_{\phi_{i}\phi_{j}}^{(2)}=\Big\langle{-}\frac{\delta_{ij}}{\phi_{i}} \frac{\partial \Delta V^{(2)}}{\partial \phi_{i}}+\frac{\partial^{2} \Delta V^{(2)}}{\partial \phi_{i} \partial \phi_{j}}\Big\rangle=\sum_{q=t,b}\sum_{k=1,2}\sum_{i,j=d,u,\eta,\bar\eta,S}\Big\{-\frac{\delta_{ij}}{v_{i}}\Big[\frac{\partial \Delta V^{(2)}}{\partial m_{q}^{2}} \Big\langle\frac{\partial m_{q}^{2}}{\partial \phi_{i}}\Big\rangle
\nonumber\\
&&+\frac{\partial \Delta V^{(2)}}{\partial m_{\tilde{q}_{k}}^{2}} \Big\langle\frac{\partial m_{\tilde{q}_{k}}^{2}}{\partial \phi_{i}}\Big\rangle
+\frac{\partial \Delta V^{(2)}}{\partial C_{2 \bar{\theta}_{q}}^{2}} \Big\langle\frac{\partial C_{2 \bar{\theta}_{q}}^{2}}{\partial \phi_{i}}\Big\rangle\Big]
+\frac{\partial^{2} \Delta V^{(2)}}{\partial m_{q}^{2} \partial m_{q}^{2}}\Big\langle\frac{\partial m_{q}^{2}}{\partial \phi_{i}}\Big\rangle\Big\langle\frac{\partial m_{q}^{2}}{\partial \phi_{j}}\Big\rangle
\nonumber\\
&&+\frac{\partial \Delta V^{(2)}}{\partial m_{q}^{2}} \Big\langle\frac{\partial^{2} m_{q}^{2}}{\partial \phi_{i} \partial \phi_{j}}\Big\rangle
+\frac{\partial^{2} \Delta V^{(2)}}{\partial m_{\tilde{q}_{k}}^{2} \partial m_{\tilde{q}_{k}}^{2}}
\Big\langle\frac{\partial m_{\tilde{q}_{k}}^{2}}{\partial \phi_{i}}\Big\rangle\Big\langle\frac{\partial m_{\tilde{q}_{k}}^{2}}{\partial \phi_{j}}\Big\rangle
+\frac{\partial \Delta V^{(2)}}{\partial m_{\tilde{q}_{k}}^{2}} \Big\langle\frac{\partial^{2} m_{\tilde{q}_{k}}^{2}}{\partial \phi_{i} \partial \phi_{j}}\Big\rangle
\nonumber\\
&&+\frac{\partial^{2} \Delta V^{(2)}}{\partial C_{2 \bar{\theta}_{q}}^{2} \partial C_{2 \bar{\theta}_{q}}^{2}}\Big\langle\frac{\partial C_{2 \bar{\theta}_{q}}^{2}}{\partial \phi_{i}}\Big\rangle\Big\langle\frac{\partial C_{2 \bar{\theta}_{q}}^{2}}{\partial \phi_{j}}\Big\rangle
+\frac{\partial \Delta V^{(2)}}{\partial C_{2 \bar{\theta}_{q}}^{2}} \Big\langle\frac{\partial^{2} C_{2 \bar{\theta}_{q}}^{2}}{\partial \phi_{i} \partial \phi_{j}}\Big\rangle
+ \frac{\partial^{2} \Delta V^{(2)}}{\partial m_{q}^{2} \partial m_{\tilde{q}_{k}}^{2}} \Big\langle\frac{\partial m_{q}^{2}}{\partial \phi_{i}}\Big\rangle\Big\langle \frac{\partial m_{\tilde{q}_{k}}^{2}}{\partial \phi_{j}}\Big\rangle
 \nonumber\\
&&+ \frac{\partial^{2} \Delta V^{(2)}}{\partial m_{q}^{2} \partial C_{2 \bar{\theta}_{q}}^{2}} \Big\langle\frac{\partial m_{q}^{2}}{\partial \phi_{i}} \Big\rangle\Big\langle\frac{\partial C_{2 \bar{\theta}_{q}}^{2}}{\partial \phi_{j}}\Big\rangle + \frac{\partial^{2} \Delta V^{(2)}}{\partial m_{\tilde{q}_{k}}^{2} \partial C_{2 \bar{\theta}_{q}}^{2}} \Big\langle\frac{\partial m_{\tilde{q}_{k}}^{2}}{\partial \phi_{i}}\Big\rangle\Big\langle \frac{\partial C_{2 \bar{\theta}_{q}}^{2}}{\partial \phi_{j}}\Big\rangle\Big\},
\nonumber\\
&&{\cal M}_{\sigma_{i}\sigma_{j}}^{(2)}=\Big\langle{-}\frac{\delta_{ij}}{\phi_{i}} \frac{\partial \Delta V^{(2)}}{\partial \sigma_{i}}+\frac{\partial^{2} \Delta V^{(2)}}{\partial \sigma_{i} \partial \sigma_{j}}\Big\rangle=\Big\langle{-}\frac{\delta_{ij}}{v_{i}} \frac{\partial \Delta V^{(2)}}{\partial \phi_{i}}+\frac{\partial^{2} \Delta V^{(2)}}{\partial \phi_{i} \partial \phi_{j}}\Big\rangle|_{\phi\rightarrow \sigma},
\nonumber\\
&&{\cal M}_{\phi_{i}\sigma_{j}}^{(2)}=\Big\langle{-}\frac{1-\delta_{ij}}{\phi_{i}} \frac{\partial \Delta V^{(2)}}{\partial \phi_{i}}+\frac{\partial^{2} \Delta V^{(2)}}{\partial \phi_{i} \partial \sigma_{j}}\Big\rangle=\Big\langle{-}\frac{1-\delta_{ij}}{v_{i}} \frac{\partial \Delta V^{(2)}}{\partial \phi_{i}}+\frac{\partial^{2} \Delta V^{(2)}}{\partial \phi_{i} \partial \phi_{j}}\Big\rangle|_{\phi_j\rightarrow \sigma_j},
\nonumber\\
&&{{\cal M}_h^2}^{(2)}=\left(
    \begin{array}{cc}
        {\cal M}_{\phi_{i}\phi_{j}}^{(2)}& {\cal M}_{\phi_{i}\sigma_{j}}^{(2)} \\
       ({\cal M}_{\phi_{i}\sigma_{j}}^{(2)})^T &  {\cal M}_{\sigma_{i}\sigma_{j}}^{(2)} \\
    \end{array}
\right).
\end{eqnarray}
The results of $\frac{\partial \Delta V^{(2)}}{\partial m_{q}^{2}},~\frac{\partial \Delta V^{(2)}}{\partial m_{\tilde{q}_{k}}^{2}},~\frac{\partial \Delta V^{(2)}}{\partial C_{2 \bar{\theta}_{q}}^{2}},...$ and some relating derivatives of $C_{2 \bar{\theta}}^2$ over $\phi_i$ ($i= d, u, \eta$ and $\bar{\eta}$) can be found in the Appendix of Refs.\cite{2deltaV1,2deltaV2,2deltaV3,2deltaV4}. According to these derivatives, we can get some other new derivatives in the NB-LSSM with $i= d, u, \eta$, $\bar{\eta}$ and $s$ here, which can be written as
\begin{eqnarray}
&&\Big\langle\frac{\partial C_{2 \bar{\theta}_{q}}^{2}}{\partial \phi_{s}}\Big\rangle=\Big\langle\frac{\partial C_{2 \bar{\theta}_{q}}^{2}}{\partial \phi_{d}}\Big\rangle|_{\phi_{d}\rightarrow \phi_{s}},\;\;
\Big\langle\frac{\partial^2 C_{2 \bar{\theta}_{q}}^{2}}{\partial \phi_{i}\partial \phi_{s}}\Big\rangle=\Big\langle\frac{\partial^2 C_{2 \bar{\theta}_{q}}^{2}}{{\partial \phi_{d}}\partial \phi_{u}}\Big\rangle|_{\phi_{d,u}\rightarrow \phi_{i,s}},
\nonumber\\
&&\Big\langle\frac{\partial C_{2 \bar{\theta}_{q}}^{2}}{\partial \sigma_{i}}\Big\rangle=\Big\langle\frac{\partial C_{2 \bar{\theta}_{q}}^{2}}{\partial \phi_{d}}\Big\rangle|_{\phi_{d}\rightarrow \sigma_{i}},\;\;
\Big\langle\frac{\partial^2 C_{2 \bar{\theta}_{q}}^{2}}{\partial \sigma_{i}\partial \sigma_{j}}\Big\rangle=\Big\langle\frac{\partial^2 C_{2 \bar{\theta}_{q}}^{2}}{{\partial \phi_{d}}\partial \phi_{u}}\Big\rangle|_{\phi_{d,u}\rightarrow \sigma_{i,j}},
\nonumber\\
&&\Big\langle\frac{\partial^2 C_{2 \bar{\theta}_{q}}^{2}}{\partial \phi_{i}\partial \sigma_{j}}\Big\rangle=\Big\langle\frac{\partial^2 C_{2 \bar{\theta}_{q}}^{2}}{{\partial \phi_{d}}\partial \phi_{u}}\Big\rangle|_{\phi_{d,u}\rightarrow \phi_i,\sigma_j}.
\end{eqnarray}
\section{Derivatives }
\subsection{Derivatives of quark mass to neutral Higgs fileds}
The derivatives involving neutral Higgs bosons can be shown as follows:
\begin{eqnarray}
&&\left\langle \frac{\partial{{m}_t^2}}{\partial\phi_u}\right\rangle=|Y_t|^2 v_u , \hspace{2.3cm} \left\langle \frac{\partial{{m}_b^2}}{\partial\phi_d}\right\rangle=|Y_b|^2 v_d.
\nonumber\\
&&
\left\langle \frac{\partial^2{{m}_t^2}}{\partial\phi_u^2}\right\rangle=\left\langle \frac{\partial^2{{m}_t^2}}{\partial\sigma_u^2}\right\rangle=|Y_t|^2,~~
\left\langle \frac{\partial^2{{m}_b^2}}{\partial\phi_d^2}\right\rangle=\left\langle \frac{\partial^2{{m}_b^2}}{\partial\sigma_d^2}\right\rangle=|Y_b|^2.
\end{eqnarray}
Note that the other derivatives that we do not list here are equal to zero.

\subsection{Derivatives of squark mass to neutral Higgs fileds}
We first define several expressions to simplify the equations below:
\begin{eqnarray}
&&IH_t=IH_T	-|Y_t|^2|\lambda|^2 v_s^2,\;\;\;IH_T=\sqrt{2}|Y_t|^2 \Im\Big(\lambda A_t e^{i(\theta_1+\theta_3)}\Big)v_s \tan\beta;\nonumber\\
&&IH_b=IH_B-|Y_b|^2|\lambda|^2 v_s^2,\;\;\;IH_B=\sqrt{2} |Y_b|^2\Im\Big(\lambda A_b e^{i(\theta_1+\theta_3)}\Big)v_s \cot\beta\nonumber\\
&&IJ_t=	IJ_T-2|Y_t|^2|A_t|^2,\;\;\;IJ_T=\sqrt{2}|Y_t|^2 \Im\Big(\lambda A_t e^{i(\theta_1+\theta_3)}\Big)v_s \cot\beta;\nonumber\\
&&IJ_b=IJ_B-2|Y_b|^2|A_b|^2,\;\;\; IJ_B=\sqrt{2}|Y_b|^2 \Im\Big(\lambda A_b e^{i(\theta_1+\theta_3)}\Big)v_s \tan\beta;\nonumber\\
&&H_t=H_T-|Y_t|^2|\lambda|^2
v_s^2,\;\;\;H_T=\sqrt{2}|Y_t|^2 \Re\Big(\lambda A_t e^{i(\theta_1+\theta_3)}\Big)v_s \tan\beta;\nonumber\\
&&H_b=H_B-|Y_b|^2|\lambda|^2 v_s^2,\;\;\;H_B=\sqrt{2}|Y_b|^2\Re\Big(\lambda A_b e^{i(\theta_1+\theta_3)}\Big)v_s \cot\beta;\nonumber\\
&&J_t=J_T- 2|Y_t|^2|A_t|^2,\;\;\;J_T=\sqrt{2}|Y_t|^2 \Re\Big(\lambda A_t e^{i(\theta_1+\theta_3)}\Big)v_s \cot\beta;\nonumber\\
&&J_b=J_B-2|Y_b|^2|A_b|^2  ,\;\;\;J_B=\sqrt{2}|Y_b|^2 \Re\Big(\lambda A_b e^{i(\theta_1+\theta_3)}\Big)v_s \tan\beta ;\nonumber\\
&&X_t=\tilde M_{Q_3}^2 - \tilde M_t^2 +\frac{1}{2} x_t v^2 \cos 2\beta - \frac{1}{2} a_t u^2 \cos 2\beta';\nonumber\\
&&X_b=\tilde M_{Q_3}^2 - \tilde M_b^2 -\frac{1}{2} x_b v^2 \cos 2\beta - \frac{1}{2} a_b u^2 \cos 2\beta'.
\end{eqnarray}

The first-order derivatives involving squark and neutral Higgs bosons are shown as follows:
\begin{eqnarray}
&&\left\langle\frac{\partial {m}_{\tilde t_k}^2}{\partial\phi_d}\right\rangle=\frac{1}{8} G^2 v_d \pm \frac{v_d}{2 \Big(m_{\tilde t_1}^2 - m_{\tilde t_2}^2 \Big) } \Big(x_t X_t- H_t\Big),\nonumber\\
&&\left\langle\frac{\partial {m}_{\tilde b_k}^2}{\partial\phi_d}\right\rangle=|Y_b|^2 v_d - \frac{1}{8} G^2 v_d \mp \frac{v_d}{2 \Big(m_{\tilde b_1}^2 - m_{\tilde b_2}^2 \Big) } \Big(x_b X_b + J_b\Big),\nonumber\\
&&\left\langle\frac{\partial {m}_{\tilde t_k}^2}{\partial\phi_u}\right\rangle=|Y_t|^2 v_u - \frac{1}{8} G^2 v_u \mp \frac{v_u}{2 \Big(m_{\tilde t_1}^2 - m_{\tilde t_2}^2 \Big) } \Big(x_t X_t + J_t\Big),\nonumber\\
&&\left\langle\frac{\partial {m}_{\tilde b_k}^2}{\partial\phi_u}\right\rangle=\frac{1}{8} G^2 v_u \pm \frac{v_u}{2 \Big(m_{\tilde b_1}^2 - m_{\tilde b_2}^2 \Big) } \Big(x_b X_b- H_b\Big),\nonumber\\
&&\left\langle\frac{\partial {m}_{\tilde t_k}^2}{\partial\phi_\eta}\right\rangle=\frac{1}{4}g_Bg_{YB} v_\eta \mp \frac{1}{2 \Big(m_{\tilde t_1}^2 - m_{\tilde t_2}^2 \Big) } a_t v_\eta X_t,\nonumber\\
&&\left\langle\frac{\partial {m}_{\tilde b_k}^2}{\partial\phi_\eta}\right\rangle= - \frac{1}{4}g_Bg_{YB} v_\eta \mp \frac{1}{2 \Big(m_{\tilde b_1}^2 - m_{\tilde b_2}^2 \Big) } a_b v_\eta X_b,\nonumber\\
&&\left\langle\frac{\partial {m}_{\tilde t_k}^2}{\partial\phi_{\bar\eta}}\right\rangle= -\frac{1}{4}g_Bg_{YB} v_{\bar{\eta}} \pm \frac{1}{2 \Big(m_{\tilde t_1}^2 - m_{\tilde t_2}^2 \Big) } a_t v_{\bar{\eta}} X_t,\nonumber\\
&&\left\langle\frac{\partial {m}_{\tilde b_k}^2}{\partial\phi_{\bar{\eta}}}\right\rangle= \frac{1}{4}g_Bg_{YB} v_{\bar{\eta}} \pm \frac{1}{2 \Big(m_{\tilde b_1}^2 - m_{\tilde b_2}^2 \Big) } a_b v_{\bar\eta}X_b,\nonumber\\
&&\left\langle\frac{\partial {m}_{\tilde t_k}^2}{\partial\phi_s}\right\rangle=\mp\frac{v_d^2/v_s}{2\Big(m_{\tilde t_1}^2 - m_{\tilde t_2}^2 \Big) } H_t,\;
\left\langle\frac{\partial {m}_{\tilde b_k}^2}{\partial\phi_s}\right\rangle=\mp\frac{v_u^2/v_s}{2\Big(m_{\tilde b_1}^2 - m_{\tilde b_2}^2 \Big) } H_b,
\end{eqnarray}
\begin{eqnarray}
&&\left\langle\frac{\partial {m}_{\tilde t_k}^2}{\partial\sigma_d}\right\rangle=\pm\frac{v_d}{2\Big(m_{\tilde t_1}^2 - m_{\tilde t_2}^2 \Big) } IH_T,\;
\left\langle\frac{\partial {m}_{\tilde b_k}^2}{\partial\sigma_d}\right\rangle=\pm\frac{v_d}{2\Big(m_{\tilde b_1}^2 - m_{\tilde b_2}^2 \Big) } IJ_B,\nonumber\\
&&\left\langle\frac{\partial {m}_{\tilde t_k}^2}{\partial\sigma_u}\right\rangle=\pm\frac{v_u}{2\Big(m_{\tilde t_1}^2 - m_{\tilde t_2}^2 \Big) } IJ_T,\;
\left\langle\frac{\partial {m}_{\tilde b_k}^2}{\partial\sigma_u}\right\rangle=\pm\frac{v_u}{2\Big(m_{\tilde b_1}^2 - m_{\tilde b_2}^2 \Big) } IH_B,\nonumber\\
&&\left\langle\frac{\partial {m}_{\tilde t_k}^2}{\partial\sigma_s}\right\rangle=\pm\frac{v_u^2/v_s}{2\Big(m_{\tilde t_1}^2 - m_{\tilde t_2}^2 \Big) } IJ_T,\;
\left\langle\frac{\partial {m}_{\tilde b_k}^2}{\partial\sigma_s}\right\rangle=\pm\frac{v_u^2/v_s}{2\Big(m_{\tilde b_1}^2 - m_{\tilde b_2}^2 \Big) } IH_B,\nonumber\\
&&\left\langle\frac{\partial {m}_{\tilde t_k}^2}{\partial\sigma_\eta}\right\rangle=\left\langle\frac{\partial{{m}_{\tilde t_k}}^2}{\partial\sigma_{\bar{\eta}}}\right\rangle=
\left\langle\frac{\partial {m}_{\tilde b_k}^2}{\partial\sigma_\eta}\right\rangle=\left\langle\frac{\partial{{m}_{\tilde b_k}}^2}{\partial\sigma_{\bar{\eta}}}\right\rangle=0.
\end{eqnarray}

Then, the second-order derivatives involving squark and neutral Higgs bosons $\left\langle \partial^2 m_{\tilde q_k}^2/\partial \sigma_m \partial \sigma_n\right\rangle$ with $\tilde q_k=\tilde t_k,\tilde b_k, k=1,2$ and $m,n=d,u,\eta,\bar\eta,s$ are deduced as follows:
\begin{eqnarray}
&&\left\langle\frac{\partial^2{{m}_{\tilde t_k}^2}}{\partial\sigma_d^2}\right\rangle=\frac{1}{8} G^2 \pm \frac{1}{2 \Big(m_{\tilde t_1}^2 - m_{\tilde t_2}^2 \Big) } \Big[x_t X_t +  |Y_t\lambda^*|^2 v_s^2 \Big] \mp \frac{v_d^2}{2 \Big(m_{\tilde t_1}^2 - m_{\tilde t_2}^2 \Big)^3} IH_T^2,\nonumber\\
&&\left\langle\frac{\partial^2{ {m}_{\tilde b_k}^2}}{\partial\sigma_d^2}\right\rangle= | Y_b |^2 - \frac{1}{8} G^2 \mp \frac{1}{2 \Big(m_{\tilde b_1}^2 - m_{\tilde b_2}^2 \Big) } \Big[x_b X_b  - 2 |A_b|^2 | Y_b |^2 \Big]\mp \frac{v_d^2}{2\Big(m_{\tilde b_1}^2 - m_{\tilde b_2}^2 \Big)^3} IJ_B^2,\nonumber\\
&&\left\langle\frac{\partial^2{ {m}_{\tilde t_k}^2}}{\partial\sigma_u^2}\right\rangle= | Y_t |^2 - \frac{1}{8} G^2 \mp \frac{1}{2 \Big(m_{\tilde t_1}^2 - m_{\tilde t_2}^2 \Big) } \Big[x_t X_t- 2 |A_t|^2 | Y_t |^2 \Big]\mp \frac{v_u^2}{2\Big(m_{\tilde t_1}^2 - m_{\tilde t_2}^2 \Big)^3} IJ_T^2\nonumber,\\
&&\left\langle\frac{\partial^2{ {m}_{\tilde b_k}^2}}{\partial\sigma_u^2}\right\rangle=  \frac{1}{8} G^2 \pm \frac{1}{2 \Big(m_{\tilde b_1}^2 - m_{\tilde b_2}^2 \Big) } \Big[x_b X_b + |Y_b\lambda^*|^2 v_s^2 \Big]\mp \frac{v_u^2}{\Big(m_{\tilde b_1}^2 - m_{\tilde b_2}^2 \Big)^3}IH_B^2,
\nonumber\\
&&\left\langle\frac{\partial^2{ {m}_{\tilde t_k}^2}}{\partial\sigma_\eta^2}\right\rangle=\frac{1}{4} g_B g_{YB}
\mp \frac{1}{2 \Big(m_{\tilde t_1}^2-m_{\tilde t_2}^2 \Big)} a_t X_t,\;
\left\langle\frac{\partial^2{ {m}_{\tilde b_k}^2}}{\partial\sigma_\eta^2}\right\rangle= -\frac{1}{4} g_B g_{YB} \mp \frac{1}{2 \Big(m_{\tilde b_1}^2 - m_{\tilde b_2}^2 \Big) } a_b X_b,\nonumber\\
&&\left\langle\frac{\partial^2{ {m}_{\tilde t_k}^2}}{\partial\sigma_{\bar{\eta}}^2}\right\rangle=-\frac{1}{4} g_B g_{YB}\pm \frac{1}{2 \Big(m_{\tilde t_1}^2-m_{\tilde t_2}^2 \Big)} a_t X_t,\;
\left\langle\frac{\partial^2{ {m}_{\tilde b_k}^2}}{\partial\sigma_{\bar{\eta}}^2}\right\rangle=\frac{1}{4} g_B g_{YB}
\pm \frac{1}{2 \Big(m_{\tilde b_1}^2-m_{\tilde b_2}^2 \Big)} a_b X_b,\nonumber\\
&&\left\langle\frac{\partial^2{ {m}_{\tilde t_k}^2}}{\partial\sigma_s^2}\right\rangle=\pm\frac{1}{2 \Big(m_{\tilde t_1}^2 - m_{\tilde t_2}^2 \Big) } |Y_t\lambda^*|^2 v_d^2\mp\frac{v_d^4 /v_s^2}{2 \Big(m_{\tilde t_1}^2 - m_{\tilde t_2}^2 \Big)^3 }IH_T^2,\nonumber\\
&&\left\langle\frac{\partial^2{ {m}_{\tilde b_k}^2}}{\partial\sigma_s^2}\right\rangle=\pm\frac{1}{2 \Big(m_{\tilde b_1}^2 - m_{\tilde b_2}^2 \Big) } |Y_b\lambda^*|^2 v_u^2\mp\frac{v_d^4 /v_s^2}{2 \Big(m_{\tilde b_1}^2 - m_{\tilde b_2}^2 \Big)^3 }IJ_B^2,
\nonumber\\
&&\left\langle\frac{\partial^2{ {m}_{\tilde t_k}^2}}{\partial\sigma_d \partial\sigma_u}\right\rangle= \pm \frac{H_T /\tan\beta}{2 \Big(m_{\tilde t_1}^2-m_{\tilde t_2}^2 \Big)}\mp\frac{v_d v_u IH_T IJ_T}{2 \Big(m_{\tilde t_1}^2-m_{\tilde t_2}^2 \Big)^3},\;\left\langle\frac{\partial^2{ {m}_{\tilde b_k}^2}}{\partial\sigma_d \partial\sigma_u}\right\rangle= \pm \frac{H_B /\cot\beta}{2 \Big(m_{\tilde b_1}^2-m_{\tilde b_2}^2 \Big)}\mp\frac{v_d v_u IH_B IJ_B}{2 \Big(m_{\tilde b_1}^2-m_{\tilde b_2}^2 \Big)^3},\nonumber\\
&&\left\langle\frac{\partial^2{ {m}_{\tilde t_k}^2}}{\partial\sigma_d\partial\sigma_s}\right\rangle= \pm \frac{v_d / v_s H_T}{2 \Big(m_{\tilde t_1}^2-m_{\tilde t_2}^2 \Big)}\mp\frac{v_d^3 / v_s IH_T^2}{2 \Big(m_{\tilde t_1}^2-m_{\tilde t_2}^2 \Big)^3},\;\left\langle\frac{\partial^2{ {m}_{\tilde b_k}^2}}{\partial\sigma_d\partial\sigma_s}\right\rangle= \pm \frac{v_d / v_s J_B}{2 \Big(m_{\tilde b_1}^2-m_{\tilde b_2}^2 \Big)}\mp\frac{v_d^3 / v_s IJ_B^2}{2 \Big(m_{\tilde b_1}^2-m_{\tilde b_2}^2 \Big)^3},\nonumber\\
&&\left\langle\frac{\partial^2{ {m}_{\tilde t_k}^2}}{\partial\sigma_u\partial\sigma_s}\right\rangle= \pm \frac{v_u / v_s J_T}{2 \Big(m_{\tilde t_1}^2-m_{\tilde t_2}^2 \Big)}\mp\frac{v_u^3 / v_s IJ_T^2}{2 \Big(m_{\tilde t_1}^2-m_{\tilde t_2}^2 \Big)^3},\;\left\langle\frac{\partial^2{ {m}_{\tilde b_k}^2}}{\partial\sigma_u\partial\sigma_s}\right\rangle=\pm \frac{v_u / v_s H_B}{2 \Big(m_{\tilde b_1}^2-m_{\tilde b_2}^2 \Big)}\mp\frac{v_u^3 / v_s IH_B^2}{2 \Big(m_{\tilde b_1}^2-m_{\tilde b_2}^2 \Big)^3},
\nonumber\\
&&\left\langle\frac{\partial^2{ {m}_{\tilde t_k}^2}}{\partial\sigma_{\bar{\eta}} \partial\sigma_s}\right\rangle\hspace{-0.1cm}=\hspace{-0.1cm} \left\langle\frac{\partial^2{ {m}_{\tilde t_k}^2}}{\partial\sigma_{\eta} \partial\sigma_s}\right\rangle\hspace{-0.1cm}=\hspace{-0.1cm}
\left\langle\frac{\partial^2{ {m}_{\tilde t_k}^2}}{\partial\sigma_d \partial\sigma_{\eta}}\right\rangle\hspace{-0.1cm}=\hspace{-0.1cm} \left\langle\frac{\partial^2{ {m}_{\tilde t_k}^2}}{\partial\sigma_d \partial\sigma_{\bar{\eta}}}\right\rangle\hspace{-0.1cm}=\hspace{-0.1cm}\left\langle\frac{\partial^2{ {m}_{\tilde t_k}^2}}{\partial\sigma_u \partial\sigma_\eta}\right\rangle\hspace{-0.1cm}=\hspace{-0.1cm}\left\langle\frac{\partial^2{ {m}_{\tilde t_k}^2}}{\partial\sigma_u \partial\sigma_{\bar{\eta}}}\right\rangle\hspace{-0.1cm}=\hspace{-0.1cm}\left\langle\frac{\partial^2{
{m}_{\tilde t_k}^2}}{\partial\sigma_\eta\partial\sigma_{\bar{\eta}}}\right\rangle\hspace{-0.1cm}=\hspace{-0.1cm}0,\nonumber\\
&&\left\langle\frac{\partial^2{ {m}_{\tilde b_k}^2}}{\partial\sigma_{\bar{\eta}} \partial\sigma_s}\right\rangle\hspace{-0.1cm}=\hspace{-0.1cm} \left\langle\frac{\partial^2{ {m}_{\tilde b_k}^2}}{\partial\sigma_\eta \partial\sigma_s}\right\rangle\hspace{-0.1cm}=\hspace{-0.1cm}\left\langle\frac{\partial^2{ {m}_{\tilde b_k}^2}}{\partial\sigma_d \partial\sigma_\eta}\right\rangle\hspace{-0.1cm}=\hspace{-0.1cm} \left\langle\frac{\partial^2{ {m}_{\tilde b_k}^2}}{\partial\sigma_d \partial\sigma_{\bar{\eta}}}\right\rangle\hspace{-0.1cm}=\hspace{-0.1cm}\left\langle\frac{\partial^2{ {m}_{\tilde b_k}^2}}{\partial\sigma_u \partial\sigma_\eta}\right\rangle\hspace{-0.1cm}=\hspace{-0.1cm}\left\langle\frac{\partial^2{ {m}_{\tilde b_k}^2}}{\partial\sigma_u \partial\sigma_{\bar{\eta}}}\right\rangle\hspace{-0.1cm}=\hspace{-0.1cm}\left\langle\frac{\partial^2{ {m}_{\tilde b_k}^2}}{\partial\sigma_\eta \partial\sigma_{\bar{\eta}}}\right\rangle\hspace{-0.1cm}=\hspace{-0.1cm}0,
\end{eqnarray}
In addition, the terms of $\left\langle\partial^2 m_{\tilde q_k}^2/\partial \phi_m \partial \sigma_n\right\rangle$ are given by
\begin{eqnarray}
&&\left\langle\frac{\partial^2{m}_{\tilde t_k}^2}{\partial\phi_d\partial\sigma_d}\right\rangle= \mp\frac{v_d^2 IH_T \Big(x_t X_t-H_t\Big)}{2\Big(m_{\tilde t_1}^2-m_{\tilde t_2}^2 \Big)^3},\;\left\langle\frac{\partial^2{m}_{\tilde b_k}^2}{\partial\phi_d\partial\sigma_d}\right\rangle= \pm \frac{v_d^2 IJ_B \Big(x_b X_b+J_b\Big)}{2\Big(m_{\tilde b_1}^2-m_{\tilde b_2}^2 \Big)^3},\nonumber\\
&&\left\langle\frac{\partial^2{m}_{\tilde t_k}^2}{\partial\phi_d\partial\sigma_{u}}\right\rangle= \pm \frac{v_u / v_d IJ_T}{2 \Big(m_{\tilde t_1}^2-m_{\tilde t_2}^2 \Big)}\Big[1 - \frac{v_d^2}{\Big(m_{\tilde t_1}^2-m_{\tilde t_2}^2 \Big)^2}\Big(x_t X_t - H_t\Big) \Big],\nonumber\\
&&\left\langle\frac{\partial^2{m}_{\tilde b_k}^2}{\partial\phi_d\partial\sigma_{u}}\right\rangle= \pm \frac{v_u / v_d IH_B}{2 \Big(m_{\tilde b_1}^2-m_{\tilde b_2}^2 \Big)}\Big[1 + \frac{v_d^2}{\Big(m_{\tilde b_1}^2-m_{\tilde b_2}^2 \Big)^2}\Big(x_b X_b + J_b\Big) \Big],
\nonumber\\
&&\left\langle\frac{\partial^2{m}_{\tilde t_k}^2}{\partial\phi_d\partial\sigma_{s}}\right\rangle= \pm \frac{v_u / v_d IJ_T}{2 \Big(m_{\tilde t_1}^2-m_{\tilde t_2}^2 \Big)}\Big[1 - \frac{v_d^2}{\Big(m_{\tilde t_1}^2-m_{\tilde t_2}^2 \Big)^2}\Big(x_t X_t - H_t\Big) \Big]\frac{v_{u}}{v_s},\nonumber\\
&&\left\langle\frac{\partial^2{m}_{\tilde b_k}^2}{\partial\phi_d\partial\sigma_{s}}\right\rangle= \pm \frac{v_u / v_d IH_B}{2 \Big(m_{\tilde b_1}^2-m_{\tilde b_2}^2 \Big)}\Big[1 + \frac{v_d^2}{\Big(m_{\tilde b_1}^2-m_{\tilde b_2}^2 \Big)^2}\Big(x_b X_b + J_b\Big) \Big]\frac{v_{u}}{v_s},
\nonumber\\
&&\left\langle\frac{\partial^2{m}_{\tilde t_k}^2}{\partial\phi_u\partial\sigma_{d}}\right\rangle= \pm \frac{v_d / v_u IH_T}{2 \Big(m_{\tilde t_1}^2-m_{\tilde t_2}^2 \Big)}\Big[1 + \frac{v_u^2}{\Big(m_{\tilde t_1}^2-m_{\tilde t_2}^2 \Big)^2}\Big(x_t X_t+J_t\Big) \Big],\nonumber\\
&&\left\langle\frac{\partial^2{m}_{\tilde b_k}^2}{\partial\phi_u\partial\sigma_d}\right\rangle=\pm \frac{v_d / v_u IJ_B}{2 \Big(m_{\tilde b_1}^2-m_{\tilde b_2}^2 \Big)}\Big[1 - \frac{v_u^2}{\Big(m_{\tilde b_1}^2-m_{\tilde b_2}^2 \Big)^2}\Big(x_b X_b - H_b\Big) \Big],\nonumber\\
&&\left\langle\frac{\partial^2{m}_{\tilde t_k}^2}{\partial\phi_u\partial\sigma_u}\right\rangle= \pm \frac{v_u^2 IJ_T \Big(x_t X_t+J_t\Big)}{2 \Big(m_{\tilde t_1}^2-m_{\tilde t_2}^2 \Big)^3},\;\left\langle\frac{\partial^2{m}_{\tilde b_k}^2}{\partial\phi_u\partial\sigma_u}\right\rangle=\mp  \frac{v_u^2 IH_B \Big(x_b X_b-H_b\Big)}{2 \Big(m_{\tilde b_1}^2-m_{\tilde b_2}^2 \Big)^3},\nonumber\\
&&\left\langle\frac{\partial^2{m}_{\tilde t_k}^2}{\partial\phi_u\partial\sigma_{s}}\right\rangle= \pm \frac{v_d / v_u IH_T}{2 \Big(m_{\tilde t_1}^2-m_{\tilde t_2}^2 \Big)}\Big[1 + \frac{v_u^2}{\Big(m_{\tilde t_1}^2-m_{\tilde t_2}^2 \Big)^2}\Big(x_t X_t+J_t\Big) \Big]\frac{v_{d}}{v_s},\nonumber\\
&&\left\langle\frac{\partial^2{m}_{\tilde b_k}^2}{\partial\phi_u\partial\sigma_s}\right\rangle=\pm \frac{v_d / v_u IJ_B}{2 \Big(m_{\tilde b_1}^2-m_{\tilde b_2}^2 \Big)}\Big[1 - \frac{v_u^2}{\Big(m_{\tilde b_1}^2-m_{\tilde b_2}^2 \Big)^2}\Big(x_b X_b - H_b\Big) \Big]\frac{v_{d}}{v_s},
\nonumber\\
&&\left\langle\frac{\partial^2{m}_{\tilde t_k}^2}{\partial\phi_\eta\partial\sigma_d}\right\rangle= \pm\frac{v_d v_\eta IH_T a_t X_t}{2\Big( m_{\tilde {t}_1}^2-m_{\tilde {t}_2}^2\Big)^3},\;\left\langle\frac{\partial^2{m}_{\tilde b_k}^2}{\partial\phi_\eta\partial\sigma_d}\right\rangle = \pm\frac{v_d v_\eta IJ_B a_b X_b}{2\Big( m_{\tilde {b}_1}^2-m_{\tilde {b}_2}^2\Big)^3},\nonumber\\
&&\left\langle\frac{\partial^2{m}_{\tilde t_k}^2}{\partial\phi_\eta\partial\sigma_{u}}\right\rangle=\pm\frac{v_u v_\eta IJ_T a_t X_t}{2\Big( m_{\tilde {t}_1}^2-m_{\tilde {t}_2}^2\Big)^3},\;\left\langle\frac{\partial^2{m}_{\tilde b_k}^2}{\partial\phi_\eta\partial\sigma_{u}}\right\rangle= \pm\frac{v_u v_\eta IH_B a_b X_b}{2\Big( m_{\tilde {b}_1}^2-m_{\tilde {b}_2}^2\Big)^3},\nonumber\\
&&\left\langle\frac{\partial^2{m}_{\tilde t_k}^2}{\partial\phi_\eta\partial\sigma_{s}}\right\rangle=\pm\frac{v_u v_\eta IJ_T a_t X_t}{2\Big( m_{\tilde {t}_1}^2-m_{\tilde {t}_2}^2\Big)^3}\frac{v_{u}}{v_s},\;\left\langle\frac{\partial^2{m}_{\tilde b_k}^2}{\partial\phi_\eta\partial\sigma_{s}}\right\rangle= \pm\frac{v_u v_\eta IH_B a_b X_b}{2\Big( m_{\tilde {b}_1}^2-m_{\tilde {b}_2}^2\Big)^3}\frac{v_{u}}{v_s},
\nonumber\\
&&\left\langle\frac{\partial^2{m}_{\tilde t_k}^2}{\partial\phi_{\bar{\eta}}\partial\phi_d}\right\rangle=\mp\frac{v_d v_{\bar{\eta}} IH_T a_t X_t}{2 \Big( m_{\tilde t_1}^2-m_{\tilde t_2}^2 \Big)^3} ,\;\left\langle\frac{\partial^2{m}_{\tilde b_k}^2}{\partial\phi_{\bar{\eta}}\partial\sigma_d}\right\rangle= \mp \frac{v_d v_{\bar{\eta}} IJ_B a_b X_b}{2 \Big( m_{\tilde b_1}^2-m_{\tilde b_2}^2 \Big)^3}, \nonumber\\
&&\left\langle\frac{\partial^2{m}_{\tilde t_k}^2}{\partial\phi_{\bar{\eta}}\partial\sigma_{u}}\right\rangle=\mp \frac{v_u v_{\bar{\eta}} IJ_T a_t X_t}{2 \Big( m_{\tilde t_1}^2-m_{\tilde t_2}^2 \Big)^3},\; \left\langle\frac{\partial^2{m}_{\tilde b_k}^2}{\partial\phi_{\bar{\eta}}\partial\sigma_{u}}\right\rangle=\mp \frac{v_u v_{\bar{\eta}} IH_B a_b X_b}{2 \Big( m_{\tilde b_1}^2-m_{\tilde b_2}^2 \Big)^3}, \nonumber\\
&&\left\langle\frac{\partial^2{m}_{\tilde t_k}^2}{\partial\phi_{\bar{\eta}}\partial\sigma_{s}}\right\rangle=\mp \frac{v_u v_{\bar{\eta}} IJ_T a_t X_t}{2 \Big( m_{\tilde t_1}^2-m_{\tilde t_2}^2 \Big)^3}\frac{v_{u}}{v_s},\; \left\langle\frac{\partial^2{m}_{\tilde b_k}^2}{\partial\phi_{\bar{\eta}}\partial\sigma_{s}}\right\rangle=\mp \frac{v_u v_{\bar{\eta}} IH_B a_b X_b}{2 \Big( m_{\tilde b_1}^2-m_{\tilde b_2}^2 \Big)^3}\frac{v_{u}}{v_s},
\nonumber\\
&&\left\langle\frac{\partial^2{m}_{\tilde t_k}^2}{\partial\phi_s\partial\sigma_{s}}\right\rangle= \pm\frac{v_d^4 / v_s^2 H_T IH_T}{2\Big( m_{\tilde {t}_1}^2-m_{\tilde {t}_2}^2\Big)^3},\;\left\langle\frac{\partial^2{m}_{\tilde b_k}^2}{\partial\phi_s\partial\sigma_{s}}\right\rangle= \pm\frac{v_d^4 / v_s^2 J_B IJ_B}{2\Big( m_{\tilde {b}_1}^2-m_{\tilde {b}_2}^2\Big)^3},\nonumber\\
&&\left\langle\frac{\partial^2{m}_{\tilde t_k}^2}{\partial\phi_s\partial\sigma_{d}}\right\rangle= \pm\frac{v_d^3 / v_s H_T IH_T}{2\Big( m_{\tilde {t}_1}^2-m_{\tilde {t}_2}^2\Big)^3},\;\left\langle\frac{\partial^2{m}_{\tilde b_k}^2}{\partial\phi_s\partial\sigma_{d}}\right\rangle= \pm\frac{v_d^3 / v_s J_B IJ_B}{2\Big( m_{\tilde {b}_1}^2-m_{\tilde {b}_2}^2\Big)^3},\nonumber\\
&&\left\langle\frac{\partial^2{m}_{\tilde t_k}^2}{\partial\phi_s\partial\sigma_{u}}\right\rangle= \pm\frac{v_d^4 / (v_uv_s) H_T IH_T}{2\Big( m_{\tilde {t}_1}^2-m_{\tilde {t}_2}^2\Big)^3},\;\left\langle\frac{\partial^2{m}_{\tilde b_k}^2}{\partial\phi_s\partial\sigma_{u}}\right\rangle= \pm\frac{v_d^4 / (v_uv_s) J_B IJ_B}{2\Big( m_{\tilde {b}_1}^2-m_{\tilde {b}_2}^2\Big)^3},
\nonumber\\
&&\left\langle\frac{\partial^2{m}_{\tilde t_k}^2}{\partial\phi_d\partial\sigma_\eta}\right\rangle=\left\langle\frac{\partial^2{m}_{\tilde b_k}^2}{\partial\phi_d\partial\sigma_\eta}\right\rangle=\left\langle\frac{\partial^2{m}_{\tilde t_k}^2}{\partial\phi_u\partial\sigma_\eta}\right\rangle=\left\langle\frac{\partial^2{m}_{\tilde b_k}^2}{\partial\phi_u\partial\sigma_\eta}\right\rangle=0,\nonumber\\
&&\left\langle\frac{\partial^2{m}_{\tilde t_k}^2}{\partial\phi_d\partial\sigma_{\bar{\eta}}}\right\rangle=\left\langle\frac{\partial^2{m}_{\tilde b_k}^2}{\partial\phi_d\partial\sigma_{\bar{\eta}}}\right\rangle=\left\langle\frac{\partial^2{m}_{\tilde t_k}^2}{\partial\phi_u\partial\sigma_{\bar{\eta}}}\right\rangle=\left\langle\frac{\partial^2{m}_{\tilde b_k}^2}{\partial\phi_u\partial\sigma_{\bar{\eta}}}\right\rangle=0,
\nonumber\\&&\left\langle\frac{\partial^2{m}_{\tilde t_k}^2}{\partial\phi_\eta\partial\sigma_\eta}\right\rangle=\left\langle\frac{\partial^2{m}_{\tilde b_k}^2}{\partial\phi_\eta\partial\sigma_\eta}\right\rangle=
\left\langle\frac{\partial^2{m}_{\tilde t_k}^2}{\partial\phi_\eta\partial\sigma_{\bar{\eta}}}\right\rangle=\left\langle\frac{\partial^2{m}_{\tilde b_k}^2}{\partial\phi_\eta\partial\sigma_{\bar{\eta}}}\right\rangle=0,
\nonumber\\&&\left\langle\frac{\partial^2{m}_{\tilde t_k}^2}{\partial\phi_{\bar{\eta}}\partial\sigma_\eta}\right\rangle=
\left\langle\frac{\partial^2{m}_{\tilde b_k}^2}{\partial\phi_{\bar{\eta}}\partial\sigma_\eta}\right\rangle=
\left\langle\frac{\partial^2{m}_{\tilde t_k}^2}{\partial\phi_{\bar{\eta}}\partial\sigma_{\bar{\eta}}}\right\rangle=
\left\langle\frac{\partial^2{m}_{\tilde b_k}^2}{\partial\phi_{\bar{\eta}}\partial\sigma_{\bar{\eta}}}\right\rangle= 0,\nonumber\\
&&\left\langle\frac{\partial^2{m}_{\tilde t_k}^2}{\partial\phi_s\partial\sigma_\eta}\right\rangle=\left\langle\frac{\partial^2{m}_{\tilde b_k}^2}{\partial\phi_s\partial\sigma_\eta}\right\rangle=\left\langle\frac{\partial^2{m}_{\tilde b_k}^2}{\partial\phi_s\partial\sigma_{\bar{\eta}}}\right\rangle=
\left\langle\frac{\partial^2{m}_{\tilde t_k}^2}{\partial\phi_s\partial\sigma_{\bar{\eta}}}\right\rangle=0.
\end{eqnarray}
Finally, the derivatives $\left\langle \partial^2 m_{\tilde q_k}^2/\partial \phi_m \partial \phi_n\right\rangle$ are calculated as
\begin{eqnarray}
&&\left\langle\frac{\partial^2{m}_{\tilde t_k}^2}{\partial\phi_d^2}\right\rangle= \frac{1}{8} G^2 \pm \frac{ x_t X_t +  |Y_t \lambda^*|^2 v_s^2+x_t^2 v_d^2  }{2 \Big( m_{\tilde t_1}^2-m_{\tilde t_2}^2 \Big)}\mp\frac{v_d^2}{2 \Big( m_{\tilde t_1}^2-m_{\tilde t_2}^2 \Big)^3} \Big(x_t X_t - H_t\Big)^2,\nonumber\\&&\left\langle\frac{\partial^2{m}_{\tilde b_k}^2}{\partial\phi_d^2}\right\rangle= |Y_b|^2 - \frac{1}{8} G^2 \mp \frac{ x_b X_b - 2 |A_b|^2 |Y_b|^2-x_b^2 v_d^2 }{2 \Big( m_{\tilde b_1}^2-m_{\tilde b_2}^2 \Big)}\mp \frac{v_d^2}{2 \Big( m_{\tilde b_1}^2-m_{\tilde b_2}^2 \Big)^3} \Big (x_b X_b + J_b\Big)^2,\nonumber\\&&\left\langle\frac{\partial^2{m}_{\tilde t_k}^2}{\partial\phi_u^2}\right\rangle= |Y_t|^2-\frac{1}{8} G^2 \mp  \frac{ x_t X_t - 2 |A_t|^2 |Y_t|^2-x_t^2 v_u^2 }{2 \Big( m_{\tilde t_1}^2-m_{\tilde t_2}^2 \Big)}\mp \frac{v_u^2}{2 \Big( m_{\tilde t_1}^2-m_{\tilde t_2}^2 \Big)^3} \Big ( x_t X_t + J_t\Big)^2,\nonumber\\&&\left\langle\frac{\partial^2{m}_{\tilde b_k}^2}{\partial\phi_u^2}\right\rangle= \frac{1}{8} G^2 \pm \frac{ x_b X_b+ |Y_b \lambda^*|^2 v_s^2+x_b^2 v_u^2}{2 \Big( m_{\tilde b_1}^2-m_{\tilde b_2}^2 \Big)} \mp \frac{v_u^2}{2 \Big( m_{\tilde b_1}^2-m_{\tilde b_2}^2 \Big)^3} \Big (x_b X_b - H_b\Big)^2,\nonumber\\
&&\left\langle\frac{\partial^2{m}_{\tilde t_k}^2}{\partial\phi_\eta^2}\right\rangle= \frac{1}{4} g_B g_{YB} \mp \frac{1}{2 \Big( m_{\tilde t_1}^2-m_{\tilde t_2}^2 \Big)} \Big(a_tX_t-a_t^2 v_\eta^2\Big)  \mp \frac{1}{2 \Big( m_{\tilde t_1}^2-m_{\tilde t_2}^2 \Big)^3} v_\eta^2 a_t^2 X^2_t,\nonumber\\
&&\left\langle\frac{\partial^2{m}_{\tilde b_k}^2}{\partial\phi_\eta^2}\right\rangle=  - \frac{1}{4} g_B g_{YB} \mp \frac{1}{2 \Big( m_{\tilde b_1}^2-m_{\tilde b_2}^2 \Big)}  \Big(a_bX_b-a_b^2 v_\eta^2\Big) \mp \frac{1}{2 \Big( m_{\tilde t_1}^2-m_{\tilde t_2}^2 \Big)^3} v_\eta^2 a_b^2 X^2_b,\nonumber\\
&&\left\langle\frac{\partial^2{m}_{\tilde t_k}^2}{\partial\phi_{\bar{\eta}}^2}\right\rangle= - \frac{1}{4} g_B g_{YB} \pm \frac{1}{2 \Big( m_{\tilde t_1}^2-m_{\tilde t_2}^2 \Big)} \Big(a_tX_t+a_t^2 v_{\bar{\eta}}^2\Big)\mp\frac{1}{2 \Big( m_{\tilde t_1}^2-m_{\tilde t_2}^2 \Big)^3} v_{\bar{\eta}}^2 a_t^2 X^2_t,\nonumber\\
&&\left\langle\frac{\partial^2{m}_{\tilde b_k}^2}{\partial\phi_{\bar{\eta}}^2}\right\rangle= \frac{1}{4} g_B g_{YB} \pm \frac{1}{2 \Big( m_{\tilde b_1}^2-m_{\tilde b_2}^2 \Big)}\Big(a_bX_b+a_b^2 v_{\bar{\eta}}^2\Big) \mp \frac{1}{2 \Big( m_{\tilde b_1}^2-m_{\tilde b_2}^2 \Big)^3} v_{\bar{\eta}}^2 a_b^2 X^2_b,\nonumber\\
&&\left\langle\frac{\partial^2{m}_{\tilde t_k}^2}{\partial\phi_s^2}\right\rangle=\mp\frac{v_d^4 /v_s^2H_t^2}{2 \Big( m_{\tilde t_1}^2-m_{\tilde t_2}^2 \Big)^3} \pm\frac{|Y_t \lambda^*|^2v_d^2}{2 \Big( m_{\tilde t_1}^2-m_{\tilde t_2}^2 \Big)},\nonumber\\
&&\left\langle\frac{\partial^2{m}_{\tilde b_k}^2}{\partial\phi_s^2}\right\rangle=\mp\frac{v_u^4 /v_s^2 H_b^2}{2 \Big( m_{\tilde b_1}^2-m_{\tilde b_2}^2 \Big)^3}\pm\frac{|Y_b \lambda^*|^2v_u^2}{2 \Big( m_{\tilde b_1}^2-m_{\tilde b_2}^2 \Big)},
\nonumber\\
&&\left\langle\frac{\partial^2{m}_{\tilde t_k}^2}{\partial\phi_d\partial\phi_u}\right\rangle= \mp \frac{\Big( \frac{ x_t^2 v^2 \sin 2 \beta}{2} + \frac{H_T}{\tan\beta}\Big)}{2 \Big( m_{\tilde t_1}^2-m_{\tilde t_2}^2 \Big)} \pm \frac{v_d v_u\Big( x_t X_t-H_t\Big)\Big(x_t X_t+ J_t\Big)}{2 \Big( m_{\tilde t_1}^2-m_{\tilde t_2}^2 \Big)^3},\nonumber\\
&&\left\langle\frac{\partial^2{m}_{\tilde b_k}^2}{\partial\phi_d\partial\phi_u}\right\rangle=\mp \frac{\Big( \frac{x_b^2 v^2 \sin 2 \beta}{2} +\frac{J_B}{\tan\beta}\Big)}{2 \Big( m_{\tilde b_1}^2-m_{\tilde b_2}^2 \Big)}\pm \frac{v_d v_u\Big( x_b X_b + J_b\Big)\Big(x_b X_b - H_b\Big)}{2 \Big( m_{\tilde b_1}^2-m_{\tilde b_2}^2 \Big)^3},\nonumber\\
&&\left\langle\frac{\partial^2{m}_{\tilde t_k}^2}{\partial\phi_d\partial\phi_\eta}\right\rangle= \mp \frac{v_d v_\eta}{2 \Big( m_{\tilde t_1}^2-m_{\tilde t_2}^2 \Big)}  x_t a_t  \pm \frac{v_d v_\eta}{2 \Big( m_{\tilde t_1}^2-m_{\tilde t_2}^2 \Big)^3} a_t X_t\Big(x_t X_t - H_t\Big),\nonumber\\
&&\left\langle\frac{\partial^2{m}_{\tilde b_k}^2}{\partial\phi_d\partial\phi_\eta}\right\rangle= \pm \frac{v_d v_\eta}{2 \Big( m_{\tilde b_1}^2-m_{\tilde b_2}^2 \Big)}  x_b a_b \mp \frac{v_d v_\eta}{2 \Big( m_{\tilde b_1}^2-m_{\tilde b_2}^2 \Big)^3} a_b X_b\Big(x_b X_b + J_b\Big),\nonumber\\
&&\left\langle\frac{\partial^2{m}_{\tilde t_k}^2}{\partial\phi_d\partial\phi_{\bar{\eta}}}\right\rangle= \pm\frac{v_d v_{\bar{\eta}}}{2 \Big( m_{\tilde t_1}^2-m_{\tilde t_2}^2 \Big)}  x_t a_t \mp\frac{v_d v_{\bar{\eta}}}{2 \Big( m_{\tilde t_1}^2-m_{\tilde t_2}^2 \Big)^3} a_t X_t\Big (x_t X_t - H_t\Big),\nonumber\\
&&\left\langle\frac{\partial^2{m}_{\tilde b_k}^2}{\partial\phi_d\partial\phi_{\bar{\eta}}}\right\rangle= \mp \frac{v_d v_{\bar{\eta}}}{2 \Big( m_{\tilde b_1}^2-m_{\tilde b_2}^2 \Big)}  x_b a_b \pm \frac{v_d v_{\bar{\eta}}}{2 \Big( m_{\tilde b_1}^2-m_{\tilde b_2}^2 \Big)^3} a_b X_b \Big(x_b X_b+ J_b\Big),\nonumber\\
&&\left\langle\frac{\partial^2{m}_{\tilde t_k}^2}{\partial\phi_d\partial\phi_s}\right\rangle=\mp\frac{v_d /v_s H_T}{2 \Big( m_{\tilde t_1}^2-m_{\tilde t_2}^2 \Big)}\pm\frac{v_d^3 /v_s\Big(x_t X_t- H_t\Big)H_t}{2 \Big( m_{\tilde t_1}^2-m_{\tilde t_2}^2 \Big)^3}, \nonumber\\
&& \left\langle\frac{\partial^2{m}_{\tilde b_k}^2}{\partial\phi_d\partial\phi_s}\right\rangle=\mp\frac{v_d/v_s J_B}{2 \Big( m_{\tilde b_1}^2-m_{\tilde b_2} \Big)^3}\mp\frac{v_d v_u^2 /v_s\Big(x_b X_b+ J_b\Big) H_b}{2 \Big( m_{\tilde b_1}^2-m_{\tilde b_2}^2 \Big)^3},\nonumber\\
 &&\left\langle\frac{\partial^2{m}_{\tilde t_k}^2}{\partial\phi_u\partial\phi_\eta}\right\rangle=\pm \frac{v_u v_\eta}{2 \Big( m_{\tilde t_1}^2-m_{\tilde t_2}^2 \Big)}  x_t a_t \mp \frac{v_u v_\eta}{2 \Big( m_{\tilde t_1}^2-m_{\tilde t_2}^2 \Big)^3} a_t X_t\Big(x_t X_t+ J_t\Big),\nonumber\\
&&\left\langle\frac{\partial^2{m}_{\tilde b_k}^2}{\partial\phi_u\partial\phi_\eta}\right\rangle= \mp \frac{v_u v_\eta}{2 \Big( m_{\tilde b_1}^2-m_{\tilde b_2}^2 \Big)}  x_b a_b \pm \frac{v_u v_\eta}{2 \Big( m_{\tilde b_1}^2-m_{\tilde b_2}^2 \Big)^3} a_b X_b\Big(x_b X_b- H_b\Big),\nonumber\\	
&&\left\langle\frac{\partial^2{m}_{\tilde t_k}^2}{\partial\phi_u\partial\phi_{\bar{\eta}}}\right\rangle= \mp \frac{v_u v_{\bar{\eta}}}{2 \Big( m_{\tilde t_1}^2-m_{\tilde t_2}^2 \Big)}  x_t a_t \pm \frac{v_u v_{\bar{\eta}}}{2 \Big( m_{\tilde t_1}^2-m_{\tilde t_2}^2 \Big)^3} a_t X_t\Big(x_t X_t+ J_t\Big),\nonumber\\
&&\left\langle\frac{\partial^2{m}_{\tilde b_k}^2}{\partial\phi_u\partial\phi_{\bar{\eta}}}\right\rangle= \pm \frac{v_u v_{\bar{\eta}}}{2 \Big( m_{\tilde b_1}^2-m_{\tilde b_2}^2 \Big)}  x_b a_b \mp \frac{v_u v_{\bar{\eta}}}{2 \Big( m_{\tilde b_1}^2-m_{\tilde b_2}^2 \Big)^3} a_b X_b \Big(x_b X_b- H_b\Big),\nonumber\\
&&\left\langle\frac{\partial^2{m}_{\tilde t_k}^2}{\partial\phi_u\partial\phi_s}\right\rangle=\mp\frac{v_u/v_s J_T}{2 \Big( m_{\tilde t_1}^2-m_{\tilde t_2}^2 \Big)}\mp\frac{v_u v_d^2 /v_s\Big(x_t X_t+ J_t\Big) H_t}{2 \Big( m_{\tilde t_1}^2-m_{\tilde t_2}^2 \Big)^3},\nonumber\\
&&\left\langle\frac{\partial^2{m}_{\tilde b_k}^2}{\partial\phi_u\partial\phi_s}\right\rangle=\pm\frac{v_u/v_s H_B}{2 \Big( m_{\tilde b_1}^2-m_{\tilde b_2}^2 \Big)}\pm\frac{v_u^3 /v_s\Big(x_b X_b- H_b\Big) H_b}{2 \Big( m_{\tilde b_1}^2-m_{\tilde b_2}^2 \Big)^3},
\nonumber\\
&&\left\langle\frac{\partial^2{m}_{\tilde t_k}^2}{\partial\phi_\eta \partial\phi_{\bar{\eta}}}\right\rangle= \mp\frac{v_\eta v_{\bar{\eta}}}{2 \Big( m_{\tilde t_1}^2-m_{\tilde t_2}^2 \Big)}a_t^2\pm \frac{v_\eta v_{\bar{\eta}}}{2 \Big( m_{\tilde t_1}^2-m_{\tilde t_2}^2 \Big)^3} a_t^2 X^2_t,\nonumber\\
&&\left\langle\frac{\partial^2{m}_{\tilde b_k}^2}{\partial\phi_\eta \partial\phi_{\bar{\eta}}}\right\rangle= \mp \frac{v_\eta v_{\bar{\eta}}}{2 \Big( m_{\tilde b_1}^2-m_{\tilde b_2}^2 \Big)}a_b^2 \pm\frac{v_\eta v_{\bar{\eta}}}{2 \Big( m_{\tilde b_1}^2-m_{\tilde b_2}^2 \Big)^3} a_b^2 X^2_b,\nonumber\\
&&\left\langle\frac{\partial^2{m}_{\tilde t_k}^2}{\partial\phi_\eta\partial\phi_s}\right\rangle=\mp\frac{v_\eta v_d^2 /v_s}{2 \Big( m_{\tilde t_1}^2-m_{\tilde t_2}^2 \Big)^3} a_t X_t H_t,\;\left\langle\frac{\partial^2{m}_{\tilde b_k}^2}{\partial\phi_{\eta}\partial\phi_s}\right\rangle=\mp\frac{v_\eta v_u^2 /v_s}{2 \Big( m_{\tilde b_1}^2-m_{\tilde b_2}^2 \Big)^3} a_b X_b H_b,\nonumber\\
&&\left\langle\frac{\partial^2{m}_{\tilde t_k}^2}{\partial\phi_{\bar{\eta}}\partial\phi_s}\right\rangle= \pm\frac{v_{\bar{\eta}} v_d^2 /v_s}{2 \Big( m_{\tilde t_1}^2-m_{\tilde t_2}^2 \Big)^3} a_t X_t H_t,\;\left\langle\frac{\partial^2{m}_{\tilde b_k}^2}{\partial\phi_{\bar{\eta}}\partial\phi_s}\right\rangle= \pm\frac{v_{\bar{\eta}} v_u^2 /v_s}{2 \Big( m_{\tilde b_1}^2-m_{\tilde b_2}^2 \Big)^3} a_b X_b H_b.
\end{eqnarray}

\section{The coupling coefficients}
In the NB-LSSM, the concrete coupling coefficients contributed to the Higgs decays are specifically discussed as follows (Here, the 95 GeV scalar excess corresponds to $n=1$ and the 125 GeV SM-like Higgs boson corresponds to $n=2$):
1. The Higgs-fermion-fermion contributions:
\begin{eqnarray}
&&g^{B-L}_{{h^n_{S}uu}}=-\sum_{i=j_2=1}^{3}\sum_{j=j_1=1}^{3}\frac{v}{m_{u_j}}\Big(-\frac{1}{\sqrt{2}}  Y^*_{u,{j_1 j_2 }} U_{R,{j j_1}}^{u}  U_{L,{i j_2}}^{u}  Z_{{n 2}}^{H}\Big), \nonumber\\
&&g^{B-L}_{{h^n_{A}uu}}=-\sum_{i=j_2=1}^{3}\sum_{j=j_1=1}^{3}\frac{v}{m_{u_j}}\Big(-\frac{1}{\sqrt{2}} Y^*_{u,{j_1 j_2}} U_{R,{j j_1}}^{u} U_{L,{i j_2}}^{u} Z_{{n 7}}^{H}\Big),
\nonumber\\
&&g^{B-L}_{{h^n_{S}dd}}=-\sum_{i=j_2=1}^{3}\sum_{j=j_1=1}^{3}\frac{v}{m_{d_j}}\Big(-\frac{1}{\sqrt{2}}  Y^*_{d,{j_1 j_2}} U_{R,{j j_1}}^{d}  U_{L,{i j_2}}^{d}  Z_{{n 1}}^{H}\Big),
\nonumber\\
&&g^{B-L}_{{h^n_{A}dd}}=-\sum_{i=j_2=1}^{3}\sum_{j=j_1=1}^{3}\frac{v}{m_{d_j}}\Big(-\frac{1}{\sqrt{2}}  Y^*_{d,{j_1 j_2}} U_{R,{j j_1}}^{u} U_{L,{i j_2}}^{u} Z_{{n 6}}^{H}\Big),
\nonumber\\
&&g^{B-L}_{h^n_{S}l l}=-\sum_{i=j_2=1}^{3}\sum_{j=j_1=1}^{3}\frac{v}{m_{l_j}}\Big(-\frac{1}{\sqrt{2}} Y^*_{e,{j_1 j_2}} U_{R,{j j_1}}^{e}  U_{L,{i j_2}}^{e}  Z_{{n 1}}^{H}\Big),\nonumber\\
&&g^{B-L}_{h^n_{A}l l}=-\sum_{i=j_2=1}^{3}\sum_{j=j_1=1}^{3}\frac{v}{m_{l_j}}\Big(-\frac{1}{\sqrt{2}} Y^*_{e,{j_1 j_2}} U_{R,{j j_1}}^{e} U_{L,{i j_2}}^{e} Z_{{n 6}}^{H}\Big),
\nonumber\\
&&g^{B-L}_{{h^n_{S}\chi_i^+\chi_j^-}}=-\sum_{i=j=1}^2\frac{2s_w}{e}\Big[-\frac{1}{\sqrt{2}} g_2 \Big(U_{{j 1}} V_{{i 2}} Z_{{n 2}}^{H}  + U_{{j 2}} V_{{i 1}} Z_{{n 1}}^{H}+\lambda U_{{j 2}} V_{{i 2}} Z_{{n 5}}^{H} \Big)\Big],\nonumber\\
&&g^{B-L}_{{h^n_{A}\chi_i^+\chi_j^-}}=-\sum_{i=j=1}^2\frac{2s_w}{e}\Big[-\frac{1}{\sqrt{2}} g_2 \Big(U_{{j 1}} V_{{i 2}} Z_{{n 7}}^{H} + U_{{j 2}} V_{{i 1}} Z_{{n 6}}^{H} -\lambda U_{{j 2}} V_{{i 2}} Z_{{n 10}}^{H}\Big)\Big].
\end{eqnarray}
2. The Higgs-$W(Z)$ boson-$W(Z)$ boson contributions:
\begin{eqnarray}
&&g^{B-L}_{h^n_{S}WW}=\cos\beta Z_{{n 1}}^{H}  + \sin\beta Z_{{n 2}}^{H} ,
\nonumber\\&&g^{B-L}_{h^n_{S}ZZ}=\frac{v}{2m_Z^2}\Big[\frac{1}{2} \Big(v_d (g_1 c'_w  s_w   + g_2 c_w  c'_w  \hspace{-0.1cm} - \hspace{-0.1cm}g_{Y B} s'_w  )^{2} Z_{{n 1}}^{H}+v_u \Big(g_1 c'_w  s_w  \hspace{-0.1cm} +\hspace{-0.1cm} g_2 c_w  c'_w  \nonumber\\
&&\hspace{1.2cm}-\hspace{-0.1cm} g_{Y B} s'_w  )^{2} Z_{{n 2}}^{H} +4 (- g_{B} s'_w )^{2} (v_{\bar{\eta}} Z_{{n 4}}^{H}  + v_{\eta} Z_{{n 3}}^{H} )\Big)\Big].
\end{eqnarray}
3. The Higgs-up squark-up squark contributions:
\begin{eqnarray}
&&g^{B-L}_{h^n_{S}\tilde{U}\tilde{U}}=-\frac{v}{2m_Z^2}\sum_{i=j=1}^6\Big[\frac{1}{12} \Big(6 \Big( v_S\lambda^*\sum_{j_2=1}^{3}Z^{U,*}_{j j_2} \sum_{j_1=1}^{3}Y_{u,{j_1 j_2}} Z_{{i 3 + j_1}}^{U}   Z_{{n 1}}^{H} \nonumber \\
&&+\hspace{-0.1cm}v_S\lambda\hspace{-0.1cm} \sum_{j_2=1}^{3}\hspace{-0.1cm}\sum_{j_1=1}^{3}\hspace{-0.1cm}Y^*_{u,{j_1 j_2}} Z^{U,*}_{j 3 + j_1}  Z_{{i j_2}}^{U}  Z_{{n 1}}^{H}+v_u\lambda^*\sum_{j_2=1}^{3}Z^{U,*}_{j j_2} \sum_{j_1=1}^{3}Y_{u,{j_1 j_2}} Z_{{i 3 + j_1}}^{U}   Z_{{n 5}}^{H}\nonumber \\
&&+\hspace{-0.1cm}v_u\lambda\hspace{-0.1cm} \sum_{j_2=1}^{3}\hspace{-0.1cm}\sum_{j_1=1}^{3}\hspace{-0.1cm}Y^*_{u,{j_1 j_2}} Z^{U,*}_{j 3 + j_1}  Z_{{i j_2}}^{U}  Z_{{n 5}}^{H}\hspace{-0.1cm}-\hspace{-0.1cm} \Big(\hspace{-0.1cm}\sqrt{2} \hspace{-0.1cm}\sum_{j_2=1}^{3}\hspace{-0.1cm}Z^{U,*}_{j j_2} \sum_{j_1=1}^{3}\hspace{-0.1cm}Z_{{i 3 + j_1}}^{U} T_{u,{j_1 j_2}}  \hspace{-0.1cm} +\hspace{-0.1cm}\sqrt{2} \hspace{-0.1cm}\sum_{j_2=1}^{3}\sum_{j_1=1}^{3}\hspace{-0.1cm}Z^{U,*}_{j 3 + j_1} T^*_{u,{j_1 j_2}}  Z_{{i j_2}}^{U}  \nonumber \\
&&+2 v_u (\sum_{j_3=1}^{3}Z^{U,*}_{j 3 + j_3} \sum_{j_2=1}^{3}\sum_{j_1=1}^{3}Y^*_{u,{j_3 j_1}} Y_{u,{j_2 j_1}}  Z_{{i 3 + j_2}}^{U}   + \sum_{j-3=1}^{3}\sum_{j_2=1}^{3}Z^{U,*}_{j j_2} \sum_{j_1=1}^{3}Y^*_{u,{j_1 j_3}} Y_{u,{j_1 j_2}}   Z_{{i j_3}}^{U} )\Big)Z_{{n 2}}^{H} \Big)\nonumber \\
&&+\hspace{-0.1cm}\sum_{j_1=1}^{3}\hspace{-0.1cm}Z^{U,*}_{j 3 + j_1} Z_{{i 3 + j_1}}^{U}  \Big(\hspace{-0.1cm}-\hspace{-0.1cm} \Big(4 g_{1}^{2} \hspace{-0.1cm}+\hspace{-0.1cm} g_{Y B}(4 g_{Y B} \hspace{-0.1cm} +\hspace{-0.1cm} g_{B})\Big)v_d Z_{{n 1}}^{H} \hspace{-0.1cm}+\hspace{-0.1cm}\Big(4 g_{1}^{2} \hspace{-0.1cm}+\hspace{-0.1cm} g_{Y B}(4 g_{Y B} \hspace{-0.1cm} +\hspace{-0.1cm} g_{B})\Big)v_u Z_{{n 2}}^{H} \nonumber \\
&&-2( 4 g_{Y B} g_{B} \hspace{-0.1cm} +\hspace{-0.1cm} g_{B}^{2})(\hspace{-0.1cm}-\hspace{-0.1cm} v_{\bar{\eta}} Z_{{n 4}}^{H} \hspace{-0.1cm} +\hspace{-0.1cm} v_{\eta} Z_{{n 3}}^{H} )\Big)\hspace{-0.1cm}+\hspace{-0.1cm}\sum_{j_1=1}^{3}\hspace{-0.1cm}Z^{U,*}_{j j_1} Z_{{i j_1}}^{U}  \Big((\hspace{-0.1cm}-\hspace{-0.1cm}3 g_{2}^{2} \hspace{-0.1cm}+\hspace{-0.1cm} g_{Y B} g_{B} \hspace{-0.1cm} +\hspace{-0.1cm} g_{1}^{2}\hspace{-0.1cm} +\hspace{-0.1cm} g_{Y B}^{2})v_d Z_{{n 1}}^{H} \nonumber \\
&&-(-3 g_{2}^{2} + g_{Y B} g_{B}  + g_{1}^{2} + g_{Y B}^{2})v_u Z_{{n 2}}^{H} +2 (g_{Y B} g_{B}  + g_{B}^{2})(- v_{\bar{\eta}} Z_{{n 4}}^{H}  + v_{\eta} Z_{{n 3}}^{H})\Big)\Big)\Big].
\end{eqnarray}
4. The Higgs-down squark-down squark contributions:
\begin{eqnarray}
&&g^{B-L}_{h^n_{S}\tilde{D}\tilde{D}}\hspace{-0.1cm}=\hspace{-0.1cm}\frac{-v}{2m_Z^2}\hspace{-0.1cm}\sum_{i=j=1}^6\hspace{-0.1cm}\Big[\frac{1}{12} \Big(\hspace{-0.1cm}-\hspace{-0.1cm}6 (\sqrt{2} \sum_{j_2=1}^{3}\hspace{-0.1cm}Z^{D,*}_{j j_2} \sum_{j_1=1}^{3}\hspace{-0.1cm}Z_{{i 3 + j_1}}^{D} T_{d,{j_1 j_2}}   Z_{{n 1}}^{H} \hspace{-0.1cm}+\hspace{-0.1cm}\sqrt{2} \sum_{j_2=1}^{3}\hspace{-0.1cm}\sum_{j_1=1}^{3}\hspace{-0.1cm}Z^{D,*}_{j 3 + j_1} T^*_{d,{j_1 j_2}}  Z_{{i j_2}}^{D}  Z_{{n 1}}^{H} \nonumber \\
&&+2 v_d \sum_{j_3=1}^{3}Z^{D,*}_{j 3 + j_3} \sum_{j_2=1}^{3}\sum_{j_1=1}^{3}Y^*_{d,{j_3 j_1}} Y_{d,{j_2 j_1}}  Z_{{i 3 + j_2}}^{D}   Z_{{n 1}}^{H} +2 v_d \sum_{j_3=1}^{3}\sum_{j_2=1}^{3}Z^{D,*}_{j j_2} \sum_{j_1=1}^{3}Y^*_{d,{j_1 j_3}} Y_{d,{j_1 j_2}}   Z_{{i j_3}}^{D}  Z_{{n 1}}^{H} \nonumber \\
&&- v_S\lambda^* \sum_{j_2=1}^{3}Z^{D,*}_{j j_2} \sum_{j_1=1}^{3}Y_{d,{j_1 j_2}} Z_{{i 3 + j_1}}^{D}   Z_{{n 2}}^{H} - v_S\lambda \sum_{j_2=1}^{3}\sum_{j_1=1}^{3}Y^*_{d,{j_1 j_2}} Z^{D,*}_{j 3 + j_1}  Z_{{i j_2}}^{D}  Z_{{n 2}}^{H} \nonumber \\
&&- v_u\lambda^* \sum_{j_2=1}^{3}Z^{D,*}_{j j_2} \sum_{j_1=1}^{3}Y_{d,{j_1 j_2}} Z_{{i 3 + j_1}}^{D}   Z_{{n 5}}^{H} - v_u\lambda \sum_{j_2=1}^{3}\sum_{j_1=1}^{3}Y^*_{d,{j_1 j_2}} Z^{D,*}_{j 3 + j_1}  Z_{{i j_2}}^{D}  Z_{{n 5}}^{H} )\nonumber \\
&&+\hspace{-0.1cm}\sum_{j_1=1}^{3}\hspace{-0.1cm}Z^{D,*}_{j 3 + j_1} Z_{{i 3 + j_1}}^{D}  \Big(\Big(2 g_{1}^{2} \hspace{-0.1cm}+\hspace{-0.1cm} g_{Y B}(2 g_{Y B}  \hspace{-0.1cm}-\hspace{-0.1cm} g_{B})\Big)v_d Z_{{n 1}}^{H} \hspace{-0.1cm}+\hspace{-0.1cm}\Big(\hspace{-0.1cm}-\hspace{-0.1cm}2 g_{1}^{2} \hspace{-0.1cm}+\hspace{-0.1cm} g_{Y B} (\hspace{-0.1cm}-\hspace{-0.1cm}2 g_{Y B} \hspace{-0.1cm} + \hspace{-0.1cm} g_{B})\Big)v_u Z_{{n 2}}^{H} \nonumber\\
&&+2(2 g_{Y B} g_{B}  \hspace{-0.1cm}-\hspace{-0.1cm} g_{B}^{2})(\hspace{-0.1cm}- \hspace{-0.1cm} v_{\bar{\eta}} Z_{{n 4}}^{H} \hspace{-0.1cm} +\hspace{-0.1cm} v_{\eta} Z_{{n 3}}^{H} )\Big)\hspace{-0.1cm}+\hspace{-0.1cm}\sum_{j_1=1}^{3}\hspace{-0.1cm}Z^{D,*}_{j j_1} Z_{{i j_1}}^{D}  \Big((3 g_{2}^{2} \hspace{-0.1cm}+\hspace{-0.1cm} g_{Y B} g_{B} \hspace{-0.1cm} +\hspace{-0.1cm} g_{1}^{2}\hspace{-0.1cm} +\hspace{-0.1cm} g_{Y B}^{2})v_d Z_{{n 1}}^{H} \nonumber \\
&&-(3 g_{2}^{2} + g_{Y B} g_{B}  + g_{1}^{2} + g_{Y B}^{2})v_u Z_{{n 2}}^{H} +2( g_{Y B} g_{B}  + g_{B}^{2})(- v_{\bar{\eta}} Z_{{n 4}}^{H}  + v_{\eta} Z_{{n 3}}^{H})\Big)\Big)\Big].
\end{eqnarray}
5. The Higgs-slepton-slepton contributions:
\begin{eqnarray}
&&g^{B-L}_{h^n_{S}\tilde{L}\tilde{L}}\hspace{-0.1cm}=\hspace{-0.1cm}\frac{-v}{2m_Z^2}\hspace{-0.1cm}\sum_{i=j=1}^6\hspace{-0.1cm}\hspace{-0.1cm}\Big[\frac{1}{4} \Big( \hspace{-0.1cm}- \hspace{-0.1cm}2 (\sqrt{2}  \hspace{-0.1cm}\sum_{j_2=1}^{3}\hspace{-0.1cm}Z^{E,*}_{j j_2}\hspace{-0.1cm} \sum_{j_1=1}^{3}\hspace{-0.1cm}Z_{{i 3 + j_1}}^{E} T_{e,{j_1 j_2}}   Z_{{n 1}}^{H}  \hspace{-0.1cm}+ \hspace{-0.1cm}\sqrt{2} \sum_{j_2=1}^{3}\sum_{j_1=1}^{3}Z^{E,*}_{j 3 + j_1} T^*_{e,{j_1 j_2}}  Z_{{i j_2}}^{E}  Z_{{n 1}}^{H} \nonumber \\
&&+2 v_d \sum_{j_3=1}^{3}Z^{E,*}_{j 3 + j_3} \sum_{j_2=1}^{3}\sum_{j_1=1}^{3}Y^*_{e,{j_3 j_1}} Y_{e,{j_2 j_1}}  Z_{{i 3 + j_2}}^{E}   Z_{{n 1}}^{H} +2 v_d \sum_{j_3=1}^{3}\sum_{j_2=1}^{3}Z^{E,*}_{j j_2} \sum_{j_1=1}^{3}Y^*_{e,{j_1 j_3}} Y_{e,{j_1 j_2}}   Z_{{i j_3}}^{E}  Z_{{n 1}}^{H} \nonumber \\
&&- v_S\lambda^* \sum_{j_2=1}^{3}Z^{E,*}_{j j_2} \sum_{j_1=1}^{3}Y_{e,{j_1 j_2}} Z_{{i 3 + j_1}}^{E}   Z_{{n 2}}^{H} - v_S\lambda \sum_{j_2=1}^{3}\sum_{j_1=1}^{3}Y^*_{e,{j_1 j_2}} Z^{E,*}_{j 3 + j_1}  Z_{{i j_2}}^{E}  Z_{{n 2}}^{H}\nonumber \\
&&- v_u\lambda^* \sum_{j_2=1}^{3}Z^{E,*}_{j j_2} \sum_{j_1=1}^{3}Y_{e,{j_1 j_2}} Z_{{i 3 + j_1}}^{E}   Z_{{n 5}}^{H} - v_u\lambda \sum_{j_2=1}^{3}\sum_{j_1=1}^{3}Y^*_{e,{j_1 j_2}} Z^{E,*}_{j 3 + j_1}  Z_{{i j_2}}^{E}  Z_{{n 5}}^{H})\nonumber \\
&&+\sum_{j_1=1}^{3}\hspace{-0.1cm}Z^{E,*}_{j 3 + j_1} Z_{{i 3 + j_1}}^{E}  \Big(\Big(2 g_{1}^{2} \hspace{-0.1cm}+\hspace{-0.1cm} g_{Y B}(2 g_{Y B} \hspace{-0.1cm} +\hspace{-0.1cm} g_{B})\Big)v_d Z_{{n 1}}^{H} \hspace{-0.1cm}- \hspace{-0.1cm}\Big(2 g_{1}^{2} \hspace{-0.1cm}+\hspace{-0.1cm} g_{Y B}(2 g_{Y B} \hspace{-0.1cm} +\hspace{-0.1cm} g_{B})\Big)v_u Z_{{n 2}}^{H} \nonumber \\
&&+2 (2 g_{Y B} g_{B}  \hspace{-0.1cm}+ \hspace{-0.1cm}g_{B}^{2})(\hspace{-0.1cm}-\hspace{-0.1cm} v_{\bar{\eta}} Z_{{n 4}}^{H} \hspace{-0.1cm} +\hspace{-0.1cm} v_{\eta} Z_{{n 3}}^{H} )\hspace{-0.1cm}\Big)\hspace{-0.1cm}+\hspace{-0.1cm}\sum_{j_1=1}^{3}\hspace{-0.1cm}Z^{E,*}_{j j_1} Z_{{i j_1}}^{E}  \Big(\hspace{-0.1cm}-\hspace{-0.1cm} (\hspace{-0.1cm}-\hspace{-0.1cm} g_{2}^{2} \hspace{-0.1cm} +\hspace{-0.1cm} g_{Y B} g_{B} \hspace{-0.1cm} +\hspace{-0.1cm} g_{1}^{2} \hspace{-0.1cm}+\hspace{-0.1cm} g_{Y B}^{2})v_d Z_{{n 1}}^{H}  \nonumber \\
&&+(- g_{2}^{2}  + g_{Y B} g_{B}  + g_{1}^{2} + g_{Y B}^{2})v_u Z_{{n 2}}^{H}-2 ( g_{Y B} g_{B}  + g_{B}^{2})(- v_{\bar{\eta}} Z_{{n 4}}^{H}  + v_{\eta} Z_{{n 3}}^{H})\Big)\Big)\Big].
\end{eqnarray}
6. The Higgs-charge Higgs-charge Higgs contributions:
\begin{eqnarray}
&&g^{B-L}_{h^n_{S}H^{\pm}H^{\pm}}=-\frac{v}{2m_Z^2}\sum_{i=j=1}^2\Big[\frac{1}{4} \Big(-2 g_{YB} g_{B}(- v_{\bar{\eta}} Z_{{n 4}}^{H} + v_{\eta} Z_{{n 3}}^{H})(Z_{{j 1}}^{+} Z_{{i 1}}^{+} - Z_{{j 2}}^{+} Z_{{i 2}}^{+})\nonumber\\
&&+\lambda_2\lambda^*(v_{\eta} Z_{{n 4}}^{H} + v_{\bar{\eta}} Z_{{n 3}}^{H})(Z_{{j 2}}^{+} Z_{{i 1}}^{+} +Z_{{j 1}}^{+} Z_{{i 2}}^{+})+Z_{{n 5}}^{H}\big(2v_S\lambda^*\lambda(Z_{{j 1}}^{+} Z_{{i 1}}^{+} \nonumber\\
&&+ Z_{{j 2}}^{+} Z_{{i 2}}^{+})+(2v_S\lambda^*\kappa+\sqrt{2}T_\lambda^*)Z_{{j 2}}^{+} Z_{{i 1}}^{+} +(2v_S\lambda\kappa^*+\sqrt{2}T_\lambda)Z_{{j 1}}^{+} Z_{{i 2}}^{+}\big)\nonumber\\
&&- Z_{{n 1}}^{H} \Big(Z_{{j 2}}^{+} \Big(-(-g_{2}^{2}+g_{1}^{2}+g_{Y B}^{2})v_d Z_{{i 2}}^{+}+(-2|\lambda|^2+g_{2}^{2}) v_u Z_{{i 1}}^{+} \Big)\nonumber \\
&&\hspace{0.8cm}+Z_{{j 1}}^{+} \Big((g_{1}^{2}+g_{Y B}^{2}+g_{2}^{2})v_d Z_{{i 1}}^{+}+(-2|\lambda|^2+g_{2}^{2}) v_u Z_{{i 2}}^{+}\Big)\Big)\nonumber\\
&&+Z_{{n 2}}^{H}\Big(Z_{{j 1}}^{+} \Big((-g_{2}^{2}+ g_{1}^{2}+g_{Y B}^{2})v_u Z_{{i 1}}^{+}-(-2|\lambda|^2+g_{2}^{2}) v_d Z_{{i 2}}^{+} \Big)\nonumber \\
&&\hspace{0.8cm}-Z_{{j 2}}^{+}\Big((g_{1}^{2}+g_{Y B}^{2}+g_{2}^{2}))v_u Z_{{i 2}}^{+}+ (-2|\lambda|^2+g_{2}^{2}) v_d Z_{{i 1}}^{+} \Big)\Big)\Big)\Big].
\end{eqnarray}

\section{the form factors}
The form factors are defined as
\begin{eqnarray}
&&A_{1/2} (x)=2\Big[x+(x-1)g(x)\Big]/x^2,\;A_0(x)=-(x-g(x))/x^2,\nonumber\\
&&A_1(x)=-\Big[2x^2+3x+3(2x-1)g(x)\Big]/x^2,\;A_2(x)=2 g(x)/x,\nonumber\\
&&g(x)=\left\{\begin{array}{l}\arcsin^2\sqrt{x},\;\;\;\;\;\;\;\;\;\;\;\;\;\;\;\;\;\;\;\;x\le1\\
-{1\over4}\Big[\ln{1+\sqrt{1-1/x}\over1-\sqrt{1-1/x}}-i\pi\Big]^2,\;x>1\;,\end{array}\right.
\nonumber\\
&&F(x)=-(1-x^2)(\frac{47}{2}x^2-\frac{13}{2}+\frac{1}{x^2})-3(1-6x^2+4x^4)\ln x \nonumber \\&&\hspace{1.5cm}+\frac{3(1-8x^2+20x^4)}{\sqrt{4x^2-1}}\arccos\Big(\frac{3x^2-1}{2x^3}\Big).
\label{g-function}
\end{eqnarray}


\begin{thebibliography}{99}
\bibitem{h0CMS}S. Chatrchyan et al. (CMS Collaboration) Phys. Lett. B {\bf716} (2012) 30-61.
\bibitem{h0ATLAS}G. Aad et al. (ATLAS Collaboration) Phys. Lett. B {\bf716} (2012) 1-29.
\bibitem{PDG2024}S. Navas et al. (Particle Data Group) Phys. Rev. D {\bf110} (2024) 030001.

\bibitem{h02gamma}G. Aad et al. (ATLAS Collaboration) JHEP {\bf07} (2023) 088.
\bibitem{h02gamma2W2Z2b2tau1}A. Tumasyanet al. (CMS Collaboration) Nature {\bf607} (2022) 7917, 60-68.
\bibitem{h02gamma2W2Z2b2tau2}G. Aad et al. (ATLAS and CMS Collaboration) JHEP {\bf08} (2016) 045.
\bibitem{h02gamma2W2b2tau}T. Aaltonen, et al. (CDF and D0 Collaborations) Phys. Rev. D {\bf88} (2013) 5, 052014.
\bibitem{h02Z}G. Aad et al. (ATLAS Collaboration) Eur. Phys. J. C {\bf80} (2020) 10, 957.
\bibitem{h02b1}G. Aad et al. (ATLAS Collaboration) Eur. Phys. J. C {\bf81} (2021) 2, 178.
\bibitem{h02b2}G. Aad et al. (ATLAS Collaboration) Eur. Phys. J. C {\bf81} (2021) 6, 537.
\bibitem{h02tau}M. Aaboud et al. (ATLAS Collaboration) Phys. Rev. D {\bf99} (2019) 072001.

\bibitem{LEP1}G. Abbiendi et al. (OPAL Collaboration) Eur. Phys. J. C {\bf27} (2003) 311.
\bibitem{LEP2}R. Barate et al. (LEP Working Group for Higgs boson searches and ALEPH and DELPHI and L3 and OPAL Collaborations) Phys. Lett. B {\bf565} (2003) 61-75.
\bibitem{LEP3}S. Schael et al. (ALEPH and DELPHI and L3 and OPAL and LEP Working Group for Higgs Boson Searches Collaborations) Eur. Phys. J. C {\bf47} (2006) 547.
\bibitem{Tevatron}CDF and D0 Collaborations FERMILAB-CONF-12-318-E, 2012.

\bibitem{95expCMS1}CMS Collaboration, Report No. CMS-PAS-HIG-20-002, 2023.
\bibitem{95expCMS2}A. Tumasyan et al. (CMS Collaboration) JHEP {\bf07} (2023) 073.
\bibitem{95expCMS3}A. M. Sirunyan et al. (CMS Collaboration) Phys. Lett. B {\bf793} (2019) 320-347.
\bibitem{95expCMS4}A. M. Sirunyan et al. (CMS Collaboration) JHEP {\bf09} (2018) 007.
\bibitem{95expATLAS}ATLAS Collaboration, Report No. ATLAS-CONF-2018-025, 2018.

\bibitem{CPMSSM1}A. Pilaftsis and C. E. M. Wagner, Nucl. Phys. B {\bf553} (1999) 3-42.
\bibitem{CPMSSM2}M. Carena, J. Ellis, A. Pilaftsisa and C. E. M. Wagner, Nucl. Phys. B {\bf586} (2000) 92-140.

\bibitem{MSSM1}H. P. Nilles, Phys. Rept. {\bf110} (1984) 1.
\bibitem{MSSM2}H. E. Haber and G. L. Kane, Phys. Rept. {\bf117} (1985) 75.
\bibitem{MSSM3}J. Rosiek, Phys. Rev. D {\bf 41} (1990) 3464.
\bibitem{MSSM4}T. F. Feng and X. Y. Yang, Nucl. Phys. B {\bf 814} (2009) 101.
\bibitem{NMSSM1}S. Moretti and S. Munir, Eur. Phys. J. C {\bf47} (2006) 791-803.
\bibitem{NMSSM2}U. Ellwanger, Phys. Lett. B {\bf698} (2011) 293-296.
\bibitem{NMSSM3}J. Cao, Z. Heng, T. Liu and J. M. Yang, Phys. Lett. B {\bf703} (2011) 462-468.
\bibitem{NMSSM4}D. Albornoz Vasquez, G. Belanger, C. Boehm, et al.,
 Phys. Rev. D {\bf86} (2012) 035023.
\bibitem{NMSSM5}U. Ellwanger and C. Hugonie, Adv. High Energy Phys. {\bf2012} (2012) 625389.
\bibitem{NMSSM6}F. Boudjema and G. D. La Rochelle, Phys. Rev. D {\bf86} (2012) 115007.
\bibitem{NMSSM7}K. Schmidt-Hoberg and F. Staub, JHEP {\bf10} (2012) 195.
\bibitem{NMSSM8}M. Badziak, M. Olechowski and S. Pokorski, JHEP {\bf06} (2013) 043.
\bibitem{NMSSM9}M. Badziak, M. Olechowski and S. Pokorski, PoS EPS-HEP {\bf2013} (2013) 257.
\bibitem{NMSSM10}R. Barbieri, D. Buttazzo, K. Kannike, et al., Phys. Rev. D {\bf88} (2013) 055011.
\bibitem{NMSSM11}J. W. Fan, J. Q. Tao, Y. Q. Shen, et al., Chin. Phys. C {\bf38} (2014) 073101.
\bibitem{NMSSM12}C. T. Potter, Eur. Phys. J. C {\bf76} (2016) 1, 44.
\bibitem{NMSSM13}U. Ellwanger and M. Rodriguez-Vazquez, JHEP {\bf02} (2016) 096.
\bibitem{NMSSM14}J. Cao, X. Guo, Y. He, et al., Phys. Rev. D {\bf95} (2017) 11, 116001.
\bibitem{NMSSM15}J. Cao, X. Jia, Y. Yue, et al., Phys. Rev. D {\bf101} (2020) 5, 055008.
\bibitem{95NMSSM1}J. J. Cao, X. L. Jia, J. W. Lian and L. Meng, Phys. Rev. D {\bf109} (2024) 7, 075001.
\bibitem{95NMSSM2}J. J. Cao, X. L. Jia and J. W. Lian, Phys. Rev. D {\bf110} (2024) 11, 115039.
\bibitem{munuSSM1}T. Biek\"{o}tter, S. Heinemeyer, and C. Mu\~{n}oz, Eur. Phys. J. C {\bf78} (2018) 504.
\bibitem{munuSSM2}T. Biek\"{o}tter, S. Heinemeyer, and C. Mu\~{n}oz, Eur. Phys. J. C {\bf79} (2019) 667.
\bibitem{munuSSM3}C. X. Liu, Y. Zhou, X. Y. Zheng, et al. Phys. Rev. D {\bf109} (2024) 056001.
\bibitem{95N2HDM1}T. Biek\"{o}tter, S. Heinemeyer and G. Weiglein, JHEP {\bf08} (2022) 201.
\bibitem{95N2HDM2}T. Biek\"{o}tter, S. Heinemeyer and G. Weiglein, Phys. Lett. B {\bf846} (2023) 138217.
\bibitem{95N2HDM3}T. Biek\"{o}tter, M. Chakraborti and S. Heinemeyer, Eur. Phys. J. C {\bf80} (2020) 1, 2.
\bibitem{95N2HDM4}T. Biek\"{o}tter and M. O. Olea-Romacho, JHEP {\bf10} (2021) 215.
\bibitem{95N2HDM5}T. Biek\"{o}tter, A. Grohsjean, S. Heinemeyer, et al., Eur. Phys. J. C {\bf82} (2022) 2, 178.
\bibitem{95N2HDM6}S. Heinemeyer, C. Li, F. Lika, et al., Phys. Rev. D {\bf106} (2022) 7, 075003.
\bibitem{95N2HDM7}T. Biek\"{o}tter, S. Heinemeyer and G. Weiglein, Eur. Phys. J. C {\bf83} (2023) no.5, 450.
\bibitem{95N2HDM8}D. Azevedo, T. Biek\"{o}tter and P. M. Ferreira, JHEP {\bf11} (2023) 017.
\bibitem{95N2HDM9}J. A. Aguilar-Saavedra, H. B. C\^{a}mara, F. R. Joaquim and J. F. Seabra, Phys. Rev. D {\bf108} (2023) 7, 075020.

\bibitem{U1X}Z. F. Ge, F. Y. Niu and J. L. Yang, Eur. Phys. J. C {\bf84} (2024) 548.
\bibitem{U1X1}S. Gao, S. M. Zhao, S. Di, et al., Nucl. Phys. B {\bf1018} (2025) 117026.

\bibitem{mB-L1}V. Barger, P. F. Perez and S. Spinner, Phys. Rev. Lett. {\bf102} (2009) 181802.
\bibitem{mB-L2}P. F. Perez and S. Spinner, JHEP {\bf04} (2012) 118.
\bibitem{mB-L3}V. Barger, P. F. Perez and S. Spinner, Phys. Lett. B {\bf696} (2011) 509-512.
\bibitem{mB-L4}P. F. Perez and S. Spinner, Phys. Lett. B {\bf673} (2009) 251-254.
\bibitem{mB-L5}P. F. Perez and S. Spinner, Phys. Rev. D {\bf80} (2009) 015004.

\bibitem{NB-LSSM1}W. Ahmed, S. Raza Q. Shafi et al. JHEP {\bf01} (2021) 161.
\bibitem{NB-LSSM2}X. Y. Han, J. Ma, S. M. Zhao et al. Eur. Phys. J. C {\bf85} (2025) 2, 1632025.

\bibitem{B-LSSM3}M. Ambroso and B. A. Ovrut, Int. J. Mod. Phys. A {\bf26} (2011) 1569.
\bibitem{B-LSSM4}P. F. Perez and S. Spinner, Phys. Rev. D {\bf83} (2011) 035004.

\bibitem{B-L R Parity}C. S. Aulakh, A. Melfo, A. Rasin and G. Senjanovic, Phys. Lett. B {\bf459} (1999) 557.

\bibitem{B-L hierarchy1}W. Abdallah, A. Hammad, S. Khalil and S. Moretti, Phys. Rev. D {\bf95} (2017) 055019.
\bibitem{B-L hierarchy2}J. L. Yang, T. F. Feng and H. B. Zhang, Eur. Phys. J. C {\bf80} (2020) 210.

\bibitem{B-LDM1}S. Khalil and H. Okada, Phys. Rev. D {\bf79} (2009) 083510.
\bibitem{B-LDM2}L. Basso, B. O'Leary, W. Porod and F. Staub, JHEP {\bf1209} (2012) 054.
\bibitem{B-LDM3}L. D. Rose, S. Khalil, S. J. D. King et al., Phys. Rev. D {\bf96} (2017) 055004.
\bibitem{B-LDM4}L. D. Rose, S. Khalil, S. J. D.King et al., JHEP {\bf07} (2018) 100.

\bibitem{CP1}J. H. Christenson, J. W. Cronin, V.L. Fitch and R. Turlay, Phys. Rev. Lett. {\bf13} (1964) 138.
\bibitem{CP2}W. Grimus, Fortschr. Phys. {\bf36} (1988) 201.
\bibitem{CP3}B. Winstein and L. Wolfenstein, Rev. Mod. Phys. {\bf65} (1993) 1113-1148.

\bibitem{signal strength ratios}A. Arbey, A. Deandrea, F. Mahmoudi and A. Tarhini, Phys. Rev. D {\bf87} (2013) 115020.

\bibitem{CPHiggs dacay}A. Djouadi, Phys. Rept. {\bf459} (2008) 1-241.
\bibitem{h2glon2gamma1}J. R. Ellis, M. K. Gaillard, D. V. Nanopoulos, Nucl. Phys. B {\bf106}
(1976) 292.
\bibitem{h2glon2gamma2}M. A. Shifman, A. I. Vainshtein, M. B. Voloshin and V. I. Zakharov, Sov. J. Nucl. Phys. {\bf30} (1979) 711-716.
\bibitem{h2glon2gamma3}L. Bergstrom, G. Hulth, Nucl. Phys. B {\bf259} (1985) 137-155.

\bibitem{h2W2Z}P. Gonzaleza, S. Palmerb, M. Wiebusch, K. Williamsd, Eur. Phys. J. C {\bf73} (2013) 2367.

\bibitem{Zpupper}G. Aad et al., (ATLAS Collaboration) Phys. Lett. B {\bf796} (2019) 68-87.

\bibitem{Zpupper1}G. Cacciapaglia, C. Csaki, G. Marandella et al., Phys. Rev. D {\bf74} (2006) 033011.
\bibitem{Zpupper2}M. Carena, A. Daleo, B. A. Dobrescu et al, Phys. Rev. D {\bf70} (2004) 093009.

\bibitem{BSgamma1}F. Mamoudi, JHEP {\bf12} (2007) 026.
\bibitem{BSgamma2}K. A. Olive and L. Velasco-Sevilla, JHEP {\bf05} (2008) 052.

\bibitem{tanB}L. Basso, Adv. High Energy Phys. {\bf2015} (2015) 980687.

\bibitem{ATLAS2024hya}G. Aad et al., ATLAS Collaboration, Phys. Rev. D {\bf111} (2025) no.7, 072006.
\bibitem{squarks1}ATLAS Collaboration, Phys. Rev. D {\bf87} (2013) 012008.
\bibitem{squarks2}CMS Collaboration, JHEP {\bf1210} (2012) 018.
\bibitem{squarks3}C. S. Un and O. Ozdal, Phys. Rev. D {\bf93} (2016) 055024.
\bibitem{LFV Higgs decay}C. Guo, X. X. Dong, S. M. Zhao, et al., Eur. Phys. J. C {\bf85} (2025) 10, 1106.
\bibitem{EDM}X. X. Dong, S. M. Zhao, H. B. Zhang and T. F. Feng, J. Phys. G {\bf47} (2020) 4, 045002.
\bibitem{2deltaV1} R. J. Zhang, Phys. Lett. B {\bf447} (1999) 89-97.
\bibitem{2deltaV2} J. R. Espinosa and R. J. Zhang, JHEP {\bf03} (2000) 026.
\bibitem{2deltaV3} G. Degrassi, P. Slavich and F. Zwirner, Nucl. Phys. B {\bf611} (2001) 403-422.
\bibitem{2deltaV4}J. L. Yang, M. H. Guo, W. H. Zhang, et al., arXiv:2406.01926 [hep-ph].

\end{thebibliography}
\end{document}